%% file: HyperonReview.tex
\documentclass[12pt]{iopart}
\usepackage{graphicx,url}
\usepackage{color,datetime2,multirow,ulem}

\newcommand{\bl}[1]{{\color{blue} #1}}

\newcommand{\rb}[1]{\raisebox{1.5ex}[-1.5ex]{#1}}

\newcommand{\rbb}[1]{\raisebox{2ex}[-2ex]{#1}}
\def\beq{\begin{equation}}
\def\eeq{\end{equation}}

\def\beqy{\begin{eqnarray}}
\def\eeqy{\end{eqnarray}}
\begin{document}

\review[A Review of Hyperon Spectroscopy]{70 Years of Hyperon Spectroscopy:\\[0.5ex] A review of strange $\Xi$, $\Omega$ baryons, and the spectrum of charmed and bottom baryons} 

\author{Volker Crede}
\address{Florida State University, Department of Physics, Tallahassee, FL 32306, USA}
\ead{\mailto{crede@fsu.edu}}

\author{John Yelton}
\address{University of Florida, Department of Physics, Gainesville, FL 32611, USA}
\ead{\mailto{yelton@ufl.edu}}
%\vspace{10pt}
%\begin{indented}
%\item[]January 2022
%\end{indented}

\date{\today}

\begin{abstract}
The first hyperon was discovered about 70 years ago, but the nature of
these particles, particularly with regard to multistrange hyperons, and many of their properties can still be considered to be literally strange. A dedicated and successful global spectroscopy program in the 1960s and 1970s using $K^-$~beams revealed many multistrange candidates, but the available evidence of their existence  
is statistically limited. For this reason,
%Experimentally, given the lack of strange probes in recent
%decades of advanced computing technology and electronics,
there is still much to learn about the systematics of the spectrum of excited hyperon states and what
they have in common with their non-strange companions, or how they
differ from the nucleon and $\Delta$~resonances. Results from photo-
and electroproduction experiments off the proton and neutron using
polarized beams and targets have provided intriguing evidence for new
nucleon excitations and shed light on the structure of some of the
known nucleon and $\Delta$~states. Recent years have also seen a great
deal of progress in the field of charmed and bottom baryon
spectroscopy. Unprecedented data from the Large Hadron Collider in
particular indicate continued rapid progress in the field of bottom
baryons. On the theoretical side, baryons with one heavy quark~$Q$ and
a light $qq$~system serve as an ideal laboratory for studying light
$qq$~(diquark) correlations and the dynamics of the light quarks in the colour
environment of a heavy quark. In this review, we discuss the status of
doubly and triply strange $\Xi$ as well as $\Omega$~baryons, and the
properties of all the known charmed and bottom states. The comparison
of the two heavy sectors reveals many similarities as predicted by
heavy-quark symmetries, together with differences in mass splittings
easily understood by potential models. The multi-strange hyperons
bridge the under-explored gap between the light- and the heavy-flavour
baryons. How do the properties of a singly charmed $Q$-$qq$~system
change with decreasing mass of the heavy quark in the transition to a
doubly strange $q$-$QQ$~system with a heavier quark-quark system relative to one
light quark? Significant progress towards understanding hyperon
resonances is expected in coming years from the ongoing experiments at
the high-energy collider facilities and planned experiments using
$K$~beams at Jefferson Laboratory and J-PARC.
\end{abstract}

\maketitle

%\section{Introduction}
\input{Introduction}

%\section{Baryon Spectroscopy}
\input{BaryonSpectroscopy}

%\section{Experimental Methods}
\input{ExperimentalMethods}

%\section{Light-quark baryons}
\input{LightQuarkBaryons}

%\section{Experiments for Studying Charmed and Bottom Baryons}
%\input{ExperimentsHeavy}

%\section{Theoretical Models for Heavy Baryons}
%\include{ModelsHeavy}

\section{Experimental Review of Charmed Baryons}
\input{CharmedBaryons}

\subsection{The $\Omega_c^0$ baryon}
\input{CharmedOmegaBaryons}

\section{An Experimental Review of B Baryons}
\input{BottomBaryons}

%\section{Discussion and Open Questions}
\input{Discussion}

\ack

We acknowledge fruitful discussions with and valuable feedback from
Simon Capstick, Sean Dobbs, Gernot Eichmann, Christian Fischer,
Eberhard Klempt, Matthias Lutz, Eulogio Oset, and Daniel Mohler. Special thanks to
Jesse Hernandez and Chandra S. Akondi for preparing figures and to
Stephanie A. Leitch for carefully proofreading the manuscript. This
work was partially supported by the Department of Energy, Office of
Science, Office of Nuclear Physics under Contract
No. DE-FG02-92ER40735, and office of High Energy Physics under
contract DE-SC0009824.
\section*{References}
\bibliography{References}{}
\bibliographystyle{iopart-num}

\end{document}

%% file: Introduction.tex
\section{Introduction}
The proton has been known as a fundamental building block of all atomic nuclei for more than a hundred 
years ever since Ernest Rutherford performed his famous early scattering experiments in the 1910s~\cite{Rutherford:1911zz}. Initially 
considered a fundamental or elementary particle, the proton is now established as a composite object made 
up of massless gluons and almost massless quarks. Protons 
and neutrons, the electrically neutral partner of the proton, are jointly referred to as nucleons. In a broader picture, nucleons are members of a family 
of particles called hadrons -- strongly interacting particles composed of quarks and gluons. Hadrons are
observed in nature as fermions (half-integer spin particles) and bosons (integer-spin particles), and classified as {\it baryons} and {\it mesons}, respectively. The 
ground-state nucleons are spin-$\frac{1}{2}$ fermions and thus, the half-integer spin makes them baryons. The modern perspective 
considers a baryon composed of three valence quarks and transitory pairs of sea quarks held together by 
the strong force which is mediated by gluons. The discovery of the top quark in the early 1990s at Fermilab~\cite{CDF:1995wbb,D0:1995jca} has completed the picture of three quark families in the Standard Model of Particle Physics, which is based on a total
of six different quark flavours: three light flavours -- u~(up), d~(down), s~(strange) -- and three heavy 
flavours -- c~(charm), b~(bottom), t~(top).

The multitude of quark flavours gives rise to a rich spectrum of excited baryons. However, a better 
understanding of the baryon as a bound state of quarks and gluons remains a fundamental challenge in hadron 
physics. Similar to the conundrum of the hydrogen atom in the early years of the 20th century, the understanding
of the structure of a bound state and of its excitation spectrum needs to be addressed simultaneously. The 
study of excited baryons is therefore complementary to understanding the structure of the nucleon in deep
inelastic scattering experiments that provide access to the properties of its fundamental constituents in 
the ground state. Such scattering experiments were first attempted in the 1960s and 1970s and provided the 
first convincing evidence of the reality of quarks. The Nobel Prize of 1990 in Physics was awarded to Kendall, 
Friedman, and Taylor ``for their pioneering investigations concerning deep inelastic scattering of electrons 
on protons and bound neutrons, which have been of essential importance for the development of the quark model 
in particle physics''~\cite{NobelPrizePhysics1990}. The findings of those years led to the development of Quantum Chromodynamics 
(QCD) as the fundamental theory of the strong nuclear force~\cite{Gross:2022hyw}, and the quark model picture of hadrons in the 
non-perturbative regime of QCD. The early quark model served as a successful classification scheme for hadrons and in 1964, the discovery of the "predicted" triply strange $\Omega^-$~baryon was a decisive step in firmly establishing SU(3)$_{\rm \,flavour}$ as the underlying symmetry in these models. From a modern perspective, the conventional light-flavour quark model treats the baryon as a system of three symmetric quark degrees of 
freedom interacting in a long-range Coulomb-like confinement potential. However, the short-range interactions 
between the quarks vary in different, more recent approaches ranging from a description in terms of one-gluon exchange
between the quarks, e.g. in Ref.~\cite{Capstick:1986ter}, to a more complex description using instanton-induced interactions, e.g. in Refs.~\cite{Loring:2001kv,Loring:2001kx,Loring:2001ky}. The exchange of pseudoscalar mesons between quarks has also been explored in these quark models. The idea was originally investigated in the early 1980s by a group at Tokyo University~\cite{Shimizu:1984iel} and was later established in the 1990s as an alternative approach, not only to study the low-energy spectrum of baryons but also to study the nucleon-nucleon interaction, see e.g. Refs.~\cite{Fernandez:1993hx,Valcarce:1995dm} for more details. Results on the spectrum of baryons using boson-exchange interactions between quarks were also reported in Refs.~\cite{Glozman:1996wq,Dziembowski:1996cv,Valcarce:2005rr}.

This picture of the baryon has been amazingly successful in many aspects but still lacks a fundamental 
understanding of its connection with QCD. The simple quark model picture is also challenged by the proton
spin crisis that precipitated in 1987. An experiment by the European Muon Collaboration (EMC)~\cite{EuropeanMuon:1987isl}, which tried 
to determine the distribution of spin within the proton, revealed that the total proton spin carried by the
quarks was far smaller than 100\,\% and thus, not consistent with the simple picture based on the proton's 
$|\uparrow\uparrow\downarrow\,\rangle$~constituent quarks. Subsequent work has shown that the proton spin 
originates not only from the quark spins, but also dominantly from the quark and gluon orbital angular momentum. 
While the high-energy scattering experiments of the 60s and 70s played a crucial role in the discovery of the
quarks, new physics concepts and accelerator technology were required for a detailed study of the behavior of 
quarks in nuclei. Nucleon structure experiments now continue for example at modern facilities, like Jefferson 
Laboratory (JLab) using the excellent quality of the beam at the Continuous Electron Beam Accelerator Facility (CEBAF), and aim at extracting structure functions in the 
deeply virtual production of mesons off the nucleon. Detailed reviews on the spin structure of the nucleon are given in Refs.~\cite{Deur:2018roz,Diehl:}.

Unfortunately, the collective degrees of freedom are lost in deep inelastic scattering experiments. For this 
reason, baryon spectroscopy provides complementary information on the existence and properties of excited 
baryon states. The main goals of recent experiments at various facilities are the determination of the excited 
baryon spectrum for light- and heavy-flavoured baryons and the identification of possibly new symmetries in 
the spectrum. A plethora of new information on $N^\ast$ and $\Delta$~resonances has been collected over recent 
years at facilities worldwide such as Jefferson Laboratory in the United States~\cite{CLAS:2003umf}, the Electron Stretcher 
Accelerator (ELSA)~\cite{Hillert:2006yb}, the Mainz Microtron (MAMI)~\cite{Mecking:2006yh}, and the Grenoble Anneau Accelerateur Laser (GRAAL) facility 
in Europe, e.g. Ref.~\cite{GRAAL:2005mor}, as well as the 8-GeV Super Photon Ring (SPring-8) in Japan hosting the Laser Electron Photon 
Experiment (LEPS)~\cite{LEPS:2013tbd}. The goal of the global $N^\ast$~program has been to perform so-called complete experiments 
that would allow for the extraction of the scattering amplitude without ambiguities in electromagnetically induced 
reactions off the proton. The accumulated data sets include cross section data and polarization observables 
for a large variety of final states, such as $\pi N$, $\eta N$, $\omega N$, $\pi\pi N$, $K \Lambda$, 
$K \Sigma$, etc. These data complement results from earlier experiments for similar final states in $\pi$- 
and $K$-induced reactions. In a very brief summary, several new excited nucleon states have been proposed based on the recent high-statistics photoproduction 
data~\cite{ParticleDataGroup:2022pth}. Table~\ref{Table:NewResonances} shows the group of six additional nucleon resonances
now listed by the Particle Data Group (PDG) and their evidence in various decay modes. The addition of the $N(1880)\,1/2^+$
state is particularly interesting since four states, $N(1880)\,1/2^+$, $N(1900)\,3/2^+$, $N(2000)\,5/2^+$,
$N(1990)\,7/2^+$, are now considered to form a previously missing quartet of nucleon resonances with spin~$\frac{3}{2}$~\cite{Anisovich:2011su,CBELSATAPS:2015kka,CBELSATAPS:2015taz} and to be members
of the $({\bf 70},\,2_2^+)$~supermultiplet (see Section~\ref{Subsection:LightBaryonSpectroscopy} for more discussion on the multiplet structure of the baryon spectrum). This group of resonances has previously been predicted by traditional quark models based on three symmetric quark degrees of freedom. However, the identification of these resonances is incompatible with the static diquark-quark picture of the nucleon, with excitations in the diquark frozen out, since the $({\bf 70},\,2_2^+)$ requires excitations between all three valence quarks. An example of a static diquark model is discussed in Ref.~\cite{Ferretti:2011zz}. Recent reviews of the progress toward understanding the baryon spectrum are available in Refs.~\cite{Gross:2022hyw,Ireland:2019uwn}, for instance.

\begin{table}[t]
%\addtolength{\extrarowheight}{4pt}
%\centering
\caption{The new nucleon resonances listed by the Particle Data Group~\cite{ParticleDataGroup:2022pth} and their status in various decay modes.\\
\makebox[1.1cm][l]{$\ast\ast\ast\,\ast$} Existence is certain, and properties are at least fairly well explored.\\ \makebox[1.1cm][l]{$\ast\ast\,\ast$} Existence ranges from very likely to certain, but further confirmation is\\ \makebox[1.1cm][l]{} desirable and/or quantum numbers, branching fractions, etc. are not well\\ \makebox[1.1cm][l]{} determined.\\ \makebox[1.1cm][l]{$\ast\,\ast$} Evidence of existence is only fair.\\ \makebox[1.1cm][l]{$\ast$} Evidence of existence is poor.}
\label{Table:NewResonances}
%\vspace{1mm}
%\footnotesize
%\begin{ruledtabular}
\begin{center}
\begin{tabular}{ccccc|ccc|ccc|ccc}
\hline\\[-2.5ex]
State & $J^P$ & overall & & $N\pi$ & $N\eta$ & & $N\eta^{\,\prime}$ & $N\rho$ & & $N\omega$ & $\Lambda K$ & & $\Sigma K$\\\hline \\[-2ex]
$N(1860)$ & $5/2^+$ & $\ast\,\ast$ & & $\ast\,\ast$ & $\ast$ & & & & & & & &\\[0.5ex]
$N(1875)$ & $3/2^-$ & $\ast\ast\ast$ & & $\ast\,\ast$ & $\ast$ & & $\ast$ & $\ast$ & & $\ast$ & $\ast$ & & $\ast$\\[0.5ex]
$N(1880)$ & $1/2^+$ & $\ast\ast\ast$ & & $\ast$ & $\ast\,\ast$ & & & & & $\ast\,\ast$ & $\ast\,\ast$ & & $\ast\,\ast$\\[0.5ex]
$N(1895)$ & $1/2^-$ & $\ast\ast\ast\,\ast$ & & $\ast$ & $\ast\ast\ast\,\ast$ & & $\ast\ast\ast\,\ast$ & $\ast$ & & $\ast$ & $\ast\,\ast$ & & $\ast\,\ast$\\[0.5ex]
$N(2080)$ & $5/2^-$ & $\ast\ast\ast$ & & $\ast\,\ast$ & $\ast$ & & $\ast$ & $\ast$ & & $\ast$ & $\ast$ & & $\ast$\\[0.5ex]
$N(2120)$ & $3/2^-$ & $\ast\ast\ast$ & & $\ast\ast\ast$ & & & $\ast$ & & & $\ast$ & $\ast\,\ast$ & & $\ast$\\[0.5ex]\hline
\end{tabular}
\end{center}
%\end{ruledtabular}
\end{table}

On the other hand, a multitude of new heavy-flavoured baryons has been observed at facilities such as the 
Large Hadron Collider (LHC) in Europe~\cite{Evans:2008zzb} and the Belle Experiment at the KEK B-Factory in Japan~\cite{Kurokawa:2001nw}, including 
the remarkable first unambiguous observation of a doubly charmed particle~\cite{LHCb:2017iph}. The identification of new heavy 
baryons is easier in one particular aspect as compared to $N^\ast$~resonances since almost all heavy-flavoured 
baryons can be observed as peaks in a mass spectrum, whereas the identification of light-flavoured baryon 
resonances is more challenging due to their broad and overlapping nature. Unlike for heavy baryons, peak-hunting is not an option in light-flavour baryon spectroscopy. 

Baryon spectroscopy has made great leaps forward in both the light and heavy-mass sectors. Many fundamental 
questions remain unanswered, though. How does QCD give rise to excited baryons? How is the mass of these 
states generated through the dynamics of the quarks and gluons? The typical mass of a baryon is about two 
orders of magnitude greater than the sum of the rest masses of the three valence quarks, while the gluons 
have zero rest mass. What is the number of the relevant degrees of freedom required to describe excited 
baryons? Is the three-valence-quark picture still valid for highly excited states or do symmetries emerge
that are inconsistent with this description? Recent results from lattice-QCD indicate that the spectrum 
exhibits the broad features expected from wave functions based on the irreducible representations of 
SU(6)\,$\otimes$\,O(3)~\cite{Edwards:2012fx}. The counting of states of each flavour and spin appears consistent with the traditional 
quark model, at least for the lowest negative- and positive-parity bands. However, a large part of the 
baryon spectrum remains experimentally unobserved, in particular for hyperons, and while recent advances 
in lattice-QCD and the availability of large-scale computing technology make numerical solutions of QCD 
now possible, the nature of these resonances and the presence of thresholds in strong-coupling QCD complicate 
the extraction of states from such calculations. The available computing power also still limits the use of 
realistic pion masses. To complete the list of motivating questions: What is the mechanism responsible for 
confinement and chiral symmetry breaking? How are the constituents (or the constituent quarks dressed with 
their clouds of gluons and quark-antiquark pairs) related to the quark and gluon fields of the underlying 
QCD Lagrangian? Or how does the chiral symmetry structure of QCD lead to the dressed quarks and produce 
the long-distance behaviour observed as the spectrum of hadrons? 

This paper builds upon other recent review articles of light-flavour $N^\ast$~states and summarizes the status and the new data for doubly, as well as triply, strange light-flavour resonances, and presents the new information on charmed and bottom baryons. 

\subsection{Guide to the literature}
\label{Subsection:Literature}
The Particle Data Group (PDG) maintains and regularly updates the listings of particles based on new 
measurements and includes mini-reviews on a large number of different topics in their Review of Particle 
Physics (RPP), which have been published biennially for many decades. Fairly detailed mini-reviews on 
$N^\ast$ and $\Delta$~resonances as well as charmed baryons can be found in the latest edition of the RPP~\cite{ParticleDataGroup:2022pth}. 
Less progress has been reported in recent years on singly- ($\Lambda$, $\Sigma$), doubly- ($\Xi$), and triply 
($\Omega$) strange resonances due to the lack of a suitable $K$-beam facility in the world.

An older article by Hey and Kelly~\cite{Hey:1982aj} still provides useful information, in particular on some 
aspects of the theoretical data analysis. A very comprehensive review of baryon spectroscopy is given by 
Klempt and Richard in their 2009 article, which also discusses the prospects for experiments using 
electromagnetic probes~\cite{Klempt:2009pi}. In fact, some of the questions raised in their review have been addressed by recent
experiments. Quark model developments have been discussed in Ref.~\cite{Capstick:2000qj} by Capstick and Roberts. Additionally, a 
historical overview of quark model approaches and other aspects of nucleon structure, as well as light- and heavy-flavour baryon spectroscopy, is given by various authors in Ref.~\cite{Gross:2022hyw} on {\it 50 Years of Quantum Chromodynamics}.

About two decades ago, Krusche and Schadmand presented a first nice summary on low-energy photoproduction~\cite{Krusche:2003ik} 
and in 2007, Drechsel and Walcher reviewed hadron structure at low~$Q^2$~\cite{Drechsel:2007sq}. The work of Tiator {\it et al.}~\cite{Tiator:2011pw}, as well as
Aznauryan and Burkert~\cite{Aznauryan:2011qj}, has provided more recent aspects on the electroexcitation of nucleon resonances. The 
list of recent reviews on baryon spectroscopy also includes a 2013 article by Crede and Roberts~\cite{Crede:2013kia}, who looked 
at the experimental developments of both light- and heavy-flavour resonances and discussed theoretical 
approaches. An article by Ireland, Pasyuk, and Strakovsky~\cite{Ireland:2019uwn} reviewed photoproduction reactions and 
non-strange baryon spectroscopy. Finally, an article by Thiel, Afzal, and Wunderlich~\cite{Thiel:2022xtb} focused on amplitude analysis used to identify light-flavour baryons and to study their properties.   

Predictions of excited charmed baryons started soon after the discovery of the $\Lambda_c^+$~resonance, with Copley, Isgur and Karl predicting the spectrum by extrapolating from the strange sector in 1979~\cite{Copley:1979wj}. The paper by Captick and Isgur~\cite{Capstick:1986ter} tabulated a large number of expected states using a relativized model, commenting that the changes from 
previous nonrelativistic models are not large. In 1996, Silvestre-Brac~\cite{Silvestre-Brac:1996myf} used several interquark potential models to predict the complete spectrum of all 
the heavy quark baryons.
In recent years, alternative models based on charmed baryons being modeled as baryon-meson 
molecular states have gained in popularity after some success in the light and strange baryon sector, for instance~\cite{Garcia-Recio:2012lts}.

The production of charmed baryons at BES\,III was discussed by Cheng in 2009~\cite{Cheng:2009zzc} and the overall experimental progress on charmed baryon physics and the theoretical developments have been reviewed by Cheng in 2015~\cite{Cheng:2015iom} and 2021~\cite{Cheng:2021qpd}. Moreover, the experimental and theoretical progress in this field of open charm and open bottom systems was also reviewed in 2016 by Hua-Xing Chen {\it et al.}~\cite{Chen:2016spr}, and they recently published an updated review of the new heavy hadron states~\cite{Chen:2022asf}. Finally, the expected spectrum of heavy baryons in the quark model has been comprehensively reviewed in 2015 by Yoshida {\it et al.}~\cite{Yoshida:2015tia}.

%% file: BaryonSpectroscopy.tex
\section{Baryon Spectroscopy}
A baryon is generally a fermion with a baryon number of $B = \frac{1}{3} + \frac{1}{3} + \frac{1}{3} = 1$ 
for a conventional $|qqq\rangle$~baryon. Other exotic baryons have been proposed, e.g. pentaquarks with a
$|qqq\bar{q}q\rangle$~quark configuration, and some experimental evidence has been reported recently by the LHCb 
collaboration based on structures observed in the $(p\,J/\psi)$~system~\cite{LHCb:2015yax,LHCb:2019kea}. Such a pentaquark would have the baryon
number $B = \frac{1}{3} + \frac{1}{3} + \frac{1}{3} + \frac{1}{3} - \frac{1}{3} = 1$ and thus, would also be 
considered a baryon. Moreover, baryons are classified according to their strangeness content. Baryons which
contain at least one strange quark (strangeness $-1$) are called hyperons and are labelled $\Lambda, \Sigma$ 
(strangeness $-1$), $\Xi$ (strangeness $-2$), and $\Omega$ (strangeness $-3$). Accordingly, antihyperons have 
strangeness $+1$, $+2$, and $+3$. The number of isospin projections gives the number of charged states. Nucleons
have isospin $I = \frac{1}{2}$ and therefore, $2I+1 = 2$ results in a positively charged and an electrically 
neutral state. The most prominent examples are the proton and the neutron. The $\Delta$~resonances have $I = 
\frac{3}{2}$, and the $\Lambda$ and $\Sigma$ baryons have $I = 0$ and $I = 1$, respectively. More relevant for 
this review are the $\Xi$~states with $I = \frac{1}{2}$ and the $\Omega$~baryon with $I = 0$. The naming scheme 
can be expanded to include heavy baryons, which are labelled with an index. The $\Lambda_c^+$ has isospin zero 
and quark content $|udc\rangle$, $\Xi_c^+$ has isospin $\frac{1}{2}$ and quark content $|usc\rangle$, and 
$\Xi_{cc}^{++}$ has isospin $\frac{1}{2}$ and quark content $|ucc\rangle$, for instance. 

\subsection{Baryons containing u, d, and s quarks}
\label{Subsection:LightBaryonSpectroscopy}
\begin{figure}[t]
\caption{The symmetric {\bf 10} (left) and mixed symmetric {\bf 8} (right) of SU(3)$_{\rm f}$. Reproduced from~\cite{Crede:2013kia}. \copyright~IOP Publishing Ltd. All rights reserved.}
\label{Figure:Multiplet10+8}
\begin{center}
\includegraphics[height=0.23\textheight]{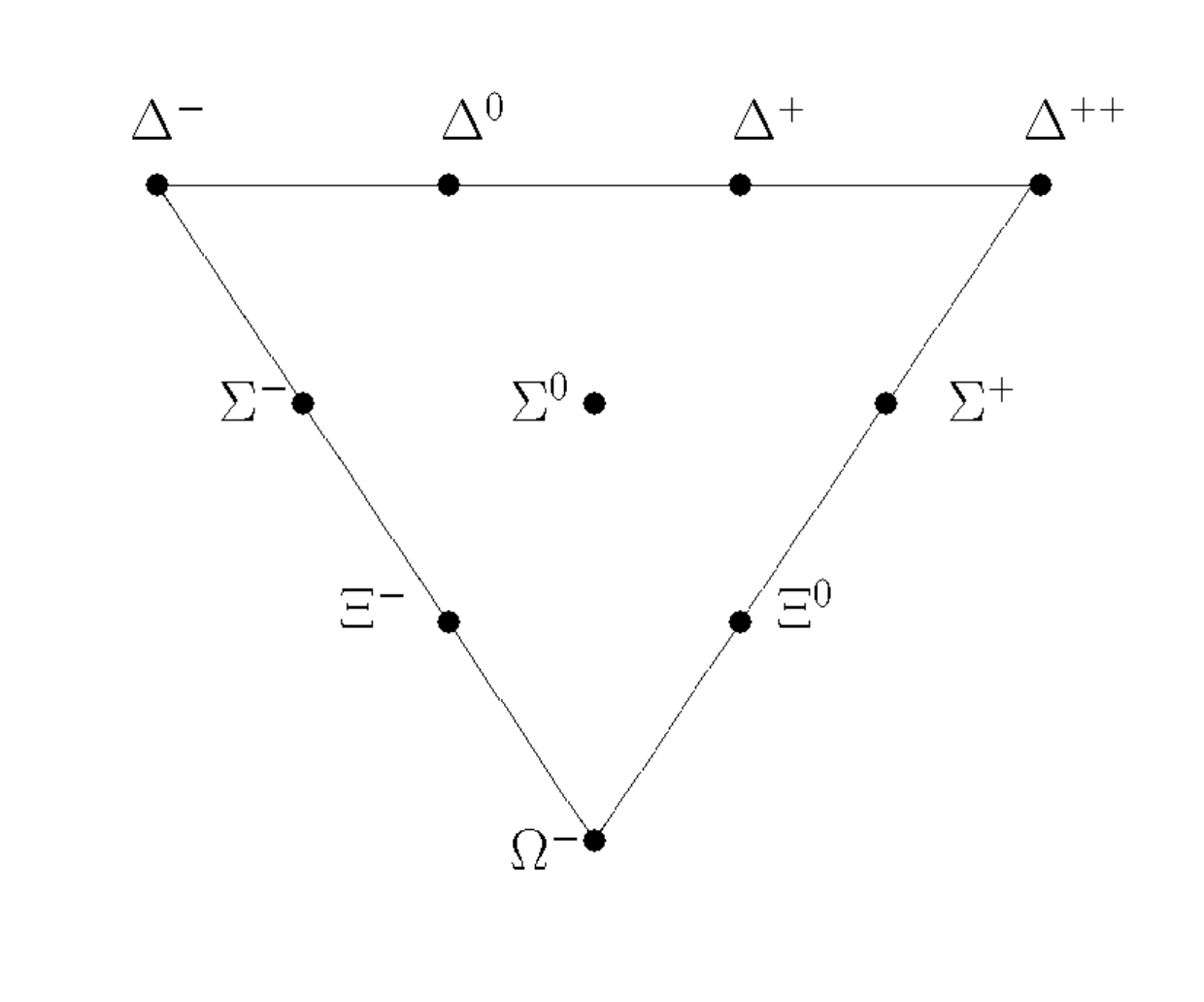}
\includegraphics[height=0.23\textheight]{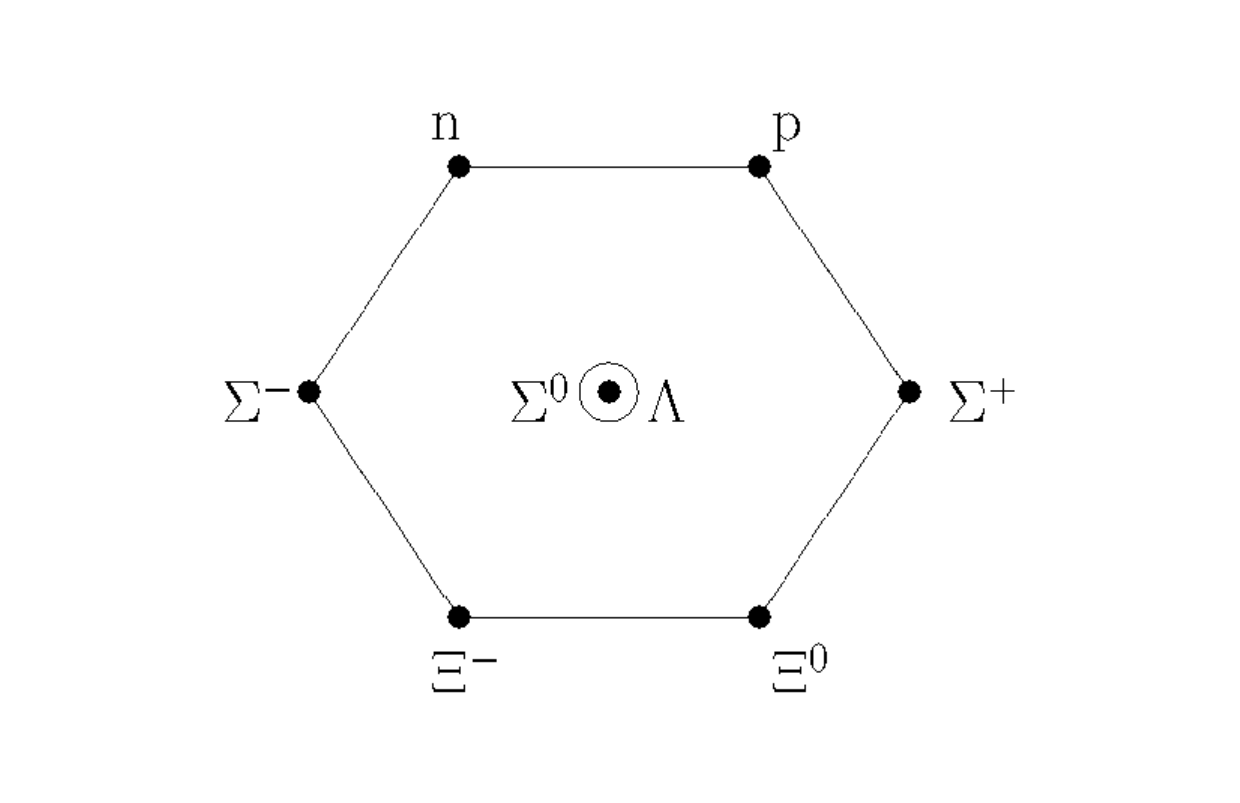}\\\vspace{-1cm}
\end{center}
\end{figure}

The multiplet structure of the light-flavour baryons is described in standard textbooks and can also be found
in some of the reviews listed in Section~\ref{Subsection:Literature}, e.g. in Ref.~\cite{Gross:2022hyw}. In a brief summary, as fermions, baryons
obey the Pauli Principle, so the total wave function
\begin{eqnarray}
|qqq\rangle_A = |{\rm colour}\rangle_A\,\times\, |{\rm space,~spin,~flavour}\rangle_S
\end{eqnarray}
must be antisymmetric (denoted by the index $A$) under the interchange of any two equal-mass quarks. Note that 
all hadrons are colour singlets and therefore, the colour component of the wave function must be completely
antisymmetric. For the light-flavour baryons, there is an approximate SU(3)$_{\rm f}$~symmetry of the strong
interaction under the exchange of the three quark flavours $u$, $d$, and $s$, which is broken by the higher mass
of the strange quark. The flavour wave functions of these baryon states can then be constructed to be members of 
SU(3)$_{\rm f}$ multiplets as
\begin{equation}
  {\bf 3}\,\otimes\, {\bf 3}\,\otimes\, {\bf 3}\,=\,{\bf 10}_S\,\oplus\, {\bf 8}_M\,\oplus\, {\bf 8}_M\,\oplus\,
  {\bf 1}_A\,.
\end{equation}
The proton and the neutron are members of both octets. The weight diagrams for the decuplet and octet representations of SU(3) are shown in Fig.~\ref{Figure:Multiplet10+8}.

Based on the approximate SU(3)$_{\rm f}$~symmetry, all
members of a multiplet are expected to have similar properties. Some differences in the masses are observed, 
though. The Gell-Mann-Okubo mass formula~\cite{Gell-Mann:1961omu,Okubo:1961jc,Okubo:1962zzc} ascribes the breaking of the symmetry in hadrons to differences in the 
hypercharge $Y = B + S$, where $B$ is the baryon number and $S$ denotes strangeness, which is now understood 
to be due to the larger mass of the strange quark. For the ground-state octet baryons, we have
\begin{equation}
  (M_N + M_\Xi)/2 = (3M_\Lambda + M_\Sigma)/4\,,
\end{equation}
which reflects the observed situation fairly accurately to a fraction of a percent. For the ground-state decuplet,
we have (equal spacing rule)
\begin{equation}
  M_{\Sigma^\ast} - M_\Delta = M_{\Xi^\ast} - M_{\Sigma^\ast} = M_\Omega - M_{\Xi^\ast}\,,
\end{equation}
with each difference about 150 MeV, which can be thought of as the difference in the strange and average light 
($u,d$) constituent-quark masses. It is worth noting that the formula is phenomenological in nature, describing an approximate relation between baryon masses, and in terms of understanding baryon properties from a theoretical point of view, has been superseded by advances in quantum chromodynamics, most notably chiral perturbation theory.
%but is now understoood in the context of chiral perturbation theory.

The simple quark model allows for two possible values of the total baryon spin since the three quark spins can 
yield a total baryon spin of either $S = \frac{1}{2}$ or $S = \frac{3}{2}$, where the spin wave function exhibits a mixed symmetry or is totally symmetric, respectively. The flavour and spin can be combined in an 
approximate spin-flavour SU(6), where the multiplets are
\begin{eqnarray}
{\bf 6}\,\otimes\, {\bf 6}\,\otimes\, {\bf 6}\,=\,{\bf 56}_S\,\oplus\,
{\bf 70}_M\,\oplus\, {\bf 70}_M\,\oplus\, {\bf 20}_A\,.
\end{eqnarray}
These can be decomposed into flavour SU(3) multiplets
\begin{eqnarray}
{\bf 56}\,=\, ^4{\bf 10}&\,\oplus\, ^2{\bf 8}\\\label{Equation:70plet} {\bf 70}\,=\,
^2{\bf 10}&\,\oplus\, ^4{\bf 8}\,\oplus\, ^2{\bf 8}\,\oplus\, ^2{\bf 1}\\\label{Equation:20plet}
{\bf 20}\,=\, ^2{\bf 8}&\,\oplus\, ^4{\bf 1}\,,
\end{eqnarray}
where the superscript $(2S+1)$ gives the spin for each particle in the SU(3)~multiplet. The proton and neutron 
belong to the ground-state {\bf 56}, in which the orbital angular momentum between any pair of quarks is zero, 
and are members of the octet with spin and parity $J^P = \frac{1}{2}^+$. The $\Xi(1320)$ is a member of the same 
ground-state octet, whereas the $\Xi(1530)$ is the doubly strange partner of the $\Delta$~resonance and a member of the ground-state decuplet with spin and parity $J^P 
= \frac{3}{2}^+$. For this reason, the number of $\Xi$~resonances is expected to be equal to the number of 
$N^\ast$ and $\Delta$~states combined.
Some excitation of the spatial part is required for the wave functions 
of the {\bf 70} and {\bf 20} to make the overall non-colour (spin\,$\times$\,space\,$\times$\,flavour) component 
of the wave function symmetric. Orbital motion is accounted for by classifying states in SU$(6)\,\otimes\,$O(3) 
supermultiplets, with the O(3)~group describing the orbital motion. 

\begin{table}[b]
\begin{center}
\caption{\label{Table:Classification}Supermultiplets, $({\bf D},\,L_N^P)$,  
  contained in the first three bands~\cite{ParticleDataGroup:2022pth} and assignments
  for the ground-state octet and decuplet to known baryons.}
%\begin{indented}
%\item[] 
\begin{tabular}{@{}l|cccccc}
\br
 N & \multicolumn{1}{c|}{SU(3)$_f$} & \multicolumn{5}{c}{Supermultiplets}\\
\mr
0 & \multicolumn{1}{c|}{} & \multicolumn{5}{c}{$({\bf 56},\,0_0^+)$}\\
   & \multicolumn{1}{c|}{$^2{\bf 8}$} & $S = \frac{1}{2}^+$ & $N(939)$ & $\Lambda(1116)$ & $\Sigma(1193)$ & $\Xi(1318)$\\[0.5ex]
   & \multicolumn{1}{c|}{$^4{\bf 10}$} & $S = \frac{3}{2}^+$ & $\Delta(1232)$ & $\Sigma(1385)$ & $\Xi(1530)$ & $\Omega(1672)$\\
\mr
  1 & & \multicolumn{5}{c}{$({\bf 70},\,1_1^-)$}\\
 %    & $^2{\bf 8}$ & $S = \frac{1}{2}^+$ & $N(939)$ & $\Lambda(1116)$ & $\Sigma(1193)$ & $\Xi(1318)$\\[0.5ex]
 %    & $^4{\bf 8}$ & $S = \frac{3}{2}^+$ & $\Delta(1232)$ & $\Sigma(1385)$ & $\Xi(1530)$ & $\Omega(1672)$\\
 %    & $^2{\bf 10}$ & $S = \frac{3}{2}^+$ & $\Delta(1232)$ & $\Sigma(1385)$ & $\Xi(1530)$ & $\Omega(1672)$\\
\mr
2 & & $({\bf 56},\,0_2^+)$ & $({\bf 70},\,0_2^+)$ & $({\bf 20},\,1_2^+)$ & $({\bf 70},\,2_2^+)$ & $({\bf 56},\,2_2^+)$\\
\br
\end{tabular}
\end{center}
%\end{indented}
\end{table}

It is finally useful to classify baryons into bands according to the harmonic oscillator model with equal 
quanta of excitation, $N = 0,1,2, ...$ Each band consists of a number of supermultiplets, specified by 
$({\bf D},\,L^P_N)$, where ${\bf D}$ is the dimensionality of the SU(6) spin-flavour representation, $L$ 
is the total quark orbital angular momentum, and $P$ is the parity. The first-excitation band contains only 
one supermultiplet, $({\bf 70},\,1_1^-)$, corresponding to states with one unit of orbital angular momentum 
and negative parity, whereas the second-excitation band contains already five supermultiplets corresponding 
to states with positive parity and either two individual units of angular momentum that can couple to 
$L = 0, 1, 2$, giving the supermultiplets $({\bf 70},\,0_2^+)$, $({\bf 20},\,1_2^+)$, $({\bf 70},\,2_2^+)$, 
respectively, two direct units of orbital angular momentum giving the supermultiplet $({\bf 56},\,2_2^+)$, 
or one unit of radial excitation, $({\bf 56},\,0_2^+)$. Table~\ref{Table:Classification} shows the supermultiplets 
contained in the first three excitation bands and the known, well-established baryons that are members of the 
ground-state {\bf 56}. Further quark-model assignments for some of the known ground-state and negative-parity light baryons to the lowest-lying SU(6)\,$\otimes$\,O(3) octets and decuplets will be discussed in 
Section~\ref{Section:Discussion}.

\begin{figure}[t]
\caption{(Colour online) Simple quark model depiction of a baryon. Reproduced from~\cite{Crede:2013kia}. 
\copyright~IOP Publishing Ltd. All rights reserved.}
\vspace{6mm}
\centerline{\includegraphics[width=0.29\textwidth]{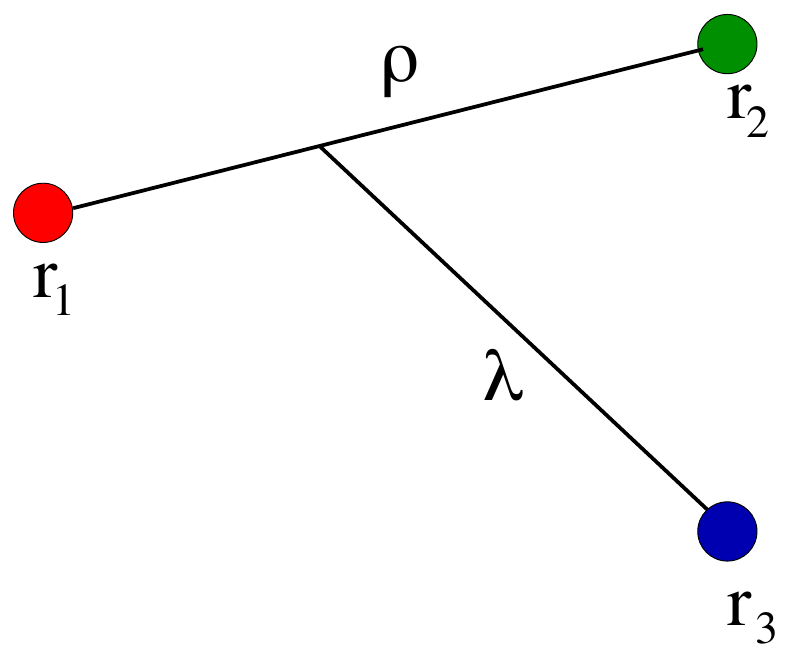}}
%\vspace*{-1.0in}
\label{Figure:JacobiCoordinates}
\end{figure}

The dynamics of the quarks in the baryon are described in terms of the Jacobi coordinates, $\vec{\rho}$ and 
$\vec{\lambda}$, which are related to the quark positions by
\beqy
\vec{\rho}&=&\frac{1}{\sqrt{2}}\left(\vec{r}_1-\vec{r}_2\right),\nonumber\\[1ex]
\vec{\lambda}&=&\frac{1}{\sqrt{6}}\left(\vec{r}_1+\vec{r}_2-2\vec{r}_3\right)\,,
\eeqy
where the $\vec{r}_i$ are the vector positions of the three quarks. The total orbital angular momentum is then $\vec{L} = \vec{l}_\rho + \vec{l}_\lambda$. Here $\vec{\rho}$ is proportional to the separation between quarks~1 and 2, and $\lambda$ is proportional to the separation between quark~3 and the centre of mass of quarks~1 and
2. Orbital excitations can therefore be described in terms of the $\lambda$~mode and the $\rho$~mode. This simple quark model picture of the baryon is depicted in
Fig.~\ref{Figure:JacobiCoordinates}.

In doubly strange $\Xi$~baryons, the $\rho$~mode is the excitation of the $|ss\rangle$~system, whereas the 
$\lambda$~mode is the excitation between the light $u$~or $d$~quark and the heavier doubly strange system. 
In this case, the two oscillators have different reduced masses~\cite{Richard:2012xw}:
\beqy
  \label{Equation:ReducedMass}
  \mu_\rho \,=\,  m_s\quad < \quad \mu_\lambda \,=\, \frac{3m_s\,m_q}{2m_s + m_q}~.
\eeqy
We refer to Section~\ref{Subsection:OpenQuestionsCharmedBottomBaryons} for more details and a similar discussion of the three-quark system with two different masses.
An interesting feature of the $\Xi$~spectrum is that there are fewer degeneracies expected than in the light-quark baryon
spectrum. If the confining potential is independent of quark flavour, the energy of spatial excitations of 
a given pair of quarks will be inversely proportional to their reduced mass. If all three quark masses are the 
same, the excitation energy of either of the two relative coordinates will be the same, which will lead to 
degeneracies in the excitation spectrum. However, with two strange and one light quark, the excitation 
energy of the relative coordinate of the strange quark pair is smaller. In this simple picture, this 
means that the lightest excitations, at least in the lowest partial waves, are between the two strange quarks, and that the 
degeneracy between excitations of the two relative coordinates is lifted.

In the absence of configuration mixing and in a spectator decay model, $\Xi$~states with the relative coordinate 
of the strange-quark pair excited cannot decay to the ground state $\Xi$ and a pion, because of the orthogonality of 
the part of the spatial wave function between the two strange quarks in the initial excited state and in the final 
ground state. Having instead to decay to final states that include Kaons rules out the decay channel with the 
largest phase space for the lightest states in each partial wave, substantially reducing their 
widths~\cite{Chao:1980em}. For this reason, we expect the lightest excitations in the first two partial waves to decouple 
from the decay into $\Xi\pi$ and rather to decay into an anti-Kaon and a singly strange $\Lambda$ or $\Sigma$~hyperon. This selection rule is modified by (configuration) mixing in the wave function; however, colour-magnetic hyperfine 
mixing is weaker in $\Xi$~states because this interaction is smaller between quarks of larger masses. The 
flavour-spin [SU(6)] coupling constants at the decay vertices for $N,\ \Delta \to N\pi,\ \Delta\pi$ are 
significantly larger than those for $\Xi,\ \Xi^\ast \to \Xi\pi,\ \Xi^\ast\pi$ decays~\cite{Zenczykowski:1985uh}, 
which also reduces these widths. The result is that the well known lower-mass resonances are expected to have 
widths $\Gamma_{\Xi^\ast}$ of about 10--20~MeV, which is 5--30 times narrower than is typical for $N^\ast$, 
$\Delta$, $\Lambda$, and $\Sigma$~states. 

These simple arguments are consistent with the quark-model calculations discussed in Ref.~\cite{Chao:1980em}. This QCD-like model is based on the two-component picture that describes hadrons by a dominant flavour-independent confinement potential and short-range forces of the type expected from one-gluon exchange. The authors present the approximate composition of the states in the first few excitation bands in terms of the underlying $|ssq\rangle$~basis, i.e. their $\rho$~and $\lambda$~mode excitations, and in terms of the SU(6)~basis, i.e. their octet and decuplet affiliation. The first excitation band with $L=1$ contains the supermultiplet $({\bf 70},\,1_1^-)$ and thus, we expect a total of seven $\Xi$~states with negative parity as members of the two octets with spin $S=\frac{1}{2}$ and $S=\frac{3}{2}$, as well as the decuplet with spin $S=\frac{1}{2}$. According to Table~II of Ref.~\cite{Chao:1980em}, we expect the first excited states in the $\Xi\,\frac{1}{2}^-$ and $\Xi\,\frac{3}{2}^-$ partial waves to be dominated by $\rho$-mode excitations, whereas the remaining five are dominated by $\lambda$-mode excitations. The first $\Xi\,\frac{5}{2}^-$ excitation is predicted to be a pure $\lambda$-mode excitation. Furthermore, the first excited states in the $\Xi\,\frac{1}{2}^-$ and $\Xi\,\frac{3}{2}^-$ partial waves exhibit a dominant $^2{\bf 8}$~octet affiliation and the first $\Xi\,\frac{5}{2}^-$ excitation must be a member of the $^4{\bf 8}$~octet.

\subsubsection{The Dyson-Schwinger Bethe-Salpeter approach}
There has been significant recent progress in understanding the physics of baryons~\cite{Eichmann:2016yit,Eichmann:2018adq} by using the Dyson-Schwinger equations of QCD and Bethe-Salpeter equations~\cite{Bashir:2012fs,Roberts:2012sv}. In this approach, baryons are relativistic bound states of three quarks, and the treatment of their interactions arising from QCD is non-perturbative, incorporating aspects of confinement and dynamical symmetry breaking. The three-body problem is solved in two different ways: direct solution of the three-body Faddeev equation, and decomposition of baryons into {\it dynamical} quark-diquark systems, with all quark pairs able to constitute the diquark. The latter path is different from the older {\it static} diquark-quark approach, and requires the calculation of diquark Bethe-Salpeter amplitudes, and diquark propagators. These depend on the quark and gluon propagators and quark-gluon vertex, which are consistent with those used for the Bethe-Salpeter equation for mesons, and with chiral symmetry. Due to the complexity of the three-body system, baryon calculations are performed using the rainbow-ladder approximation, where the $q$-$q$ kernel has the form of a single-gluon exchange with a momentum-dependent vertex strength, summed by the Bethe-Salpeter equation into Feynman diagrams, taking the form of a ladder (rainbow). This construction preserves chiral symmetry.
Using this dynamical quark-diquark approach, the proton, $\Delta(1232)\,\frac{3}{2}^+$, and Roper $N(1440)\,\frac{1}{2}^+$~resonance are described well~\cite{Eichmann:2016hgl}, as their configurations are dominated by scalar and axial-vector diquarks. However, other baryons are sensitive to different diquark channels, which are known to be too strongly bound in
this approximation, as are the corresponding scalar and axial-vector mesons. The result is that the other excited baryon masses come out too low. Reducing the strength
of the attraction in the pseudo-scalar and vector-diquark kernels simulates effects beyond the rainbow-ladder approximation, and the result is good agreement between the calculated spectrum for excited $N$, $\Delta$~\cite{Eichmann:2016hgl} and $\Lambda$, $\Sigma$, $\Xi$, $\Omega$~baryons~\cite{Fischer:2017cte} with $J^P = \frac{1}{2}^\pm,~\frac{3}{2}^\pm$, with the exception of the $\Lambda(1405)\frac{1}{2}^-$, $\Lambda(1520)\,\frac{3}{2}^-$, and to a lesser extent the Roper resonance $N(1440)\,\frac{1}{2}^+$. The authors of Ref.~\cite{Eichmann:2016hgl} point out that this is likely due to the lack of a consistent treatment of baryon-meson coupled channel effects.

\subsubsection{Lattice QCD}

\begin{figure}[t]
\caption{The spectrum of excited $\Xi$ and $\Omega$~states from the Hadron Spectrum Collaboration using $m_\pi = 391$~MeV/$c^2$~\cite{Edwards:2012fx}. The colours denote the 
flavour symmetry of dominant operators as follows: blue for {$\bf 8_f$} (flavour octet) and yellow for {$\bf 10_f$} (flavour decuplet). Symbols with thick border lines indicate states with strong hybrid content. The lowest bands of positive- and negative-parity states are highlighted within slanted boxes. Reprinted figure with permission from~\cite{Edwards:2012fx}, Copyright (2013) by the American Physical Society.}
\vspace{4mm}
\centerline{\includegraphics[width=0.97\textwidth]{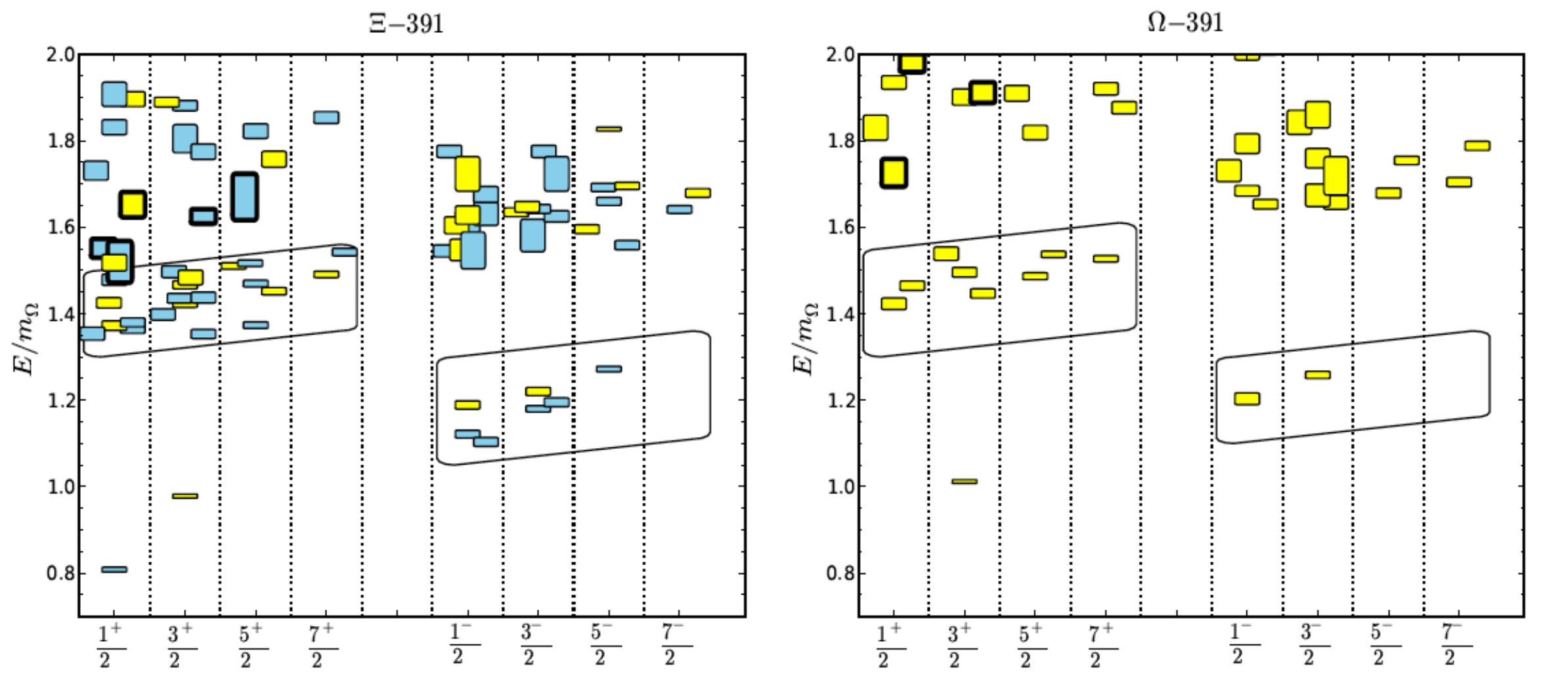}}
%\vspace*{-1.0in}
\label{Figure:LatticeJLab}
\end{figure}

A numerical approach to QCD considers the theory on a finite, discrete grid of points in a way that would become exact if the lattice spacing were taken to zero and the spatial extent of the calculation, i.e. the "box size," was made infinitely large. In practice, rather fine spacings and large boxes are used to minimize
the systematic effect of this approximation. At present, the main limitation of these calculations is the poor scaling of the numerical algorithms with decreasing quark mass. In practice, most contemporary calculations
use a range of artificially heavy light quarks and attempt to observe a trend as the light quark
mass is reduced toward the physical value. Trial calculations at the physical quark mass have begun but remain computationally demanding. The spectrum of QCD eigenstates can be extracted from correlation functions of the type $\langle 0|{\cal{O}}_f(t){\cal{O}}_i^\dagger(0)|0\rangle$, where the ${\cal{O}}^\dagger$ are composite QCD operators capable of
interpolating a meson or baryon state from the vacuum.
The time-evolution of the Euclidean correlator indicates
the mass spectrum (e$^{-m_n t}$) and information about the quark-gluon substructure can be inferred from matrix-elements $\langle n|{\cal{O}}^\dagger|0\rangle$. In a series of recent papers~\cite{Edwards:2012fx,Dudek:2009qf,Dudek:2010wm,Edwards:2011jj}, the Hadron
Spectrum Collaboration has explored the spectrum of
mesons and baryons using a large basis of composite QCD
interpolating fields, and has extracted a spectrum of meson and baryon states with definite $J^{P(C)}$~quantum numbers, including states of high internal excitation. The results have been obtained using a pion mass of 391~MeV/$c^2$ and are shown in Fig.~\ref{Figure:LatticeJLab} for the spectrum of $\Xi$ and $\Omega$~resonances. These calculations indicate that the number of states in the baryon spectrum is consistent with the traditional quark model for the lowest positive- and negative-parity bands. For the spectrum of doubly strange hyperons in particular, the lattice-QCD calculations of Ref.~\cite{Edwards:2012fx} confirm the existence of seven states in a group of negative-parity excitations above the ground states. Moreover, the calculations also indicate that the first radial excitation of the $\Xi(1320)$ octet ground state would be located above the first negative-parity excitations in the spectrum.

A lattice-QCD study of baryon ground states and low-lying excitations of nonstrange and strange baryons has also been presented by the Bern-Graz-Regensburg (BGR) Collaboration~\cite{Engel:2013ig}. The results are based on seven gauge field ensembles with two flavours of light chirally improved quarks corresponding to pion masses between 255 and 596 MeV/$c^2$ and a strange valence quark with a mass that was fixed by the $\Omega$~baryon. A variational method was applied to extract energy levels and to discuss the flavour content of states. Results for the spin-$\frac{1}{2}$ and $\frac{3}{2}$~channels for both parities, extrapolated to the physical pion mass, are shown and discussed in Ref.~\cite{Engel:2013ig}. 

\subsection{Baryons containing heavy quarks}
\label{Subsection:HeavyBaryonSpectroscopy}
The baryons containing a single charm quark can be described in terms of SU(3) flavour multiplets. However, these represent but a subgroup of the larger SU(4) group that includes all of the baryons containing zero, one, two or three charmed quarks. Furthermore, this multiplet structure is expected to be repeated for every combination of spin and parity, leading to a very rich spectrum of states. One can also construct SU(4) multiplets in which charm is replaced by beauty, as well as place the two sets of SU(4) structures within a larger SU(5) group to account for all the baryons that can be constructed from the five flavours of quark accessible at low to medium energies. It must be understood that the classification of states in SU(4) and SU(5) multiplets serves primarily for enumerating the possible states, as these symmetries are badly broken. Only at the level of the SU(3) ($u,\,\,d,\,\,s)$ and SU(2) $(u,\,\,d)$ subgroups can these symmetries be used in any quantitative way to understand the structure and properties of these states.

\begin{figure}[t]
\caption{(Colour online) (a) The symmetric {\bf 20} of flavour SU(4), showing the SU(3)
  decuplet on the lowest layer. (b) The mixed-symmetric {\bf 20}s
  and (c) the antisymmetric {\bf 4} of SU(4). The mixed-symmetric {\bf
    20}s have the SU(3) octet on the lowest layer, while the {\bf 4} has
  the SU(3) singlet at the bottom. Note that there are two $\Xi_c^+$ and
  two $\Xi_c^0$~resonances on the middle layer of the mixed-symmetric {\bf 20}. Reproduced from~\cite{Gross:2022hyw}, with permission from Springer Nature.}
\vspace{6mm}
\centerline{\includegraphics[width=0.94\textwidth]{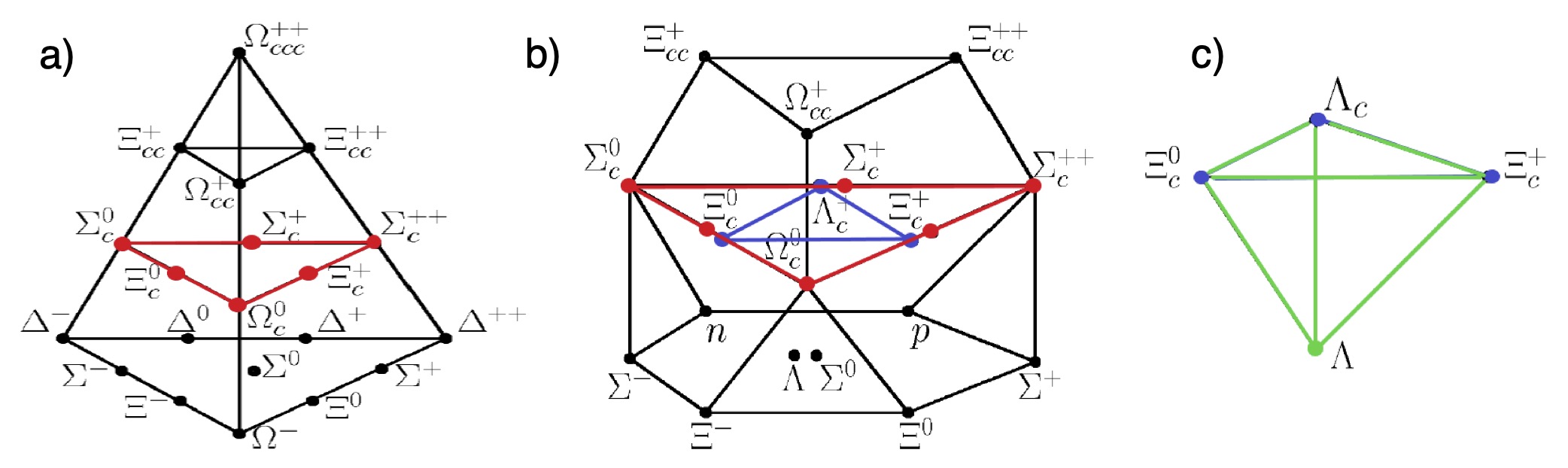}}
%\vspace*{-1.0in}
\label{Figure:Multiplets}
\end{figure}

The multiplet structure for flavour SU(4) is ${\bf 4}\,\otimes\,{\bf 4}\,\otimes {\bf
  4} = {\bf 20}_S\,\oplus\,{\bf 20}_M\,\oplus\,{\bf 20}_M\,\oplus\,{\bf 4}_A$. The symmetric {\bf 20} contains the flavour SU(3) decuplet as a subset, forming the `ground floor' of the weight diagram shown in Figure~\ref{Figure:Multiplets}~(a), and all the ground-state baryons in this multiplet have $J^P=\frac{3}{2}^+$. The mixed-symmetric {\bf 20}s shown in Figure~\ref{Figure:Multiplets}~(b) contain the flavour SU(3) octets on the lowest level, and all the ground-state baryons in this multiplet have $J^P=\frac{1}{2}^+$. The ground-floor state of the {\bf 4} shown in Figure~\ref{Figure:Multiplets}~(c) is the flavour SU(3) singlet $\Lambda$ with $J^P=\frac{1}{2}^-$.

Within the flavour SU(3) subgroups, the ground-state heavy baryons containing a single heavy quark belong either to a sextet of flavour symmetric states, or an antitriplet of flavour antisymmetric states, both of which sit on the second layer of the mixed-symmetric {\bf 20} of SU(4) in Figure~\ref{Figure:Multiplets}~(b). There is also expected to be a sextet of states with $J^P=\frac{3}{2}^+$ sitting on the second floor of the symmetric {\bf 20}. The members of the two multiplets of singly charmed baryons have flavour wave functions
\beqy
\Sigma_c^{++}&=&uuc,\qquad\Sigma_c^{+}=\frac{1}{\sqrt2}\left(ud+du\right)c,\qquad\Sigma_c^{0}=ddc\nonumber\\[1ex]
\Xi_c^{'+}&=&\frac{1}{\sqrt2}\left(us+su\right)c,\qquad\Xi_c^{'0}=\frac{1}{\sqrt2}\left(ds+sd\right)c
\nonumber\\[1ex]
\Omega_c^{0}&=&ssc
\eeqy
for the sextet and
\beqy
\Lambda_c^{+}=\frac{1}{\sqrt2}\left(ud-du\right)c\nonumber\\[1ex]
\Xi_c^{+}=\frac{1}{\sqrt2}\left(us-su\right)c,\qquad
\Xi_c^{0}=\frac{1}{\sqrt2}\left(ds-sd\right)c
\eeqy
for the antitriplet. There is a similar set of flavour wave functions for baryons containing a single $b$~quark. In conclusion, we briefly note that the triply strange $\Omega^-$~ground-state resonance can be described in flavour SU(3) as a genuinely symmetric flavour state and for this reason, the quark model predicts a single ground state, which is also observed in nature as $\Omega(1670)$. The situation is very similar to the series of $I=\frac{3}{2}$~$\Delta$~resonances. The $|ddd\rangle~\Delta^{++}$~resonance initially motivated the introduction of the colour charge and served as the basis for predicting the $|sss\rangle~\Omega$~resonance. In contrast, the situation is slightly different for the singly heavy $|ssc\rangle~\Omega_c$ (and also $|ssb\rangle~\Omega_b$)~resonance which is observed in nature as $\Omega_c\,\frac{1}{2}^+$ and $\Omega_c(2770)\,\frac{3}{2}^+$, i.e. two ground states exist. In the heavy-quark, light-diquark picture, this can be easily understood in terms of a light spin-1 quark-quark system that couples to the spin-$\frac{1}{2}$ heavy quark, as discussed below.

The baryons containing one heavy (i.e. charm or bottom) quark are particularly amenable to potential models for predictions of
their masses. The baryons may be simplified to one, approximately stationary, heavy quark and a comparatively light, tightly bound quark-quark system (diquark).
There is then a separation between this heavy quark with well-defined $J^P=\frac{1}{2}^+$~quantum numbers, and the diquark which has $S$ = 0 or 1 leading to an antisymmetric or symmetric spin wave function, respectively. The flavour wave functions of the light diquark can be constructed to be members of 
\begin{equation}
\label{Equation:diquarkStructure}
{\bf 3}\,\otimes\, {\bf 3}\,=\,{\overline{\bf 3}}_A\,\oplus\,{\bf 6}_S\,,
\end{equation}
where the light diquark\ of the ${\bf 6}$ is symmetric and the light diquark of the ${\bf \overline{3}}$ is antisymmetric. The colour part of the diquark wave function is totally antisymmetric.

The introduction of orbital angular momentum adds to the spin of the quark-quark system and 
changes the total spin-parity of the
``light degrees of freedom'', with the last step of any model being the combination of these light degrees of freedom with the
 $J^P=\frac{1}{2}^+$ of the heavy quark making for a ``hyperfine'' splitting. 
This view is hardly a new one. In the early days of the quark model, Gell-Mann
showed how the $sud$~quark combination's first states were a $\Lambda$, $\Sigma$ and $\Sigma^*$, and that the splitting between
the $\Lambda$ and spin-weighted average of the $\Sigma$ is independent of the $s$ quark constituent mass 
(though it does depend on the effective mass of the light quark-quark system), whereas the hyperfine splitting
between the $J^P=\frac{1}{2}^+$ and $J^P=\frac{3}{2}^+$ states is inversely proportional to the heavy quark mass. 
These relations can be found 
in any elementary particle physics textbook even before the discovery of charm and bottom. 
It clearly follows that the mass difference between the (spin-weighted) $\Sigma_c$ and $\Lambda_c$ should be the same as that 
of the $\Sigma_b$ and $\Lambda_b$, whereas the hyperfine splitting should be much smaller in the bottom case.
The $\Xi_c$ and $\Xi_b$ can still be treated in the same way. However, in this case, the light quark-quark system is itself more massive. This leads
to a smaller mass splitting between the ground state and the first excited states. 

\begin{figure}[b]
\caption{The excitation energy associated with a $\rho$ excitation (straight line) and a $\lambda$ excitation (curved line) as a function of heavy quark mass. Reprinted figure with permission from~\cite{Yoshida:2015tia}, Copyright (2015) by the American Physical Society.\label{fig:Yoshida}}
\centerline{\includegraphics[width=0.77\textwidth]{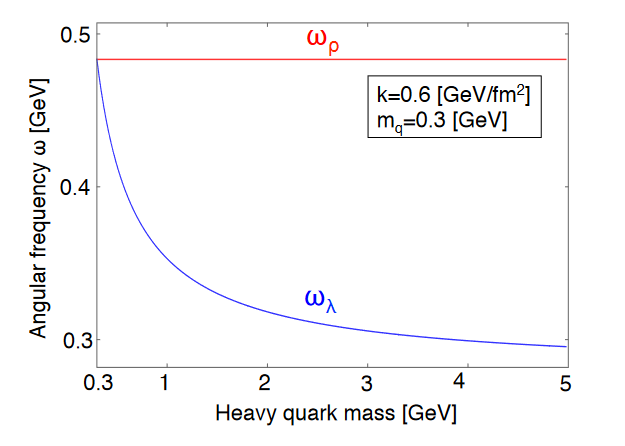}}
\end{figure}

The situation with orbital excitations is of course more complicated.  As discussed above, there are two types, the $\lambda$ and the $\rho$ excitations. Here it is clear that the former corresponds to the excitation between the (static) heavy quark and the light quark-quark system, whereas the latter is the excitation of the light quark-quark system itself.  
The mixing between these configurations appears to be small.  
The scaling of the orbital excitation energy with this reduced mass (which, in turn, depends on the relative masses of the heavy
quark and light quark-quark system) is not completely defined by the model. In Ref.~\cite{Karliner:2018bms}, Karliner and Rosner show how they
can fit the existing data to find the excitation energy as a function of reduced mass for heavy mesons and baryons. Of course
this will depend on what we consider the constituent quark masses to be. Yoshida {\it et al.}~\cite{Yoshida:2015tia} show explicitly in Fig.~\ref{fig:Yoshida}
their estimate for the $\lambda$ excitation energy as a function of heavy quark mass. As we can see, the excitation energy of the $\rho$ mode is only dependent on the light quarks, whereas there is small lowering of the excitation energy of $\lambda$ modes as we move from charm to bottom. 

Recent years have seen much activity in applying lattice QCD to heavy quark physics. Much of this has been directed towards tetraquarks and pentaquarks which are subjects of great experimental and theoretical interest, but go beyond the scope of this review. The Flavour Lattice Averaging Group Collaboration~\cite{FlavourLatticeAveragingGroupFLAG:2021npn} summarize the effective quark
masses and decay constants necessary for all calculations. A particularly useful review by Padmanath~\cite{Padmanath:2019ybu} shows nine different predictions of the $\Xi_{cc}$ mass made just before its discovery by the LHCb Collaboration~\cite{LHCb:2017iph}, and all are closer to the now-accepted experimental value than to the original SELEX results~\cite{SELEX:2002wqn,SELEX:2004lln}. It will be interesting to compare other predictions made using the same 
formulisms to the $\Xi_{cc}$ excited states and $\Omega_{cc}$ states that we can hope will be found experimentally in the next few years.

%% file: ExperimentalMethods.tex
\section{Experimental Methods and Historical Context}
The first indication for a strange particle, which is now known to be the $K^+$~meson, emerged in the early 1940s in cosmic-ray physics. A few years later in 1947, a group at the University of Manchester published two cloud chamber photographs of cosmic ray-induced events, which showed a neutral particle decaying into two charged pions and a charged particle decaying into a charged pion and something neutral. The group estimated the mass of these new particles to be approximately half the proton mass. The historical timeline is quite remarkable in that the first indication for strange particles was observed even before the experimental discovery of the pion. We refer to Ref.~\cite{Ezhela:1996xi} for an annotated chronological bibliography.

The first evidence of strange baryons emerged in cosmic-ray studies in
the early 1950s with the discovery of the neutral $V_1^0$~particle, which is
now known as the $\Lambda^0$. In rapid succession, those years
witnessed the additional discovery in 1953 of the $V_1^+$, now called the
$\Sigma^+$, in Italy~\cite{Bonetti:1953} and at the California Institute of Technology~\cite{York:1953} 
using a photographic
emulsion and a cloud chamber, respectively. The existence of a
negative hyperon, now known as the $\Xi^-$, was confirmed in 1954 at
Caltech~\cite{Cowan:1954zz}. The title of the publication was {\it ``A V-Decay Event with a Heavy
Negative Secondary, and Identification of the Secondary V-Decay Event
in a Cascade''} and gave this doubly strange baryon its colloquial
name. The neutral partner of the $\Xi^-$ was not discovered until 1959
at the Lawrence Berkeley Laboratory~\cite{Alvarez:1959zz} when data from accelerators began
to supplant those from cosmic rays. The challenge in producing the
$\Xi$ is its strangeness $-2$ flavour content that requires a minimal process
of $\pi^- p\to K^0 K^0 \,\Xi^0$ in its production by pions. A more
effective way is to start with a probe with strangeness $-1$ which was
accomplished at Berkeley using a hydrogen bubble chamber and a
1-GeV/$c$ mass-separated beam of $K^-$ mesons produced by the
Bevatron. These particles were all identified in single-event
images. In particular, the $\Xi^0$ was identified in an event $K^-
p\to K^0 \,\Xi^0$ using the excellent analytical power of the bubble
chamber technique.

\begin{figure}[t]
\caption{The discovery of the $\Omega^-$~hyperon in a bubble chamber picture. The $\Omega^-$ leaves the short, thick track in the lower left corner. An incoming $K^-$~meson interacts with the proton in the liquid hydrogen of the bubble chamber and produces an $\Omega^-$, a $K^0$, and a $K^+$~meson which all decay into other particles. Neutral particles which produce no tracks in the chamber are shown as dashed lines. Reprinted figure with permission from~\cite{Barnes:1964pd}, Copyright (1964) by the American Physical Society.}
\vspace{4mm}
\centerline{\includegraphics[width=0.8\textwidth]{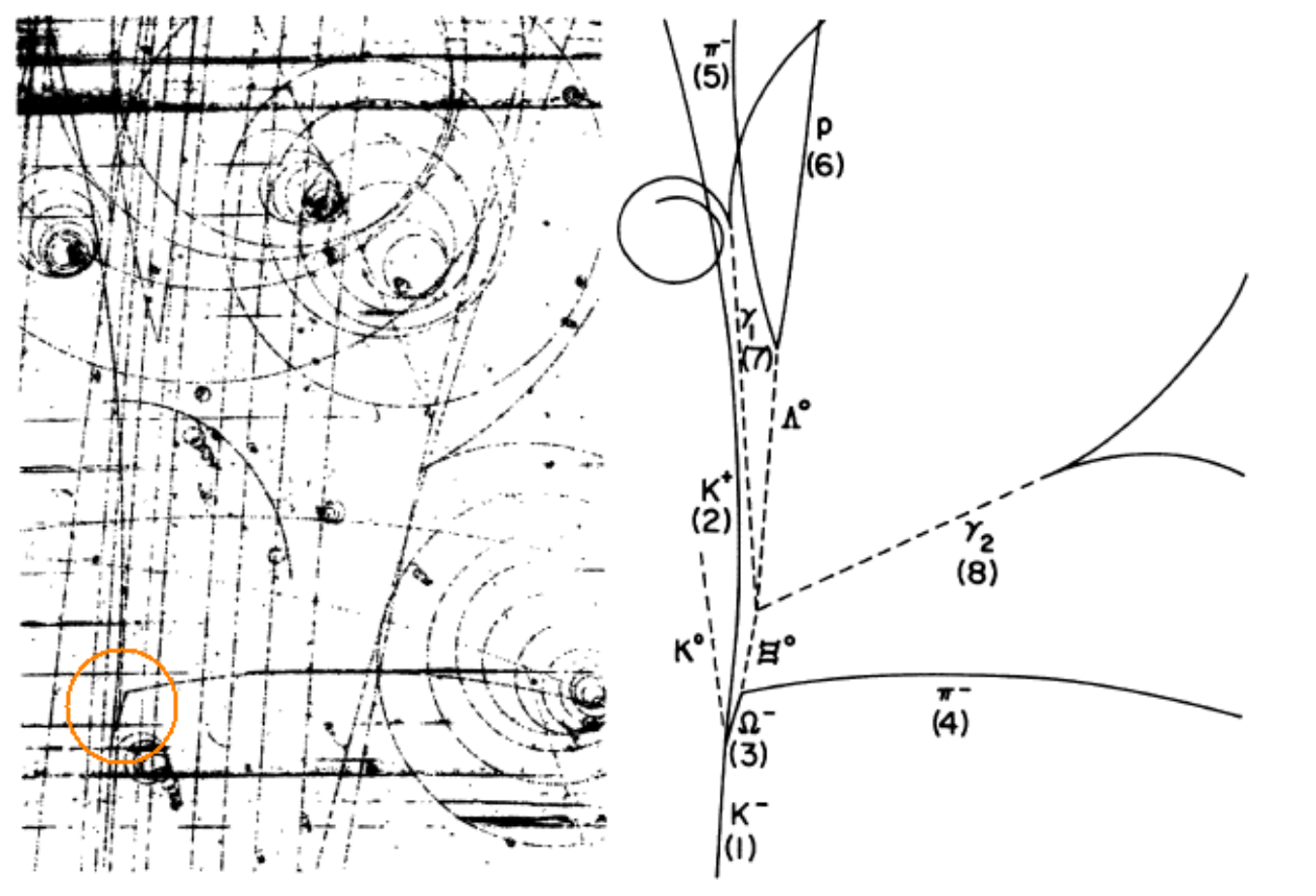}}
%\vspace*{-1.0in}
\label{Figure:OmegaDiscovery}
\end{figure}

The $\Xi(1530)$ decuplet ground state was found in 1962 independently by a group at
UCLA~\cite{Pjerrou:1962pc} and the Brookhaven-Syracuse Collaboration~\cite{Bertanza:1962zz} in the reactions $K^-
p\to (\Xi^-\pi^0)_{\Xi(1530)^-}\,K^+$ and $K^-
p\to (\Xi^-\pi^+)_{\Xi(1530)^0}\,K^0$. The experiments determined the
second reaction to be dominant and therefore, isospin $I=1/2$ was
assigned to the resonance in the $\Xi\pi$~system with a mass around 1530~MeV.
The last gound-state hyperon to be discovered experimentally
was the $\Omega^-$ in 1964~\cite{Barnes:1964pd}. The famous picture from the experiment at
Brookhaven National Laboratory is shown in Fig.~\ref{Figure:OmegaDiscovery}. In this decay
chain, the $\Xi^0$ was also observed.

In modern experiments, hyperons can only be produced as part of a final state or a decay
chain in associated strangeness production due to flavour
conservation since no strange probes have been available since the 1990s. This makes the analysis of $\Xi$ or $\Omega$~baryons
very challenging because the production cross sections are typically
also 
very small. Furthermore, the name of the {\it Cascade} baryon already
suggests that the final states are topologically complicated and
difficult to analyse. To this effect, significant contributions to
$\Xi$~spectroscopy did not happen until the 1980s when technology was
sufficiently advanced to produce data sets that went beyond the few
event numbers of the bubble chamber experiments.

\subsection{Hyperon beams}
A large amount of our experimental knowledge regarding the properties and interactions of hyperons comes from hyperon-beam experiments that started in the early seventies with beam momenta of about 25~GeV/$c$ at the Brookhaven AGS and the CERN Proton Synchrotron (PS). Hyperon beams are challenging to operate since they are characterized by very short lengths varying from about 7~m for the $\Lambda$ to just about 1.5~m for the $\Omega^-$, which results in severely limited hyperon beam fluxes available for experiments~\cite{Aleksandrov:1997kj}.

A unique facility was operational at the CERN Super Proton Synchrotron (SPS) between 1976 and 1982 where a beam 
of charged hyperons including $\Sigma^-$, $\Xi^-$, and $\Omega^-$ baryons was available with momenta between 70 and 135 GeV/$c$. Reviews of the physics activities can be found in Refs.~\cite{Bourquin:1984kcf}.
The beam was derived 
from an external proton beam of the SPS hitting a hyperon production target and the incident $\Xi^-$~baryons in particular 
were identified by a differential Cherenkov counter (DISC) in the beam. The total beam had an intensity of
$1.5\times 10^6$ particles over a 1.5~s effective spill time and there were typically about 300~$\Xi^-$
hyperons counted by the DISC with a $\Sigma^-$~background of about 3\,\%. The trajectory of an incident
$\Xi^-$ was measured by multiwire proportional chambers (MWPC). Data were recorded with momenta of 102 
and 135~GeV/$c$. A detailed description of the CERN SPS charged hyperon beam is provided in Ref.~\cite{Bristol-Geneva-Heidelberg-Orsay-Rutherford-Strasbourg:1978vvp}. More information about the experimental apparatus and relevant results on hyperon spectroscopy using the beam of $\Xi^-$~baryons can be found in Refs.~\cite{Biagi:1981cu,Biagi:1985rn,Biagi:1986zj,Biagi:1986vs}. The hyperon-beam 
facility at the CERN-SPS was later upgraded and operated from 1989 to 1994. The setup allowed a rapid changeover between hyperon and conventional hadron beam configurations and excited $\Xi$~baryons were produced by a $\Sigma^-$~beam of 345 GeV/$c$ mean momentum from copper and carbon targets. The new facility using 
the Omega spectrometer started its new physics operation in 1990 and results on $\Xi$~spectroscopy were reported by the 
WA89 Collaboration~\cite{WA89:1997vxx,WA89:1999nsc}. 

At the Fermi National Accelerator Laboratory (FNAL or Fermilab), the development of hyperon beams culminated in a 600 GeV/$c$~beam delivering $\Sigma^-$~baryons to the SELEX experiment in 1997~\cite{SELEX:1987bde}. The FNAL hyperon facility has made fewer contributions in the search for excited hyperon resonances, but has provided a significant knowledge base on the properties of the ground-state hyperons. The negatively charged secondary beam was formed by the interaction of (up to) 800 GeV/$c$ primary protons from the Tevatron, with the sign and momenta of secondaries selected by a long curved collimator within a dipole magnetic field. In a brief and incomplete summary to highlight major achievements, the HyperCP (E871) Collaboration has searched for CP~violation in charged $\Xi$- and $\Lambda$-hyperon decays, as well as for rare and forbidden hyperon and Kaon decays, see Refs.~\cite{HyperCP:2004zvh,HyperCP:2004not,HyperCP:2005sby}, for instance. The E761 Collaboration has focused on studying branching fractions and asymmetry parameters, e.g. in $\Sigma^+\to p \gamma$ and $\Xi^-\to \Sigma^- \gamma$~radiative decays~\cite{E761:1993unn}, and the E756 Collaboration has measured $\Xi^-$~decay parameters~\cite{FNALE756:2003kkj,E756:2000rge} and the magnetic moment of the $\Omega^-$~hyperon~\cite{Wallace:1995pf}, for instance.

\subsection{Meson-induced (fixed target) experiments}
Hyperons including doubly strange baryons can still effectively be produced and studied if the incident 
probe already contains at least one unit of strangeness. An incident beam of $K^-$~mesons was available 
for many decades at several laboratories worldwide until about the mid-1980s. At SLAC, the Large Aperture Superconducting Solenoid (LASS) spectrometer performed the last such experiments using an intense Kaon beam of 11~GeV/$c$. Evidence was reported for an $\Omega^\ast$~resonance with a mass of $(2474\pm 12)$~MeV/$c^2$ and a width of $(72\pm 33)$~MeV/$c^2$~\cite{Aston:1988yn}. In a brief summary, results on the spectroscopy of 
excited $\Xi$~baryons in $K^-\,p$~reactions were reported from experiments in the United States using bubble chambers at the 
Lawrence Radiation Laboratory at the University of California Berkeley for several beam momenta from the $\Xi$~production threshold to 2.7~BeV/$c$, 
see e.g. Refs.~\cite{Smith:1965zze,Berge:1966zz} for a description of the experimental setups and relevant results; using the MURA 30-in. hydrogen bubble chamber at the Argonne National Laboratory 
(ANL) and the high-purity separated 5.5~GeV/$c$ $K^-$~beam at the Argonne Zero Gradient Synchrotron (ZGS), see e.g. Ref.~\cite{Goldwasser:1970fk}; 
using bubble chambers at the Brookhaven National Laboratory (BNL) in the 1960s and early 1970s and 
incident $K^-$~mesons of momenta $\leq 5.0$~GeV/$c$, see e.g. Refs.~\cite{Alitti:1968zz,Alitti:1969rb,Apsell:1970uf,Borenstein:1972sb} for more information, and later using the BNL 
Multiparticle Spectrometer (MPS) in the 1980s at 5~GeV/$c$~\cite{Jenkins:1983pm}. The largest beam momentum was available at the Stanford Linear Accelerator Center (SLAC) using the LASS spectrometer. An 11~GeV/$c$ rf-separated $K^-$~beam was used to inclusively produce data samples of $X = \Xi^-,\,\Omega^-,\,\Xi(1530)^0$ in the reaction $K^-\,p\to X + {\rm anything}$. More details are available in Ref.~\cite{Aston:1985sn}, for instance. A neutral beam of $K_L^0$~mesons was also available at SLAC, which was produced by impinging a high-energy electron beam onto a beryllium target 56~m upstream of the bubble chamber. Details on the construction of the beamline and the determination of the $K_L$~momentum spectrum are discussed in Ref.~\cite{Brandenburg:1972pm}.

In Europe, $K^-\,p$~scattering was studied in bubble chamber experiments at Saclay for a beam in the $[\,1,2\,]$~GeV/$c$ momentum range, see for example Refs.~\cite{Berthon:1974kr} for more details, and at the European Organization for Nuclear Research (CERN) for the momentum range $[\,1.9, 4.2\,]$~GeV/$c$ using the CERN 2-m hydrogen bubble chamber (HBC) that was available in the 1970s. A more detailed description of the experiments at CERN and relevant results are available in Refs.~\cite{Bellefon:1975pvi,Amsterdam-CERN-Nijmegen-Oxford:1976ezm}, for instance. 

\subsection{Collider experiments}
Some early measurements of charmed baryons were performed using fixed target experiments at Fermilab.
However, charmed baryon spectroscopy studies was dominated for many years by $e^+e^-$ annihilation
experiments running in the $\Upsilon$ energy range. Although the main interest of these experiments
has been in $B$~meson physics, continuum $c\bar{c}$ proved a fertile hunting ground for charmed baryons.

\subsubsection{Experiments for studying charmed and bottom baryons} In the late 1980's the CLEO experiment, operating at the Cornell Electron Storage Ring (CESR), competed with the
rather similar ARGUS experiment located at the DOuble RIng Storage (DORIS) Facility at the Deutsches Elektronen-Synchrotron (DESY), Hamburg. Whereas ARGUS stopped
taking data in 1992, CLEO received a series of upgrades over a long history,
coinciding with increases in luminosity. For example, the CLEO~II experiment started operating in 1989 and combined
a precision inner tracker, time-of-flight detectors to augment the particle identification in the main
tracking system, and an important upgrade from the original CLEO experiment and its rival ARGUS, a CsI~calorimeter
which allowed good efficiency for $\pi^0\to\gamma\gamma$ detection. The final configuration used for charmed baryons
was the CLEO~III experiment which collected $\sim 15~{\rm fb}^{-1}$ of data in the $\Upsilon$~energy range. The
baton was then passed to Belle and BaBar operating at KEKB, Japan, and at the
PEP-II facility at SLAC, respectively.

A characteristic of both KEKB and PEP-II was the energy asymmetry of the two beams with the goal
to investigate CP~violation in $B$~meson physics. This, in turn, led to asymmetric detectors,
but this asymmetry played little role in studies of charmed baryons. The detectors had similar characteristics,
using combination tracking of silicon vertex detectors and wire drift chambers immersed in a 1.5~T magnetic
field. Particle identification was
largely due to Cerenkov counters, and both had CsI-based crystal electromagnetic calorimeters. Belle took
an integrated luminosity of $\sim 1~{\rm ab}^{-1}$, surpassing the BaBar total by almost a factor of two, and much higher than CLEO's.
Clearly, the two experiments made the CLEO data set obselete.

In recent years, the Belle detector was rebuilt as the Belle~II detector, with improved particle identification systems
and improved precision in charged particle measurements near the beam interaction point. The design was to
take advantage of the increased luminosity of the Super-KEKB accelerator, itself largely due to the
extremely small beam profile. So far, charmed baryon analyses from Belle~II have been limited to lifetime
measurements, but with an integrated luminosity goal of $50~{\rm ab}^{-1}$, many more
studies will be performed in the years to come.

The BESIII detector operating at the $e^+e^-$~Beijing Electron–Positron Collider (BEPC)~II generally runs below the energy threshold for
charmed baryon production. However, some running has been made at the $\Lambda_c^+\bar{\Lambda}_c^-$ threshold
leading to some interesting studies of $\Lambda_c^+$~decays. It is possible that upgrades to the
accelerator energy will allow similar investigations of higher mass states.

Bottom baryons obviously have too high a mass to be produced at $B$~factories, and $e^+e^-$ machines
at higher energy (e.g. the Large Electron–Positron (LEP) collider at CERN) had rather low luminosities. Therefore, research into $B$ baryons has centered on
the hadron colliders. The experiments D0 and CDF operating at Fermilab were both used for early 
studies of $B$~baryons, but very much as a sideline to their main purposes.

More recently, the highest energies are available at the LHC. Here, the general purpose experiments, ATLAS and CMS
are clearly capable of adding to the world's knowledge of heavy hadron decays, but in general, this has not been
their priority. 
On the other hand LHCb, operating at the LHC, was designed to take advantage of the
huge cross-sections for charmed and bottom particles and to trigger on their finite lifetimes to help suppress
backgrounds. Unlike the other LHC experiments, which have been designed to maximise the solid angle of the detector, LHCb is only 
instrumented in 4\,\% of the solid angle in the forward region on one side of the interaction point (IP) but accepts a large fraction
of the beauty and charmed hadrons. The particles are precisely tracked near the IP, 
which enables reconstruction of primary
and secondary and further vertices from weak decays. 
The charged particles are bent by a dipole magnetic field, and a particle identification
system based on Ring-Imaging CHerenkov (RICH) detectors provides differentiation between protons, Kaons and pions. 

\subsubsection{Production and study of doubly strange $\Xi$ baryons in the decay of heavy baryons} Significant contributions have come from collider experiments in recent years even for our understanding of light doubly and triply strange hyperons. Excited $\Xi$~baryons are produced and have been studied in the decay of the charmed $\Lambda_c^+$ into $(\Sigma^+ K^-)_{\Xi(1690)}\,K^+$ and $(\Lambda \overline{K}^{\,0})_{\Xi(1690)}\,K^+$ by the Belle Collaboration~\cite{Belle:2001hyr}, and into $(\Xi^- \pi^+)_{\Xi^\ast}\,K^+$ by the BaBar Collaboration~\cite{BaBar:2008myc}, but also by Belle in the decay $\Xi_c^+\to(\Xi^-\pi^+)_{\Xi^\ast}\,\pi^+$ with unprecedented statistical quality. The $\Omega(2012)^-\to \Xi^0\,K^-~(\Xi^-\,K_S^0)$ was discovered by Belle in the decays of the heavy $b\bar{b}$~mesons $\Upsilon(1S), \Upsilon(2S)$, and $\Upsilon(3S)$. The ground-state $\Omega^-$ is copiously produced in the decay $\Xi_c^0\to \Omega^-\,K^+$ and with lower statistics in $\Omega_c^0\to \Omega^-\,\pi^+$. Such data samples were used by the BaBar Collaboration for a spin measurement of the $\Omega^-$~hyperon~\cite{BaBar:2006omx}, for instance.

\subsection{Production of $\Xi$ baryons in electromagnetically induced reactions}
In addition to studying heavy and doubly strange $\Xi$~baryons at
collider facilities, active spectroscopy programs are carried out by
the nuclear physics community at Jefferson Lab. The challenges
are similar to those experienced in the study of $\pi$-induced
reactions since two strange quarks need to be produced in a fairly
complex production process, and the cross sections are typically very
small. However, the luminosity of the available photon beams and the
advanced detector and read-out technology render such studies
feasible in all-exclusive measurements.
Given the lack of a suitable beam of $K$~mesons worldwide for hadron spectroscopy
experiments, photoproduction appears to be a very promising 
alternative. Results from earlier Kaon-beam experiments indicate that
it is possible to produce the $\Xi$~ground state through the decay of
high-mass $Y^\ast$~states~\cite{Tripp:1967kj,Burgun:1968ice,Litchfield:1971ri}.
It is therefore possible to produce Cascade resonances through 
$t$-channel photoproduction of singly strange hyperon resonances using the
photoproduction reaction $\gamma p\to K K\,\Xi^{(\ast)}$. 

The CEBAF Large Acceptance Spectrometer (CLAS) at Jefferson Lab using the CEBAF 6-GeV electron beam could be operated with electron
beams and with energy-tagged photon beams. It utilized information from 
a set of drift chambers in a toroidal magnetic field and time-of-flight
information to detect and reconstruct charged particles. Each of the six 
drift-chamber sectors was instrumented as an independent spectrometer
with 34~layers of tracking chambers allowing for the full reconstruction 
of the charged particle 3-momentum vectors. A detailed description of 
the spectrometer and its various detector components is given 
in Ref.~\cite{CLAS:2003umf}. Since CLAS was an almost pure spectrometer and had only a very limited
photon-detection coverage, an undetected neutral particle was inferred 
through the overdetermined event kinematics, making use of the good 
momentum and angle resolution of $\Delta p/p\approx 1\,\%$ and
$\Delta\theta\approx 1$--$2^\circ$, respectively.

The CLAS collaboration investigated $\Xi$~photoproduction in the reactions $\gamma p\to K^+K^+\,(\Xi^-)_{\rm miss}$ as well as $\gamma p\to K^+K^+\pi^-\,(\Xi^0)_{\rm miss}$
and, among other things, determined the mass splitting of the ground state ($\Xi^-$, $\Xi^0$) doublet to be $(5.4 \pm 1.8)$~MeV/$c^2$, which is consistent with previous
measurements. Moreover, the differential cross sections
for the production of the $\Xi^-$ (and the $\Xi(1530)^-$)~have been determined in
the photon energy range from 2.75 to 3.85 (4.75)~GeV~\cite{Guo:2007dw,CLAS:2018kvn}.
The cross section results are consistent with a production
mechanism of $Y^\ast\to \Xi^- K^+$ through a $t$-channel process. The reaction $\gamma p\to K^+K^+\pi^-\,(\Xi^0)_{\rm miss}$ was also studied in search of excited $\Xi$~resonances, but no significant signal
for an excited $\Xi$~state, other than the $\Xi(1530)$, was observed. The absence of higher-mass signals 
was very likely due to the relatively low photon energies available to these experiments and the limited 
acceptance of the CLAS detector. Equipped with a Kaon-identification 
system, the GlueX experiment in Hall D and the CLAS12 experiment in Hall B at JLab will be better suited to search for and study excited $\Xi$~resonances.

The polarized tagged-photon beam in Hall D is produced off a diamond radiator via the coherent bremsstrahlung
technique. The orientation of the diamond is chosen such that the highest degree of polarization is
achieved at photon energies between 8 and 9~GeV. The liquid hydrogen target is surrounded by the
GlueX apparatus, consisting of multiple tracking devices and calorimeters as well as particle identification
detectors. The entire experimental setup provides an almost 4$\pi$ coverage of the full solid angle.
A detailed description of the GlueX
detector system and beamline is given in Ref.~\cite{GlueX:2020idb}. Baryon spectroscopy is an important
component of the GlueX physics program. The study at GlueX of excited mesons and baryons, especially
in strange final states, is summarized in Ref.~\cite{GlueX:2012idx,GlueX:2013ffx}. GlueX completed its
first phase of data taking (GlueX Phase-I) during 2017 and 2018. This data sample comprises an integrated
luminosity of about 440~pb$^{-1}$ for $6~{\rm GeV} < E_\gamma < 11.6~{\rm GeV}$. More recently, GlueX
was upgraded from its initial baseline setup and a new DIRC detector was added to improve
$\pi\,/\,K$~separation. GlueX Phase-II has now taken data since 2020 at higher beam intensities.

The Cascade octet ground states $(\Xi^0,\,\Xi^-)$ can be studied in
photoproduction {\it via} exclusive $t$-channel (meson exchange)
processes in the reactions
\begin{eqnarray}
\label{Equation:GroundOctet}
\gamma p\,\to\, K\,Y^\ast\,\to\,K^+\,(\,\Xi^-\,K^+\,),~K^+\,(\,\Xi^0\,K^0\,),~K^0\,(\,\Xi^0\,K^+\,)\,.
\end{eqnarray}
However, the $\Xi$~octet ground states ($\Xi^0$ and $\Xi^-$) can be 
challenging to study via exclusive $t$-channel
production due to the high-momentum forward-going Kaon and
the relatively low-momentum pions produced in the $\Xi$~decays.
In contrast,
the production of the $\Xi$~decuplet ground state, $\Xi(1530)$, and other 
$\Xi^\ast$ states decaying to $\Xi\pi$ results in a lower momentum Kaon at 
the upper vertex, and heavier $\Xi$~states produce higher momentum 
pions in their decays.

The Cascade decuplet ground state, $\Xi(1530)$, and other excited Cascades
can be searched for and studied in the reactions
\begin{eqnarray}
\label{Equation:GroundDecuplet}
\gamma p\,\to\, K\,Y^\ast\,\to\,K^+\,(\,\Xi\,\pi\,)\,K^0,~K^+\,(\,\Xi\,\pi\,)\,K^+,~K^0\,(\,\Xi\,\pi\,)\,K^+\,.
\end{eqnarray}
The lightest excited $\Xi$~states of a given spin and parity $J^P$ are expected to decouple from $\Xi\pi$,
and can be searched for and studied in the reactions
\begin{eqnarray}
\label{Equation:Excited}
\gamma p\,\to\,K^+\,(\,K\Lambda\,)_{\Xi^{-\ast}}\,K^+,~K^+\,(\,K\Lambda\,)_{\Xi^{0\ast}}\,K^0
  ,~K^0\,(\,K\Lambda\,)_{\Xi^{0\ast}}\,K^+\,,\\[1ex]
\gamma p\,\to\,K^+\,(\,K\Sigma\,)_{\Xi^{-\ast}}\,K^+,~K^+\,(\,K\Sigma\,)_{\Xi^{0\ast}}\,K^0
  ,~K^0\,(\,K\Sigma\,)_{\Xi^{0\ast}}\,K^+\,.
\end{eqnarray}

In a similar way, the CLAS12 Collaboration aims to study $\Xi$~resonances in photoproduction and in electroproduction~\cite{Afanasev:2012fh}, e.g. in the reaction
\begin{eqnarray}
\label{Equation:GroundOctetCLAS12}
e\,p\,\to\, e^\prime\,K^+ K^+ \,\Xi^{\ast -} \,\to\, e^\prime\,\,K^+ K^+ K^-\,(\Lambda\,/\,\Sigma^0)\,.
\end{eqnarray}
Scattered electrons are detected either in the CLAS12 Forward Detector (FD), covering a polar angle range 
of $5^\circ$ to $35^\circ$ (in electroproduction), or in the CLAS12 Forward Tagger (FT), covering a polar 
angle range of $2.5^\circ$ to $4.5^\circ$ (in quasi-real photoproduction). The CLAS12 detector with its 
nearly $4\pi$~solid-angle coverage is used to detect scattered electrons and charged Kaons in the final 
state. The $\Lambda\,/\,\Sigma^0$~hyperons are reconstructed using the missing-mass technique to study intermediate doubly strange $\Xi$~states. More information on the CLAS12 experimental setup is available 
in Ref.~\cite{Burkert:2020akg}.

\subsection{Planned future experiments}
\subsubsection{Strange hadron spectroscopy with a tertiary $K_L$ beam at Jefferson Lab}
At Jefferson Lab, experiments are planned with a tertiary beam of neutral Kaons. The first data-taking period is currently expected for 2026/27 (subject to change). This incident $K_L$~beam will be
used for strange hadron spectroscopy in Hall~D in conjunction with the GlueX experimental setup. The high-quality CEBAF electron
beam will allow for a flux on the order of $1\times 10^4~K_L$ per second, which exceeds the flux of that previously attained at SLAC~\cite{Yamartino:1974sm} by three orders of magnitude. The use of a deuteron target will provide first measurements of $K^0 \,n$~interactions. The scientific goal for baryons is to extract both differential cross sections 
and polarization observables for the $K^0$\,-\,induced reactions producing $\Lambda$, $\Sigma$, $\Xi$, and $\Omega$
hyperons in the final state. The measurements are expected to cover a large center-of-mass angular range of
about $-0.95 < {\rm cos}\,\theta < 0.95$ for $W\in [\,1490, 2500\,]$~MeV.

To prepare the incident $K_L$~beam, 12-GeV electrons at a rate of $\sim 3\times 10^{13}$ per second will scatter off the copper radiator inside a Compact Photon Source~\cite{Day:2019qdz} generating an intense beam of untagged bremsstrahlung photons with an estimated intensity of $4.7\times 10^{12}$~$\gamma$/second for $E_\gamma > 1.5$~GeV. The photon beam will then impinge on a Beryllium target of 40~cm in length and 6~cm in diameter. The main source of producing $K_L$~mesons is the $\phi$-meson decay. The photoproduction threshold for the reaction $\gamma p\to p\phi$ is $E_\gamma\approx  1.58$~GeV. The distance between the Be~target and the liquid hydrogen target inside the GlueX apparatus will be about 24~m. Given this distance, neutrons are expected to be the main source of background in the $K_L$~beam. More details about this experimental facility and planned experiments can be found in Ref.~\cite{KLF:2020gai}.

\subsubsection{Excited $\Xi$~spectroscopy at the J-PARC high-momentum beamline}
Experiments on baryon spectroscopy using a $K^-$~beam (E50 Experiment) are also planned in Japan at the high-momentum secondary beamline of J-PARC~\cite{Nagae:2008zz}. An 8-GeV/$c$ $K^-$~beam will be incident on a liquid hydrogen target with a thickness of about 4~g/cm$^2$, resulting in an expected beam intensity of about $6\times 10^5$ $K^-$/spill. The current estimate is that the commissioning of the beamline will start around 2025/26, with a longer shutdown for the upgrade of the current experimental facility in 2027--2029. First data-taking is currently expected around mid-2029.  

The scientific goal is to study the production of excited $\Xi$~hyperons with masses up to 2.3~GeV/$c^2$ in $K^-\, p$~interactions. The resonances will be identified with a significance of at least $7\sigma$ and a mass resolution of about 6.6~MeV/$c^2$ using the missing-mass technique in the reaction $K^- \,p\to K^{\ast 0}\,\Xi^{\ast 0}$. Moreover, $K^-/\pi^+$~identification in the decays $\Xi^{\ast 0} \to \Sigma^+ K^-\,/\,\Xi^- \pi^+$ will be available. A RICH-type detector has been developed which will allow for $K\,/\,\pi$~separation for momenta greater than 5~GeV/$c$.
Based on simulations and feasibility studies, the expected yields range from about $1.3\times 10^4$~events for $\Xi^\ast$~states below 2~GeV/$c^2$ and $8.4 \times 10^3$ for the $\Xi(2030)$ to about $4.5\times 10^3$~events for higher-lying states in a data-taking period of about 30~days. More details can be found in the J-PARC P97~Proposal~\cite{J-PARC-P97}. Additional experiments are planned on excited $\Omega$~spectroscopy (J-PARC Proposal P85~\cite{J-PARC-P85}) and charmed-baryon spectroscopy (J-PARC Proposal P50~\cite{J-PARC-P50}).

\subsubsection{The $\overline{P}$ANDA experiment at FAIR}
The Proton antiproton ANnihilations at DArmstadt (PANDA) Experiment at the Facility for Antiproton and Ion Research (FAIR) in Darmstadt, Germany, will be studying hadrons in $p\overline{p}$~annihilation events and is well-suited for a comprehensive baryon spectroscopy program, in particular
on the spectroscopy of (multi-)strange and possibly
also charmed baryons. The experimental setup has been designed as multipurpose detector for a broad
physics program, with a particular focus on studying the strong interaction. The antiprotons will be delivered from the planned High Energy Storage Ring (HESR), see e.g. Ref.~\cite{Toelle:2007zz}, in
a momentum range from 1.5~GeV/$c$ up to 15~GeV/$c$. 

In $p\overline{p}$ collisions, a large fraction
of the inelastic cross section is associated with
channels resulting in a baryon-antibaryon pair in
the final state in reactions like $p\overline{p}\to Y\overline{Y}$. The production cross sections for excited $\Xi$~resonances in particular are expected
to be of the same magnitude as for ground-state
$\Xi$~production, i.e. for the reaction $p\overline{p}\to\Xi\overline{\Xi}$, for which cross sections of up to $2~\mu$b have been measured~\cite{Musgrave:1965zz}.
Reactions involving baryons in the final
state mostly proceed via excited states giving access to the decay modes of the populated resonances and to the
angular distributions of the decay particles. A particular
benefit of using antiprotons in the study of
\mbox{(multi-)strange} and charmed baryons is that in $p\overline{p}$~collisions, no production of extra Kaons or $D$ mesons
is required for strangeness or charm conservation,
respectively. This substantially reduces the energy thresholds as compared to pp collisions, for instance, and thus, also the number
of background channels. In addition, the requirement
that the patterns found in baryon and
antibaryon channels have to be identical reduces
the experimental systematic uncertainties. Reactions involving strange, multi-strange
and charmed baryons are characterised by
their displaced decay vertices, which can be identified and reconstructed by the expected good tracking
capabilities of the PANDA tracking system. The current estimate for first data taking is 2030 or later. For more details on the $\overline{\rm P}$ANDA experimental setup and planned physics program, see Ref.~\cite{PANDA:2009yku}.

%% file: LightQuarkBaryons.tex
\section{Light-Quark $\Xi$ and $\Omega$ Baryons}

\subsection{The known baryons}
In the 2022 edition of the RPP~\cite{ParticleDataGroup:2022pth}, the PDG lists a total of 12 doubly strange $\Xi$~hyperons, five of which are omitted from the Summary Table, and five triply strange $\Omega^-$~hyperons, three of which are omitted from the Summary Table. Only the decuplet ground states, $\Xi(1530)$ and $\Omega^-$, and the octet ground states, $\Xi^-$ and $\Xi^0$, currently have a 4-star status\,\footnote{See caption of Table~\ref{Table:NewResonances} for an explanation of the star assignments}. The existence of the remaining states in the Summary Table ranges from likely to certain with a 3-star assignment by the PDG. With the exception of those for the $\Xi^0$, $\Xi^-$, and $\Omega^-$, no decay branching fractions have been measured and values are merely listed as upper limits or the status is given as {\it seen} for all other resonances. Of the six $\Xi$~states that have at least a 3-star rating in the RPP, only two are listed with weak experimental evidence for their spin-parity ($J^P$) quantum numbers: $\Xi(1530)\,\frac{3}{2}^+$ and $\Xi(1820)\,\frac{3}{2}^-$. All other $J^P$ assignments, including the $J^P$ for the $\Xi(1320)$ ground state, are based on quark-model predictions. The quantum numbers for the $\Omega(1670)$ follow from the assignment of the particle to the baryon decuplet. The BaBar collaboration reported that the $\Omega^-$ spin is consistent with $J = \frac{3}{2}$~\cite{BaBar:2006omx}, dependent on the spins of the $\Xi_c^0$ and $\Omega_c^0$ being $J = \frac{1}{2}$. The spins of all other $\Omega$~resonances are unknown. 

\subsection{Recent results and the status of $\Xi$ baryons}
Almost all information on doubly and triply strange hyperons presented in the RPP comes from $K^-$- 
or hyperon $(\Sigma,\,\Xi)$\,-\,induced reactions. These experiments were performed until the 1990s and the $\Xi$~minireview in the 2022 edition of the RPP~\cite{ParticleDataGroup:2022pth} says that "nothing of significance on $\Xi$~resonances 
has been added since our 1988 edition." However, charmed baryon decays have emerged as a very powerful 
tool in the study of light-flavoured hyperons and the RPP~review appears outdated. Several $\Xi$~resonances 
have also been observed in photoproduction experiments at Jefferson Lab. While the current evidence for excited $\Xi$~states beyond the ground states remains suggestive in photoproduction, significant contributions to $\Xi$~physics are on the horizon. The exception to the suggestive observation in photoproduction is a clear signature for the $\Xi(1820)^-$~state in its decay to $\Lambda K^-$ observed by the GlueX Collaboration~\cite{Crede:2023ncq}. In the following, the current status of $\Xi$~resonances is discussed.

\subsubsection{The octet and decuplet ground-state $\Xi$ resonances}

\subsubsection*{\label{Subsubsection:Xi1320}The $\Xi(1320)$ Resonance} is based on the $(\Xi^-,\Xi^0)$~doublet of states, both are listed with a 4-star status in the RPP. The assigned $J^P$~quantum numbers of $\frac{1}{2}^+$ make these resonances the partners of the proton and the neutron in the ground-state octet of baryons. The positive parity is expected beyond any doubt, but has not been measured, yet. Some ideas for such measurements are discussed in Section~\ref{Subsubsection:PlannedExperiments}. The properties of the ground-state~$\Xi$ are reasonably well known. The mass splitting is listed as $M_{\Xi^-} - M_{\Xi^0} = (6.3\pm 0.7)$~MeV/$c^2$ (RPP~average)~\cite{ParticleDataGroup:2022pth}. The fairly large uncertainty is dominated by the uncertainty in the $\Xi^0$~mass. Only a few measurements have been performed and a single measurement of the $\Xi^0$~mass is based on more than 50~events~\cite{NA48:1999dxg}.  The dominant decay mode is $\Lambda\pi$ for both states with branching fractions for $\Xi^0\to \Lambda\pi^0$ and $\Xi^-\to\Lambda\pi^-$ of $(99.525\pm 0.012)$\,\% and $(99.887\pm 0.035)$\,\%, respectively. The radiative decay of $\Xi^0$ into $\Sigma^0\gamma$ is about an order of magnitude larger than the decay of $\Xi^-$ into $\Sigma^-\gamma$ with values of $(3.33\pm 0.10)\times 10^{-3}$ and $(1.27\pm 0.23)\times 10^{-4}$, respectively. The neutral state has of course also a radiative decay mode into $\Lambda\gamma$ with a branching fraction of $(1.17\pm 0.07)\times 10^{-3}$. For a discussion of the magnetic moments, we refer to the "Note on Baryon Magnetic Moments" in the $\Lambda$~listings of the RPP~\cite{ParticleDataGroup:2022pth}, and for the decay parameters to the "Note on Baryon Decay Parameters" in the neutron listings. A nice signal for the $\Xi^-$ in photoproduction from GlueX is shown in Fig.~\ref{Figure:Photoproduction}, left side.   

\subsubsection*{\label{Subsubsection:Xi1530}The $\Xi(1530)$ Resonance}
\begin{figure}[t]
\caption{\label{Figure:Xi1530} Observation of the $\Xi(1530)^0$ in the decay $\Lambda_c^+\to \Xi^-\pi^+ K^+$ reported by the BaBar Collaboration~\cite{BaBar:2008myc}. Left: The uncorrected $\Lambda_c^+$-mass-sideband-subtracted $\Xi^-\pi^+$ invariant mass distribution for $\Xi^- \pi^+ K^+$~candidates. Right: The cos$\,\theta_{\,\Xi^-}$~distribution (helicity frame) for $\Lambda_c^+\to \Xi^-\pi^+ K^+$ events in the $\Xi(1530)^0 \to \Xi^-\pi^+$ signal region after efficiency correction. The solid (dashed) curve corresponds to the parametrisation of the $\Xi(1530)$ angular distribution for the assumption of pure spin $\frac{3}{2}~(\frac{5}{2})$. See text for more details on the description of the distribution. Reprinted figure with permission from~\cite{BaBar:2008myc}, Copyright (2008) by the American Physical Society.}
\vspace{2mm}
\begin{minipage}{1.0\textwidth}
\begin{tabular}{cc}
\includegraphics[height=0.215\textheight]{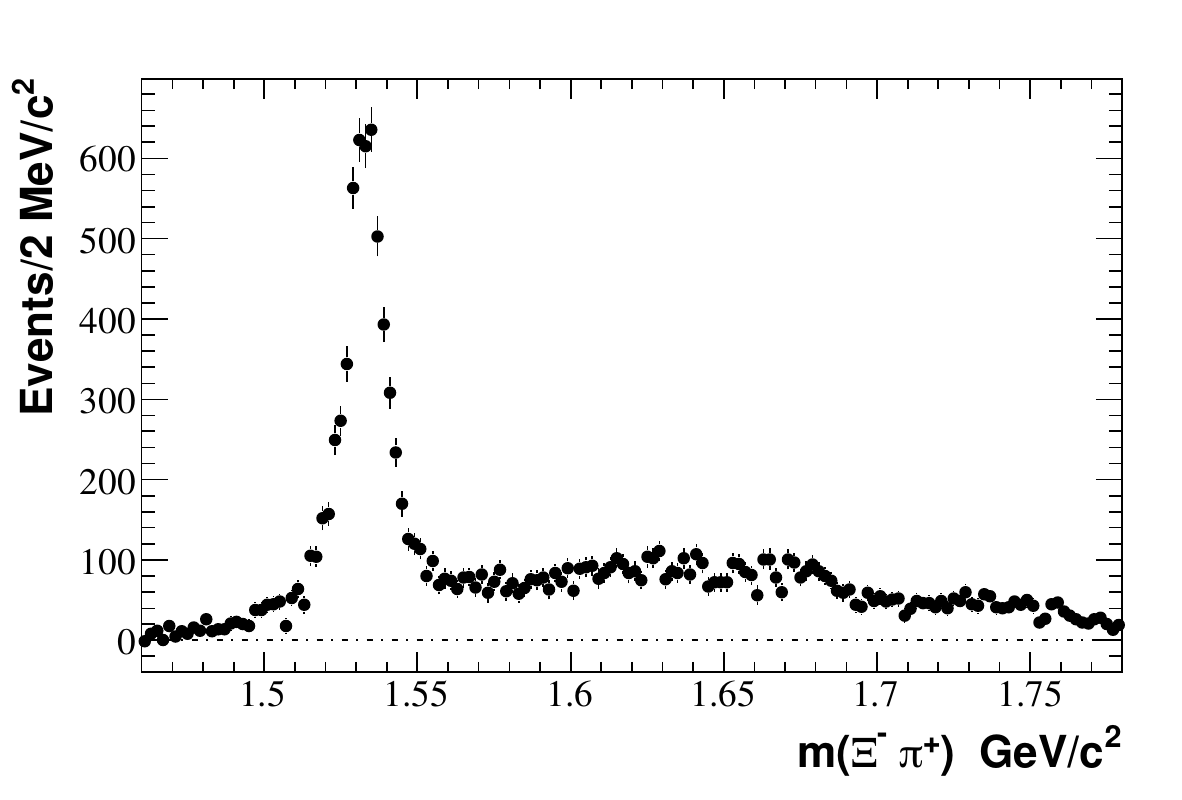} &
\includegraphics[height=0.215\textheight]{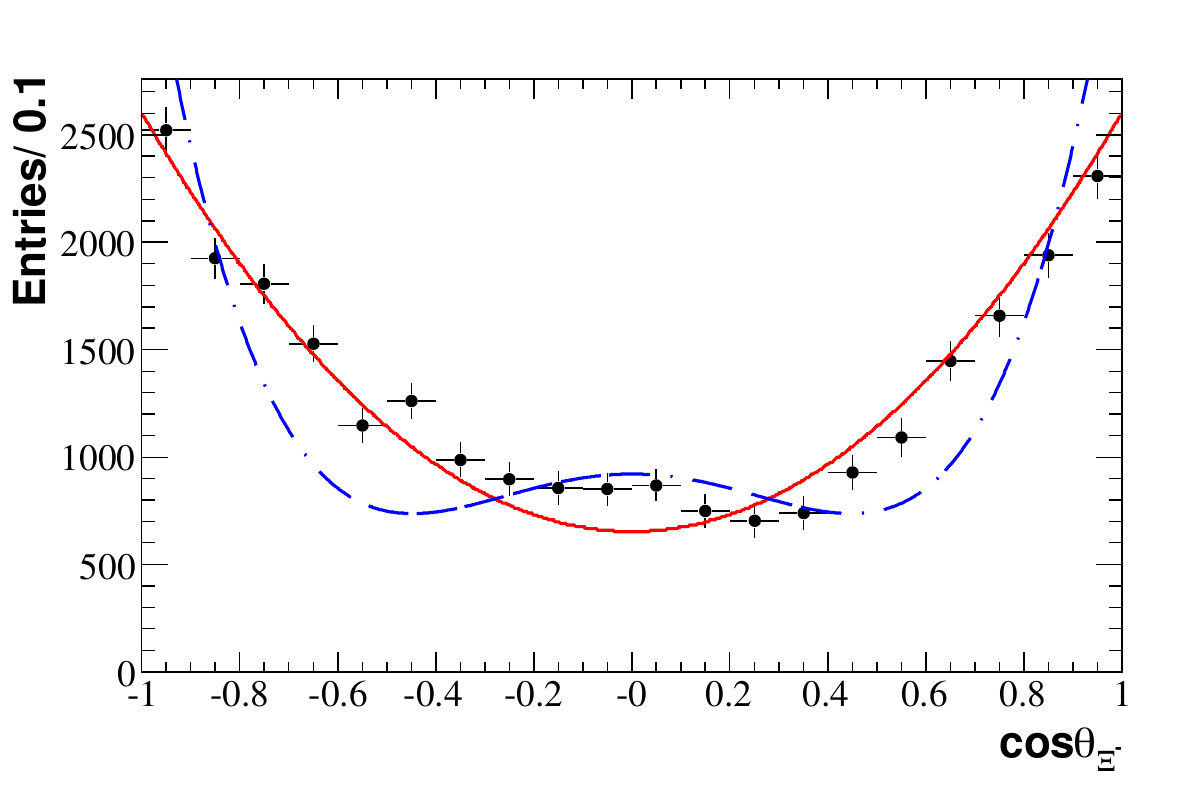}
\end{tabular}
\end{minipage}
\end{figure}

was discovered in the invariant $\Xi\pi$~mass spectrum and this decay mode is listed with a branching fraction of 100\,\%~\cite{ParticleDataGroup:2022pth}. A search for the radiative decay of the $\Xi(1530)^-$ into $\Xi^-\gamma$ was discussed in Ref.~\cite{Kalbfleisch:1974nb}, but the study was based on only $60\pm 10$~$(\Xi(1530)^-\to \Xi^-\pi^0)$~events to begin with. The group determined an upper limit of $<4$\,\%. However, the true branching fraction is undoubtedly much smaller. The $\Xi(1530)$~state is also the only excited $\Xi$~resonance whose mass splitting has been measured by several groups. The current value is $M_{\Xi(1530)^-} - M_{\Xi(1530)^0} = (2.9\pm 0.9)$~MeV/$c^2$ (RPP~average)~\cite{ParticleDataGroup:2022pth}.
The most recent observation of the $\Xi(1530)^0$~resonance was reported by the BaBar Collaboration in the decay $\Lambda_c^+\to (\Xi^-\pi^+)_{\Xi^{\ast 0}} \,K^+$~\cite{BaBar:2008myc} and by the Belle Collaboration in $\Xi_c^+\to (\Xi^-\pi^+)_{\Xi^{\ast 0}} \,\pi^+$~\cite{Belle:2001hyr}. The signals in the $\Xi^-\pi^+$~spectrum from BaBar and Belle are shown in Fig.~\ref{Figure:Xi1530} (left side) and in Fig.~\ref{Figure:1620-1690} (right side), respectively. The corresponding isospin-related negative $\Xi(1530)^-$~resonance was clearly observed recently in photoproduction by the CLAS Collaboration~\cite{Guo:2007dw,CLAS:2018kvn} and less significantly in the very first report based on photoproduction data discussed in Ref.~\cite{CLAS:2004gjf}. 

The $\Xi(1530)$ 4-star~state is a well-established $\Xi$~resonance that has been clearly observed in various production mchanisms ranging from $K^-$\,-, $\Xi^-$\,-, and $\gamma$-induced reactions to the decay of charmed baryons. The properties are all fairly well known and based on its mass, the assignment of the state to the ground-state decuplet of $J^P = \frac{3}{2}^+$~baryons is straightforward. However, there is even experimental evidence for the spin and parity of this resonance. In 2008, the BaBar reported on a spin measurement in a study of $\Lambda_c^+\to \Xi^-\pi^+ K^+$~decays and found that the spin of the $\Xi(1530)^0$ was $J = \frac{3}{2}$. The efficiency-corrected cos$\,\theta_{\,\Xi^-}$~distribution in the helicity frame for $\Lambda_c^+\to \Xi^-\pi^+ K^+$ events in the $\Xi(1530)^0 \to \Xi^-\pi^+$ signal region is shown in Fig.~\ref{Figure:Xi1530} (right side). The helicity angle $\theta_{\,\Xi^-}$ is defined as the angle between the direction of the $\Xi^-$ in the rest frame of the $\Xi(1530)^0$ and the direction of the $\Xi(1530)^0$ in the $\Lambda_c^+$~rest frame. Based on this choice, the  
$\Xi(1530)^0$ inherits the spin projection of the $\Lambda_c^+$ since any orbital angular momentum in the $\Lambda_c^+$~decay has no projection in this direction. The fit that assumes $J^P = \frac{1}{2}^+$ for the $\Lambda_c^+$ and pure $J = \frac{3}{2}$ for the $\Xi(1530)$ clearly describes the distribution better than the fit that assumes pure $J = \frac{5}{2}$ for the $\Xi(1530)$. However, the authors of Ref.~\cite{BaBar:2008myc} also discussed that the description of this angular distribution in terms of a single resonant structure was an {\it over-simplification} since deviations of the data points from the fit curve are apparent (solid line in Fig.~\ref{Figure:Xi1530}, right side). Therefore, the group also looked into a more complex description of the $\Xi^-\pi^+$~system and inferred the presence of a resonant $\frac{1}{2}^-$~amplitude associated with the $\Xi(1690)$~state adding coherently to the $\Xi(1530)\,\frac{3}{2}^+$~amplitude. Although the $\Xi(1690)$ is not observed directly in the $\Xi^-\pi^+$~mass spectrum (see Fig.\ref{Figure:Xi1530}, left side), the BaBar analysis found conclusively that the spin of the $\Xi(1530)^0$ is $J = \frac{3}{2}$ and provided some weaker evidence that the spin-parity of the $\Xi(1690)^0$ is $J^P = \frac{1}{2}^-$. Since the BaBar Collaboration has established $J=\frac{3}{2}$, the previous studies of Refs.~\cite{Schlein:1963zza,Button-Shafer:1966buh} confirm the positive parity for the $\Xi(1530)$.

\subsubsection{The 1600-1700~MeV mass region}
This mass range appears rich in resonant $\Xi^\ast$~structures. Two states are currently
listed in the RPP, $\Xi(1620)\,\ast$ and $\Xi(1690)\,\ast\ast\,\ast$. The lower-mass state was added in the 1976 edition as
$\Xi(1630)$ and the higher-mass state made its debut in 1980 as $\Xi(1680)$\,S$_{11}$. They have appeared with
their current names from the 1988 edition of the RPP~\cite{ParticleDataGroup:1988htj}. In a brief summary of their experimental properties, the
$\Xi(1620)$ is strongly coupled to $\Xi\pi$ with a surprisingly large width compared to other known $\Xi$~states, 
whereas the $\Xi(1690)$ is observed to couple strongly to $Y\overline{K}$ with a presumably narrow width of $\Gamma\approx 10$~MeV/$c^2$ comparable to the one of the $\Xi(1530)$.
The large width of the lower-mass state may hint at the presence of more than one state or at a more complex, exotic 
interpretation of the $\Xi(1620)$. The quantum numbers have not been measured beyond $I=1/2$. On the other hand, the
BaBar Collaboration has reported some evidence that the $\Xi(1690)$ has $J^P = 1/2^-$~\cite{BaBar:2008myc}.

\subsubsection*{\label{Subsubsection:1690} The $\Xi(1690)$ Resonance} The first indication for a new $\Xi^\ast$~resonance with a mass around 1690~MeV/$c^2$ emerged in the late 
1970s at CERN in experiments studying $K^- p$~interactions at 4.2~GeV/$c$ using the CERN 2-m hydrogen
bubble chamber~\cite{Amsterdam-CERN-Nijmegen-Oxford:1978chc}. The Amsterdam-CERN-Nijmegen-Oxford Collaboration reported on the 
observation of a threshold enhancement in the invariant $\Sigma\overline{K}$~spectrum. Both 
the neutral $(\Sigma^+ K^-)$ and the negatively $(\Sigma^0 K^-)$ charged state were observed. The
authors emphasized that the data from the $\Sigma\overline{K}$ channels alone could not distinguish between
a resonance and a large scattering length. However, similar structures in the $\Lambda\overline{K}$~system
and a coupled-channel analysis made the $\Xi^\ast$~resonance interpretation more plausible. The evidence
was weak, though, and based on fewer than 25~events for each $\Lambda\overline{K}$~charged state. 

The new $\Xi^\ast$~resonance was later confirmed in experiments using the CERN SPS hyperon beam. An
enhancement at 1700~MeV/$c^2$ in the $\Lambda K^-$~mass spectrum was first observed in diffractive
production in the reaction $\Xi^- N\to \Lambda K^-\,X$ and reported in Ref.~\cite{Biagi:1981cu}. In this
experiment, the ($\Lambda K^-$)~momentum was close to the two available beam momenta of 102~and 135~GeV/$c$. A 
follow-up experiment by the same group using incident $\Xi^-$~hyperons of mean momentum 116~GeV/$c$
is described in Ref.~\cite{Biagi:1986zj} and definitely confirmed the existence of the resonance in 
diffractive dissociation of the $\Xi^-$ into $\Lambda K^-$. A Breit-Wigner description of the narrow 
structure in the $\Lambda K^-$~channel yielded a mass of $M = (1691.1\pm 2.7)$~MeV/$c^2$ giving the 
new resonance its name. The authors claimed a statistical significance of $6.7\sigma$ based on $(104
\pm 24)$~events. A possible alternate decay mode into $\Xi\pi^+\pi^-$, possibly originating from
$\Xi(1530)$~decays remained very suggestive.

The WA89 Collaboration at CERN in 1998 unambigously confirmed the neutral member of the $\Xi(1690)$
doublet in 345~GeV/$c$ $\Sigma^-$~interactions with copper and carbon targets; details of this 
experiment can be found in Ref.~\cite{WA89:1997vxx}. A narrow peak of $(1400\pm 300)$~events was 
observed in the $\Xi^-\pi^+$~mass spectrum with mass and width values of $M = (1686\pm 4)$~MeV/$c^2$ 
and $\Gamma = (10\pm 6)$~MeV/$c^2$, respectively, consistent with the earlier measurements. The 
lower mass for the neutral $\Xi(1690)$~state is also in line with the isospin splittings observed 
in the $\Xi$~octet and decuplet ground states of $(6.3\pm 0.7)$~MeV/$c^2$ and $(2.9\pm 0.9)$~MeV/$c^2$, RPP averages~\cite{ParticleDataGroup:2022pth}, respectively.

\begin{figure}[t]
\caption{\label{Figure:1620-1690}(Colour online) Evidence for the $\Xi(1620)$ and $\Xi(1690)$ from the Belle Collaboration. Left: Invariant mass spectrum of $\Sigma^+ K^-$ combinations from the $\Lambda_c^+\to \Sigma^+ K^+ K^-$ signal area (data points) and from the $\Lambda_c^+$ sidebands~\cite{Belle:2001hyr}. The $\phi\to K^+ K^-$ signal region was excluded for this mass distribution. 
Reprinted from~\cite{Belle:2001hyr}, Copyright (2002), with permission from Elsevier.
Right: The invariant $\Xi^-\pi^+$ mass spectrum in the $\Xi_c^+\to\Xi^-\pi^+\pi^+$ signal region (data points), together with the fit result (solid blue curve) including the following components: $\Xi(1530)^0$ signal (dashed red curve), $\Xi(1690)^0$ signal (dot-dashed pink curve), $\Xi(1620)^0$ signal and non-resonant contribution (dot-dashed black curve), and the combinatorial backgrounds (dotted black curve)~\cite{Belle:2018lws}. The bottom plot shows the normalized residuals of the fits. This distribution is the strongest evidence for the $\Xi(1620)$~structure so far. Reprinted figure with permission from~\cite{Belle:2018lws}, Copyright (2019) by the American Physical Society.}
\vspace{3mm}
\begin{minipage}{1.0\textwidth}
\begin{tabular}{cc}
\includegraphics[width=0.6\textwidth, height=0.31\textheight]{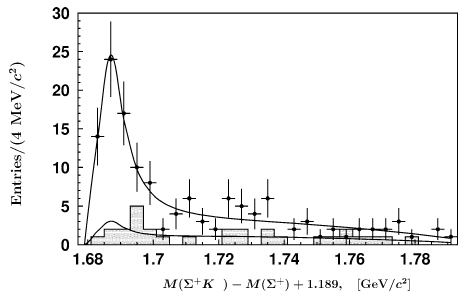} &
\includegraphics[height=0.30\textheight]{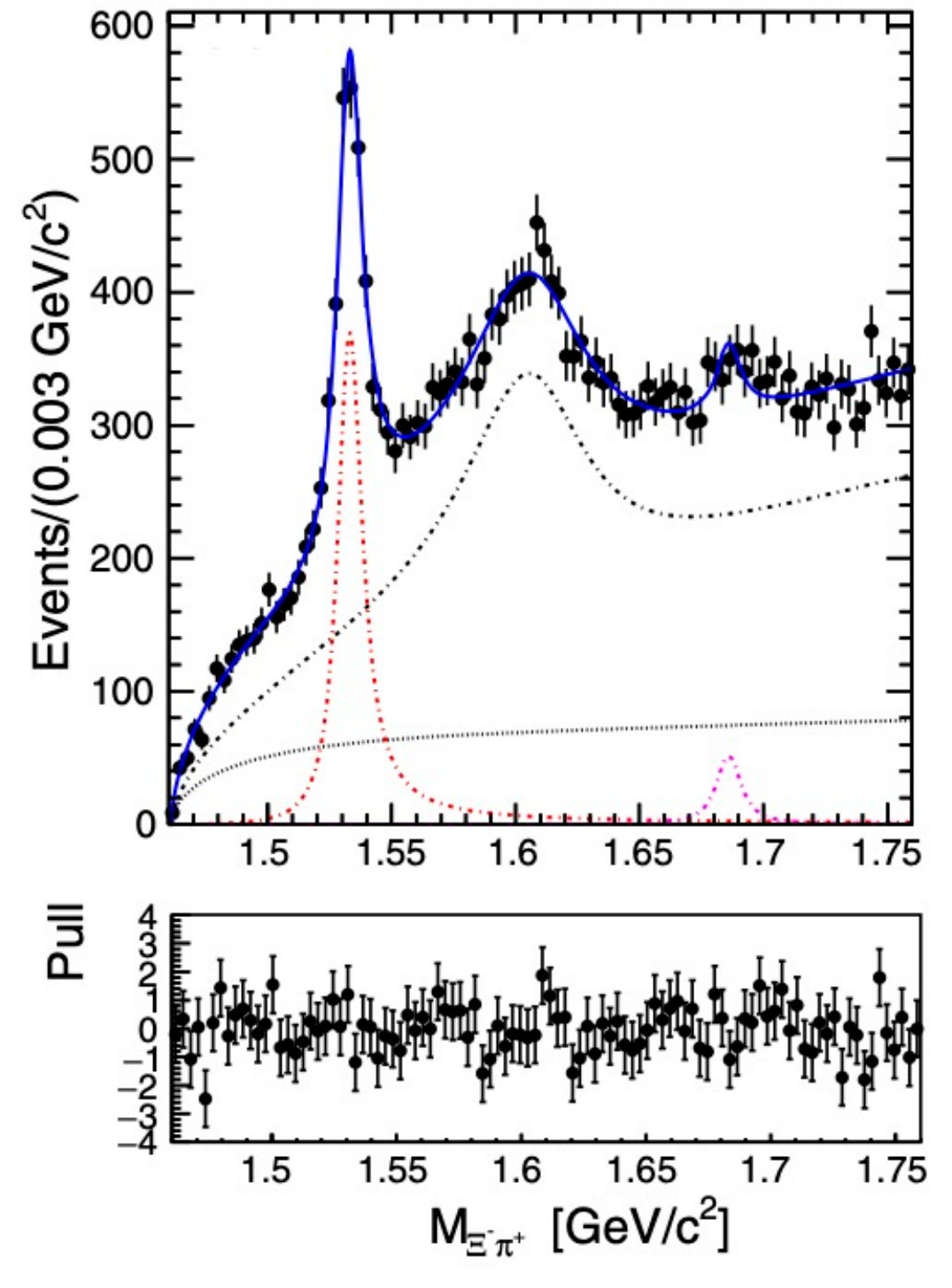}
\end{tabular}
\end{minipage}
\end{figure}

The most recent evidence for the $\Xi(1690)^0$~resonance comes from the $e^+ e^-$~collider experiments including the decay of charmed baryons.
The CLEO~Collaboration first observed the decay $\Lambda_c^+\to \Sigma^+K^+K^-$ which proceeds 
dominantly via W-exchange diagrams~\cite{CLEO:1993fhs}. About a decade later, the Belle Collaboration studied 
this decay with more accuracy and in 2002, reported first evidence for the decay $\Lambda_c^+\to 
\Xi(1690)^0 K^+$~\cite{Belle:2001hyr}. The invariant $\Sigma^+ K^-$~mass distribution showed a significant peak
and a fit yielded $(82\pm 15)$~events, and values for mass and width of $(1688\pm 2)$~MeV/$c^2$ and 
$(11\pm 4)$~MeV/$c^2$, respectively. The $\Sigma^+ K^-$~spectrum from Belle is shown in Fig.~\ref{Figure:1620-1690} (left side). This measurement confirmed the fairly narrow resonance width 
of just about 10~MeV/$c^2$. In the same study, the Belle Collaboration also observed a significant
signal for the decay $\Lambda_c^+ \to\Xi(1690) K^+\to (\Lambda \overline{K}^{\,0})\,K^+$ based on $(93\pm 
26)$~events and determined the ratio of the decay rates to be
\begin{eqnarray}
\frac{{\cal{B}}(\Xi(1690)^0\to\Sigma^+ K^-)}{{\cal{B}}(\Xi(1690)^0\to\Lambda\overline{K}^{\,0})} = 0.50\pm 0.26\,.
\end{eqnarray}
No signal was found for the $\Xi\pi$~decay of the $\Xi(1690)$~resonance in the $\Lambda_c^+\to (\Xi^-\pi^+)\,K^+$
decay mode.

The BaBar Collaboration reported on a study of the decay $\Lambda_c^+\to \Xi^-\pi^+K^+$ and found
the spin of the $\Xi(1530)^0$ to be $S=3/2$~\cite{BaBar:2008myc}. Moreover, the analysis of the Legendre 
polynomial moments of the $\Xi^-\pi^+$~system and an attempt to quantitatively describe the $\Xi(1530)$
line shape suggested the interference with a $S=1/2$~wave at higher mass. Therefore, the inference
of $J^P = 1/2^-$ for the $\Xi(1690)$ was concluded by the BaBar Collaboration in Ref.~\cite{BaBar:2008myc} as discussed earlier in Section~\ref{Subsubsection:Xi1530}.
No structure indicating the presence of the $\Xi(1690)$~resonance was directly observed in the
invariant $\Xi^-\pi^+$~mass spectrum. In 2019, the Belle Collaboration published the $\Xi^-\pi^+$
mass spectrum originating from the decay $\Xi_c^+ \to \Xi^-\pi^+\pi^+$ and observed a $4.0
\sigma$~evidence for the $\Xi(1690)^0$. Figure~\ref{Figure:1620-1690} (right side) shows the invariant $\Xi^-\pi^+$~mass from 
Belle and the structure around 1690~MeV/$c^2$ affiliated with the $\Xi(1690)$~resonance is clearly
visible. In the fit, the mass and width of the $\Xi(1690)$ were fixed to the values determined by the 
WA89 Collaboration~\cite{WA89:1997vxx} which gives a good qualitative description of the mass spectrum. 

A fairly significant $>10\sigma$~$\Xi(1690)$~signal has just recently been claimed by the BESIII~Collaboration in a study of $\psi(3686)$~decays~\cite{BESIII:2023mlv}. They have observed the negatively charged $\Xi(1690)^-$~state in the $K^- \Lambda$~spectrum in the reaction $e^+ e^-\to \psi(3686) \to (K^- \Lambda)_{\Xi^\ast}\,\overline{\Xi}^+ + c.c.$ with the subsequent decay $\overline{\Xi}^+ \to \bar{\Lambda}\pi^+ \to (\overline{p}\pi^+)\,\pi^+$. The group performed a partial wave analysis and the spin-parity quantum numbers were determined to be $J^P = \frac{1}{2}^-$, consistent with the standard $J^P$~interpretation, based on $(464\pm 43)$ observed $\Xi(1690)$~events. However, the mass of $M = 1685^{+3}_{-3}$~MeV/$c^2$ is at the lower end of previous mass measurements for the negatively charged state and the reported width of $\Gamma = 81^{+10}_{-9}$~MeV/$c^2$ is unusually large. The announcement is interesting since it adds yet another independent confirmation of the $\Xi(1690)$~resonance in a new production mechanism.

\subsubsection*{The $\Xi(1620)$ Resonance} The evidence for the second, lower-mass $\Xi$~resonance in the 1600--1700~MeV/$c^2$ mass region has been very weak
until recently. Initially listed as $\Xi(1630)$ by the Particle Data Group, some indications for the $\Xi(1620)$ 
already emerged in the late 60s and 70s, but the evidence was inconclusive. The strongest support was reported
in 1972 by the Oxford bubble chamber group in the reaction $K^- p\to \Xi^-\pi^+ (\pi^- K^+)_{K(890)^0}$ at the 
$K^-$~momenta 3.13, 3.30, and 3.58~GeV/$c$ using the CERN 2-m hydrogen bubble chamber~\cite{Ross:1972bf}. However,
the authors emphasized in their paper {\it that (1) the data of this experiment could not be taken as evidence for 
or against $\Xi(1630)$ and that (2) the possibly resonant state was the same as that reported earlier.} Additional
weak evidence from incident-$K^- p$~interactions was published in Refs.~\cite{Bellefon:1975pvi,Briefel:1977bp}. But other 
groups also searched for this excited $\Xi$~resonance and did not find any effect~\cite{Borenstein:1972sb,Hassall:1981fs}.

The first convincing evidence for a $\Xi^-\pi^+$~structure around 1620~MeV/$c^2$ was finally reported by the Belle 
Collaboration in the decay $\Xi_c^+\to\Xi^-\pi^+\pi^+$ based on a 980~fb$^{-1}$ data sample collected at the KEKB 
asymmetric-energy $e^+e^-$~collider~\cite{Belle:2018lws}. The mass and width have been measured to be $[1610.4\pm 6.0\,{\rm (stat.)}\,^{+6.1}_{-4.2}\,({\rm syst.})]$~MeV/$c^2$ and $[59.9\pm 4.8\,{\rm (stat.)}\,^{+2.8}_{-7.1}\,{\rm (syst.)]}$~MeV/$c^2$, respectively. Figure~\ref{Figure:1620-1690} (right side) shows the $\Xi^-\pi^+$~spectrum from Belle~\cite{Belle:2018lws}. A comparatively broad structure for the $\Xi(1620)^0$ is visible.

\subsubsection{The 1700-1900~MeV mass region}
Only one excited $\Xi$~resonance is currently known in this mass range. This state was already mentioned in the 1961 edition of the RPP as a $Y^\ast$~resonance with a mass of 1815~MeV/$c^2$, later in 1964 as $\Xi^\ast(1810)$, and is now listed as $\Xi(1820)\,\ast\ast\,\ast$ with $J^P$~quantum numbers of $\frac{3}{2}^-$. The $\Xi(1820)$ is the best studied excited $\Xi$~resonance and has been confirmed in various production mechanisms. It exhibits a very dominant $\Lambda \overline{K}$~decay mode and is the only excited doubly strange resonance listed in the RPP whose $J^P$~quantum numbers are fairly well known. 

\subsubsection*{The $\Xi(1820)$ Resonance} 
\begin{figure}[t]
\caption{\label{Figure:1820} Clearest (published) evidence for the $\Xi(1820)$ from the Amsterdam-CERN-Nijmegen-Oxford Collaboration at CERN~\cite{Amsterdam-CERN-Nijmegen-Oxford:1976ezm}. Shown are invariant squared-mass distributions. The cut in parentheses, $u^\prime = u - u_{\rm min}$, refers to the squared four-momentum transfer from the incident $K^-$ to the outgoing system which is shown in the figures. Left: Invariant $M^2_{\Lambda K^-}$ distribution for the reaction $K^- \,p\to \Lambda K^+ K^-$. The signal for the $\Xi(1820)^-$ at $M^2 \approx 3.3$~(GeV/$c^2$)$^2$ is clearly visible and is the most significant evidence for the state to date. The fit also describes the structure around $M^2 \approx 4.5$~(GeV/$c^2$)$^2$ associated with the $\Xi(2120)^-$. Right: Invariant $M^2_{\Xi^-\pi^+\pi^-}$ distribution for the reaction $K^- \,p\to \Xi^- K^+ \pi^+\pi^-$. Reprinted from~\cite{Amsterdam-CERN-Nijmegen-Oxford:1976ezm}, Copyright~(1976), with permission from Elsevier.}
\begin{minipage}{1.0\textwidth}
\begin{tabular}{cc}
\includegraphics[scale=0.83]{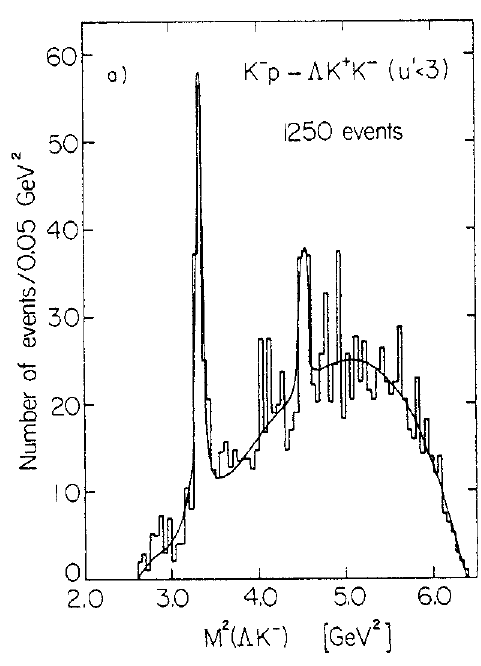} &
\includegraphics[scale=0.83]{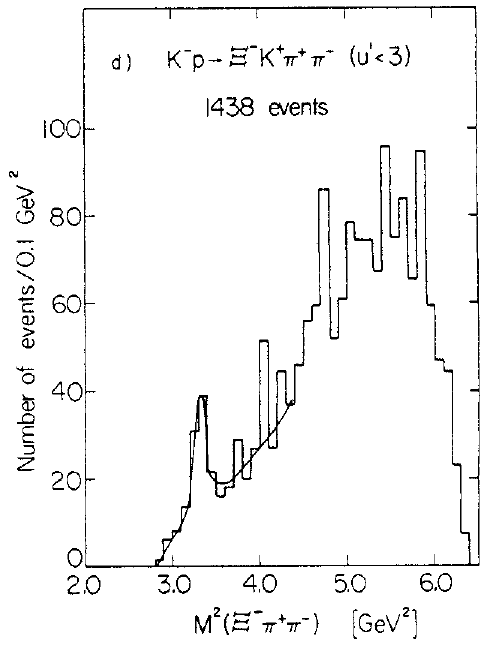}
\end{tabular}
\end{minipage}
\end{figure}

The first weak evidence for this resonance emerged in the early 1960s at the Lawrence Radiation Laboratory in Berkeley using an incident $K^-$~beam of momenta between 2.4 and 2.8 BeV/$c$~\cite{Smith:1965zze}. The resonance was observed in the $\Xi^-\pi^{+,0}$ and in the $\Lambda \overline{K}^{\,-,0}$ mass spectra with significantly weaker evidence for the neutral state. Subsequently, reports on the new resonance were also published by groups at CERN~\cite{Amsterdam-CERN-Nijmegen-Oxford:1976ezm,Badier:1972az} and BNL~\cite{Alitti:1969rb,Apsell:1970uf,Crennell:1970sd,DiBianca:1975ey}, which confirmed the initial observations in decay modes ranging from the hyperon channels, $\Lambda\overline{K}$ and $\Sigma\overline{K}$, to the $\Xi$~channels, $\Xi\pi$ and $\Xi\pi\pi$, including $\Xi(1530)\pi$. All these experiments used $K^-$~beams and bubble chambers of various sizes. The most significant evidence to date comes from the Amsterdam-CERN-Nijmegen-Oxford Collaboration in $K^- p$~reactions at 4.2~GeV/$c$ using the CERN 2-m hydrogen bubble chamber~\cite{Amsterdam-CERN-Nijmegen-Oxford:1976ezm}. The reported $8\sigma$~peak in the $\Lambda K^-$ mass spectrum is shown in Fig.~\ref{Figure:1820} (left side). 

In the 1980s, new hardware developments and electronic experiments allowed for more significant contributions and bubble chamber experiments became obsolete. The first observation of the $\Xi(1820)$ using the CERN SPS charged hyperon beam was reported in Ref~\cite{Biagi:1981cu}. The resonance was studied in $\Xi^- N$~interactions decaying to $\Lambda K^-$ and $\Lambda K_S$. Subsequent diffractive-production experiments at the CERN SPS confirmed the definite existence of the $\Xi(1820)^-$ in the $\Lambda K^-$ mass and hints for its existence in the $\Xi(1530)\pi$~mass spectrum~\cite{Biagi:1986zj}. A follow-up 
publication confirmed the strong $\Lambda \overline{K}^{\,0}$ decay mode of the $\Xi(1820)^0$ and the weaker $\Sigma^0\overline{K}^{\,0}$ decay. The experimental apparatus at the CERN SPS included a magnetic spectrometer, a threshold Cherenkov counter for $\pi/K$ separation, and two neutral particle detectors, and is described in Ref.~\cite{Biagi:1985rn}. In 1999, the WA89 Collaboration at CERN reported on a 
measurement of the differential and total cross sections of the inclusive production of $\Xi^\ast$~resonances in $\Sigma^-$-nucleus collisions at 
345~GeV/$c$~\cite{WA89:1999nsc}. The authors found the $\Xi(1820)^- \to \Xi(1530)^0\pi^-$ cross section to be an order of magnitude smaller than the one for inclusive $\Xi(1530)$~production.

Further evidence for the $\Xi(1820)$ was reported from Brookhaven using the BNL multiparticle spectrometer in the reaction $K^- p\to K^+\,X^-$ at 5~GeV/$c$~\cite{Jenkins:1983pm} and from SLAC in 11-GeV/$c$ $K^- p$~interactions using the LASS spectrometer~\cite{Aston:1985sn}. Surprisingly, the $\Xi(1820)$~state is also the dominant excited $\Xi$~resonance clearly observed in photoproduction~\cite{Crede:2023ncq}.   

The discussion on a possible $J=\frac{3}{2}$~assignment started shortly after the first discovery of the $\Xi(1820)$. The authors of the initial Berkeley experiment claimed that the resonant (plus background) events required a spin greater than $\frac{1}{2}$ in all channels, and that these events gave better $\chi^2$~values for $\frac{3}{2}^-$, than for $\frac{3}{2}^+$ parity~\cite{Smith:1965zze}. An explicit spin analysis based on data described in Ref.~\cite{Amsterdam-CERN-Nijmegen-Oxford:1976ezm} also favours $J=\frac{3}{2}$, but cannot make a parity discrimination~\cite{Teodoro:1978bu}. More recently, the study of the angular distributions of the decay chain $\Xi(1820)\to \Lambda \overline{K}^{\,0}$, $\Lambda\to p\pi^-$ recorded at the CERN SPS~\cite{Biagi:1986vs} using a double moment formalism determined that the spin was consistent with $J=3/2$ and that the decay angular distribution clearly supported negative parity for this $J$~value. Finally, the BESIII~Collaboration has recently reported on the observation of a $\approx 18\sigma$~$\Xi(1820)^-$~signal in the reaction $e^+ e^-\to \psi(3686) \to (K^- \Lambda)_{\Xi(1820)}\,\overline{\Xi}^+ + c.c.$~\cite{BESIII:2023mlv}, concurrently with a significant signal for the $\Xi(1690)^-$~state. The results of a partial wave analysis also confirm $J^P = \frac{3}{2}^-$ as the most likely spin-parity assignment for the $\Xi(1820)$ and $J^P = \frac{1}{2}^-$ for the $\Xi(1690)$. 

\subsubsection{The 1900-2100~MeV mass region}
Two structures shape the mass region around 2~GeV/$c^2$. The first is currently listed in the RPP as $\Xi(1950)\,\ast\ast\,\ast$ with unknown $J^P$~quantum numbers. This entry was upgraded to a three-star {\it resonance} from $\Xi(1940)\,\ast\ast$ in the 1988 edition of the RPP, a decision likely based on the new emerging evidence at the time from the hyperon-beam experiments at the CERN SPS using a magnetic spectrometer~\cite{Biagi:1986zj,Biagi:1986vs}. Many authors had reported enhancements in this mass region earlier, particularly in the $\Xi\pi$~channel, with conflicting resonance parameters, though. If just a single $\Xi$~state, the resonance would be unusually broad compared with all other well-established excited states. Therefore, the interpretation of this structure as a superposition of more than one state needs to be explored in future experiments. 
The second structure in this mass region is listed as $\Xi(2030)\,\ast\ast\,\ast$ with likely quantum numbers $J^P \geq 5/2$ based on a moments analysis described in Ref.~\cite{Amsterdam-CERN-Nijmegen-Oxford:1977bvi}. This comparatively narrow state earned three-star status from the beginning. Note that the star assignments were introduced in the 1984 edition of the RPP. The observed decay modes are dominated by the $\Sigma\overline{K}$~final state with almost negligible contributions from the typical $\pi$~channels.

\subsubsection*{The $\Xi(1950)$ Resonance} 
\begin{figure}[t]
\caption{\label{Figure:1950} Strongest evidence for the $\Xi(1950)$ to date. Left: The $\Xi^-\pi^+\pi^-$ effective mass distribution in different $x_F$~regions from the WA89 Collaboration based on \mbox{$\Sigma^-$-induced} reactions. The open circles denote an estimate of the background shape from {\it event mixing} (see Ref.~\cite{WA89:1999nsc} for more details on the procedure), and the stars represent the background shape from “wrong sign” combinations. The lower parts display the ratio of the observed spectra and the backgrounds from event mixing. Reproduced from~\cite{WA89:1999nsc}, with permission from Springer Nature.
Right: The $\Lambda \overline{K}^{\,0}$ effective mass distribution from the experiment 
at the CERN SPS charged hyperon beam providing the best evidence so far for the $\Xi(1950)\to Y\,\overline{K}$ in $\Xi^- N$~interactions~\cite{Biagi:1986vs}. Shown is the mass distribution for events where the $\overline{K}^{\,0}$ 
momentum is greater than 27.5~GeV/$c$ and the $\Lambda \overline{K}^{\,0}$~momentum is greater than 75~GeV/$c$. 
The smooth curve is the result of an unbinned maximum likelihood fit with a background plus two Breit-Wigner functions (and the experimental resolution folded in). Reproduced from~\cite{Biagi:1986vs}, with permission from Springer Nature.}
\vspace{2mm}
\begin{minipage}{1.0\textwidth}
\begin{tabular}{cc}
\includegraphics[height=0.28\textheight]{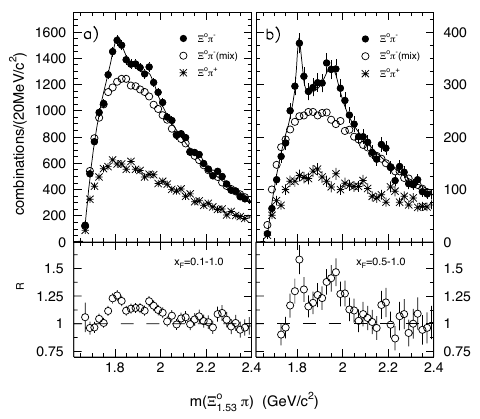} &
\includegraphics[height=0.28\textheight]{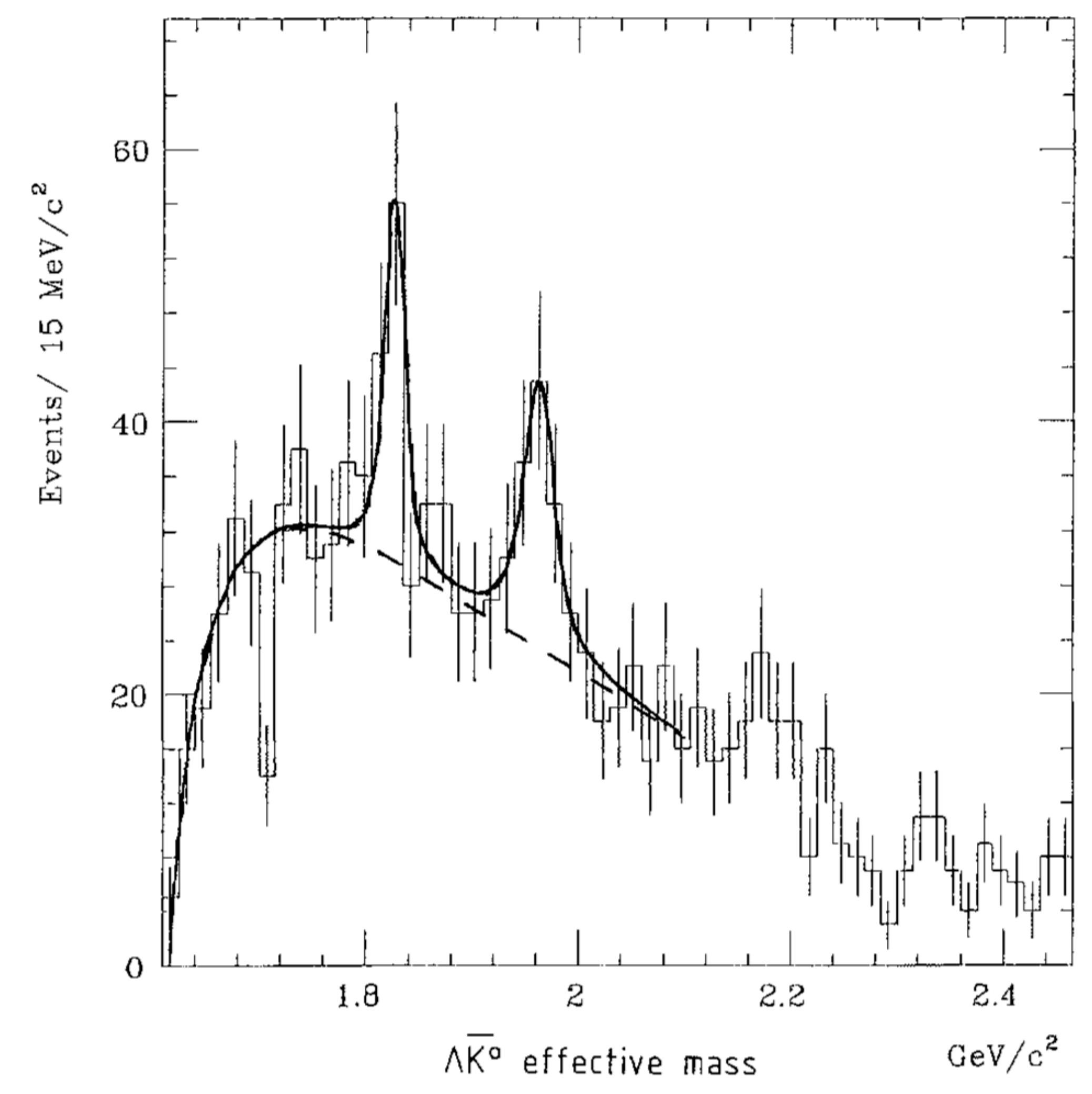}
\end{tabular}
\end{minipage}
\end{figure}

The best evidence for this structure was reported fairly recently by the WA89 Collaboration in Ref.~\cite{WA89:1999nsc} based on $\Sigma^-$\,-\,induced reactions at 345~GeV/$c$ using nuclear targets (one copper and three carbon blocks arranged in a row along the beam) and an incident hyperon beam at the CERN SPS. The $\Xi^-\pi^+\pi^-$~effective mass distribution is dominated by $\Xi(1530)^0\pi^-$ and shows a clear signal at 1820~MeV/$c^2$ and a convincing broader peak at about 1950 MeV/$c^2$ (see Fig.~\ref{Figure:1950}, left side). The mass and width of the $\Xi(1950)$ was determined to be $M = (1955\pm 6)$~MeV/$c^2$ and $\Gamma = (68\pm 22)$~MeV/$c^2$. The authors also studied the dependence of the differential cross sections of inclusive $\Xi(1820)^-$ and $\Xi(1950)^-$ production by $\Sigma^-$~baryons on the fractional longitudinal momentum $x_F$, which measures the momentum transfer in the forward direction. The excited $\Xi$~signals become much clearer for $x_F > 0.5$ and show significantly harder $x_F$ and $p_t$ (transfer momentum) distributions than for $\Xi^-$ and $\Xi(1530)^0$ hyperons. Both the $\Xi(1820)^-\to\Xi(1530)^0\pi^-$ and $\Xi(1950)^-\to\Xi(1530)^0\pi^-$ cross sections are observed to be about one order of magnitude smaller than for direct $\Xi(1530)^0$~production~\cite{WA89:1999nsc}.

In the early 1980s, the $\Xi(1950)^0$~structure was clearly observed in $\Xi^- N$~interactions using the CERN charged hyperon beam in its decay to $\Xi^-\pi^+$. The width of $\Gamma = (60\pm 8)$~MeV/$c^2$~\cite{Biagi:1981cu} was determined based on about 150~events, which is consistent with the measurement of Ref.~\cite{WA89:1999nsc} in $\Sigma^-$\,-\,nucleus collisions. The resolution of the magnetic spectrometer was about 20~MeV/$c^2$. The findings were confirmed a few years later in a subsequent experiment described in Ref.~\cite{Biagi:1986zj} based on about 130~events. A broad bump near 1950~MeV/$c^2$ was visible in the $\Xi^-\pi^+$~effective mass with an even bigger width of $\Gamma = (100\pm 31)$~MeV/$c^2$~\cite{Biagi:1986zj}. In that same experiment, the observation of $\Xi(1950)^-$ into $\Xi^-\pi^+\pi^-$ remained highly speculative. However, the same group observed a clear signal around 1960~MeV/$c^2$ in the $\Lambda\overline{K}^{\,0}$ effective mass~\cite{Biagi:1986vs}, which is shown in Fig.~\ref{Figure:1950} (right side). The fit yields a mass of $M = (1963\pm 5\pm 2)$~MeV/$c^2$ and a slightly smaller width of $\Gamma = (25\pm 15\pm 1.2)$~MeV/$c^2$, where the uncertainties are statistical and systematic, respectively. 

Finally, it is worth noting that the $\Xi(1950)$~structure has only been convincingly observed in hyperon-induced interactions. It is the only well-established $\Xi$~resonance, which was not observed in $K^- p\to K^+\,X^-$~reactions at the BNL experiment using the multiparticle spectrometer~\cite{Jenkins:1983pm}, for instance. The squared-mass distribution is shown in Fig.~\ref{Figure:2030} (right side) and no signal is visible at about $M^2 = 3.8$~(GeV/$c^2)^2$. The introduction in the RPP states that the {\it accumulated evidence for a $\Xi$ near 1950~MeV seems strong enough to include a $\Xi(1950)$ in the main Baryon Table, but not much can be said about its properties. In fact, there may be more than one $\Xi$ near this mass}~\cite{ParticleDataGroup:2022pth}. 

\subsubsection*{The $\Xi(2030)$ Resonance} 
\begin{figure}[t]
\caption{\label{Figure:2030} Left: Strongest evidence for the $\Xi(2030)$ to date from the Amsterdam-CERN-Nijmegen-Oxford Collaboration~\cite{Amsterdam-CERN-Nijmegen-Oxford:1977bvi} in the $\Sigma \,\overline{K}$~spectrum. a) $\Sigma^-\,\overline{K}^{\,0}$ mass spectrum for the reaction $K^- p \to (\Sigma^-\,\overline{K}^{\,0})\,K^+$, (b) same $\Sigma^-\,\overline{K}^{\,0}$ mass spectrum with different 
restrictions on the forward-going $K^+$~\cite{Amsterdam-CERN-Nijmegen-Oxford:1977bvi}, (c) $\Sigma^0 K^-$ mass spectrum for the reaction $K^- p\to (\Sigma^0 K^-)\,K^+$. Events with $M(K^+ K^-) < 1.06$~GeV/$c^2$ are excluded. (d) Total $\Sigma\overline{K}$ mass spectrum. The fit is based on an incoherent sum of simple Breit-Wigner functions and a polynomial background.
Reprinted from~\cite{Amsterdam-CERN-Nijmegen-Oxford:1977bvi}, Copyright (1977), with permission from Elsevier. Right: Missing squared-mass for $K^- p\to K^+ \,X$ at BNL using the multiparticle spectrometer~\cite{Jenkins:1983pm}. The acceptance is shown in (a). Two different $K^+$~detectors, $K_{\rm A}$ and $K_{\rm B}$, were used and results 
are shown in (b) and (c), respectively. In (b), the cross hatched area are events with detected $\Lambda\to p\pi$.
Reprinted figure with permission from~\cite{Jenkins:1983pm}, Copyright (1983) by the American Physical Society.}
\vspace{2mm}
\begin{minipage}{1.0\textwidth}
\begin{tabular}{ccccc}
 & & \includegraphics[height=0.33\textheight]{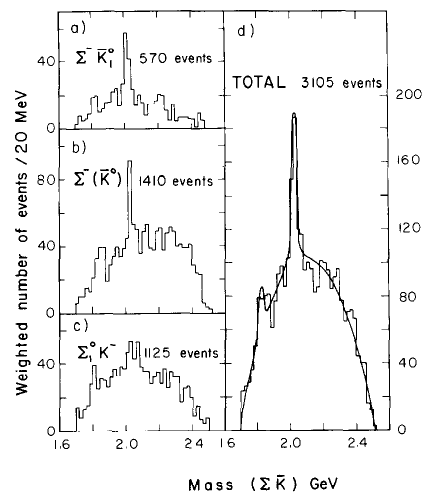} & &
\includegraphics[width=0.36\textwidth,height=0.33\textheight]{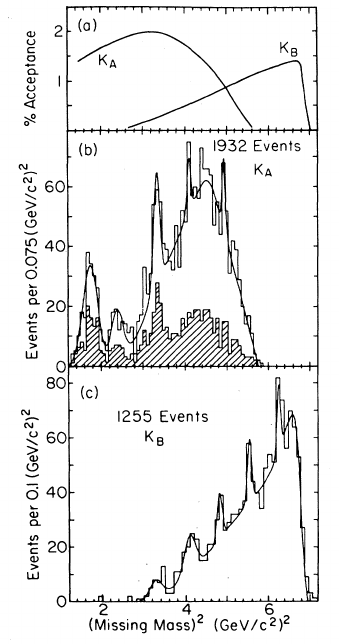}
\end{tabular}
\end{minipage}
\end{figure}

The case for the highest-mass $\Xi$~resonance with a three-star assignment is significantly easier to make than for the $\Xi(1950)$. The state was nicely and convincingly observed in the 1970s by the Amsterdam-CERN-Nijmegen-Oxford Collaboration in $K^-$\,-\,induced reactions. The results reported in Ref.~\cite{Amsterdam-CERN-Nijmegen-Oxford:1977bvi} were based on an analysis of a high-statistics exposure of $K^-\, p$~interactions in the CERN 2-m bubble chamber at 4.2 GeV/$c$ nominal incident momentum. An $8\sigma$~enhancement was observed in the invariant $\Sigma\overline{K}$~mass in the reaction $K^- \,p\to (\Sigma^- \overline{K}^{\,0})_{\Xi^\ast}\,K^+$, which is shown in Fig.~\ref{Figure:2030} (left side). Moreover, the $(\Lambda K^-)_{\Xi^\ast}\,K^+$~spectrum showed some weaker evidence for the $\Xi(2030)^-$ and no decay into $\Xi\pi\pi$ or $(\Lambda/\Sigma)\overline{K}\pi$ was reported by the group. All other reports of decays into non-$(\Lambda/\Sigma)\overline{K}$~channels remain controversial and the branching fractions are small at best. In a 1983~publication, the state was confirmed in $K^- \,p\to K^+\,X^-$~reactions at the BNL experiment using the multiparticle spectrometer~\cite{Jenkins:1983pm}. No further evidence has been reported since then. The most significant width measurement was reported by the Amsterdam-CERN-Nijmegen-Oxford Collaboration in 1977 and determined to be $\Gamma = (16\pm 5)$~MeV/$c^2$.

The Amsterdam-CERN-Nijmegen-Oxford Collaboration also performed a moments analysis of the decay angular distribution of the $(\Sigma^- \,\overline{K}^{\,0})$~system~\cite{Amsterdam-CERN-Nijmegen-Oxford:1977bvi} and found "significant $L=4$~moments" providing the strong indication that the spin of the $\Xi(2030)$~is at least $J = 5/2$. 

\subsubsection{The mass region above 2100~MeV} Four excited $\Xi$~states with masses above 2100~MeV/$c^2$ are currently listed in the RPP~\cite{ParticleDataGroup:2022pth}. Two of them have a one-star assignment, $\Xi(2120)$ and $\Xi(2500)$, whereas the other two have a two-star assignment, $\Xi(2250)$ and $\Xi(2370)$. The existence of any of these four states is either questionable or urgently requires an independent confirmation. Moreover, the $J^P$~quantum numbers are entirely unknown. For these reasons, all four states have been omitted from the RPP Summary Table.

\subsubsection*{The $\Xi(2120)\,\ast$ Resonance} The best and probably only evidence for this state is described in Ref.~\cite{Amsterdam-CERN-Nijmegen-Oxford:1976ezm}. The data were recorded by the Amsterdam-CERN-Nijmegen-Oxford Collaboration in the mid-1970s at a "new generation $K^- \,p$ experiment at
4.2~GeV/$c$, involving 3~million pictures from the CERN 2-m hydrogen bubble chamber"~\cite{Amsterdam-CERN-Nijmegen-Oxford:1976ezm}. Using about 2/3 of the accumulated events, the group observed a clear signal for the $\Xi(1820)^-$ in the invariant $\Lambda K^-$ mass and a previously unobserved peak at $M=2120$~MeV/$c^2$ (or $M^2 \approx 4.5$~(GeV/$c^2)^2$), with a statistical significance of about $4\sigma$~(standard deviations), shown in Fig.~\ref{Figure:1820} (left side). The observation of this structure in $\Lambda K^-$ clearly required $I = 1/2$. A maximum likelihood fit described in Ref.~\cite{Amsterdam-CERN-Nijmegen-Oxford:1976ezm} gave a mass of $(2123\pm 7)$~MeV/$c^2$ and a width of $(25\pm 12)$~MeV/$c^2$. The state is therefore fairly narrow. No other decay mode was observed in the analysis.

A subsequent publication in 1977 did not confirm the resonance in the $\Lambda K^-$~spectrum although more events were used in the analysis. The authors made the interesting point that a tighter cut on the squared four-momentum transfer was applied from the incident $K^-$ to the outgoing $\Lambda K^-$~system and suggested an anomalous production mechanism at large momentum transfer. The evidence for the state's existence is very poor but if real, the production of the $\Xi(2120)$ at a large momentum transfer in $K^- \,p$~interactions may be connected to the production of the $\Xi(1820)$ and $\Xi(1950)$ states at large values of $x_F$ (fractional longitudinal momentum) and $p_t$ (transfer momentum) in $\Sigma^-$\,-\,induced reactions reported by the WA89 Collaboration at CERN~\cite{WA89:1999nsc}. 

\subsubsection*{The $\Xi(2500)\,\ast$ Resonance} This state was first claimed in the late 1960s based on less than 50~events by groups at BNL~\cite{Alitti:1969rb} and CERN~\cite{Aachen-Berlin-CERN-London-Vienna:1969bau} in $K^- \,p$~interactions at 4.6--5.0~GeV/$c$ and 10~GeV/$c$, respectively. The reported resonance parameters were conflicting, with the BNL mass almost 100~MeV/$c^2$ lower and thus, the PDG states in the introduction that "the peak observed at BNL might be instead the $\Xi(2370)$ or might be neither the $\Xi(2370)$ nor the $\Xi(2500)$." In the 1980s, the state was also observed at BNL using the multiparticle spectrometer with a mass of $(2505\pm 10)$~MeV/$c^2$~\cite{Jenkins:1983pm}, consistent with the earlier CERN measurement. If real, the width of the peak is fairly unknown.

\subsubsection*{\mbox{The $\Xi(2250)\,\ast\ast$ Resonance}} 
\begin{figure}[t]
\caption{\label{Figure:2250-2370} Most significant evidence for the $\Xi(2250)\,\ast\ast$ and $\Xi(2370)\,\ast\ast$. Left: Invariant mass distribution for $\Xi^-\pi^+\pi^-$~combinations
in four- and five-body final states from Argonne~\cite{Goldwasser:1970fk}. The singly hatched distribution is for those combinations which do not have conflicting $K^\ast(890)$ or $\Xi(1530)^0$ combinations.
The doubly hatched distribution is for those selected combinations (from the singly hatched distribution) which have a $\Xi^-\pi^+$ combination in the $\Xi(1530)^0$ region. Reprinted figure with permission from~\cite{Goldwasser:1970fk}, Copyright (1970) by the American Physical Society.
Right: Combined $(Y\overline{K}\pi)^{0,-}$ and $(\Omega \overline{K})^{0,-}$ mass distribution in $K^- p$~interactions using the CERN 2-m bubble chamber at 8.25~GeV/$c$~\cite{Birmingham-CERN-Glasgow-MichiganState-Paris:1979oxu} (Birmingham-CERN-Glasgow-Michigan State-Paris LPNHE Collaboration). Reprinted from~\cite{Birmingham-CERN-Glasgow-MichiganState-Paris:1979oxu}, Copyright (1980), with permission from Elsevier.}
\vspace{2mm}
\begin{minipage}{1.0\textwidth}
\begin{tabular}{ccc}
\includegraphics[scale=0.95]{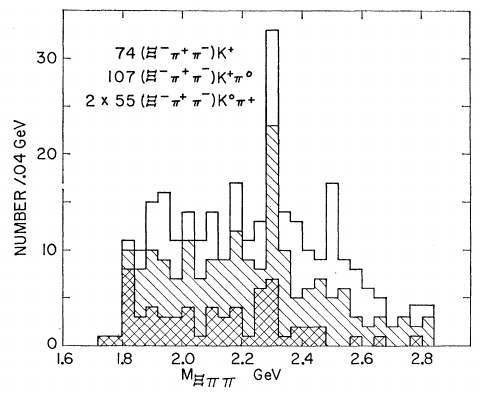} & &
\includegraphics[scale=0.80]{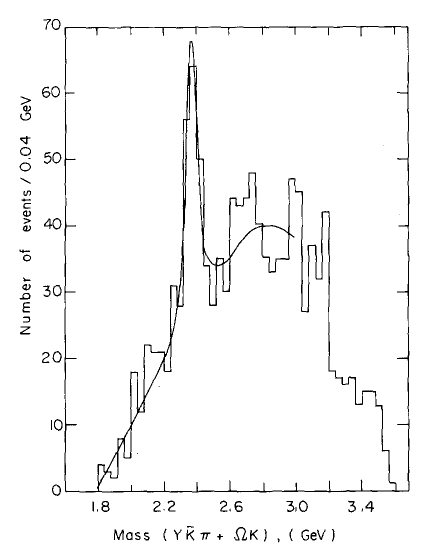}
\end{tabular}
\end{minipage}
\end{figure}

This state was first observed in the late 1960s at CERN in 10~GeV/$c$ $K^- \,p$~interactions. More details of the experiment can be found in Ref.~\cite{Aachen-Berlin-CERN-London-Vienna:1969bau}. The mass of $M = (2244\pm 52)$~MeV/$c^2$ was determined from a fit to the combined $Y\overline{K}\pi$ and $\Xi\pi\pi$~mass distribution using a Breit-Wigner description for the $\Xi^\ast$~signal. One year later, the initial claim for the $\Xi(2250)$ was confirmed at Argonne National Laboratory using the MURA 30-in. hydrogen bubble chamber, exposed to the high-purity separated 5.5-GeV/$c$ $K^-$~beam at the Argonne ZGS~\cite{Goldwasser:1970fk}. At a slightly higher mass of $(2295\pm 15)$~MeV/$c^2$, the group observed a remarkably narrow $3\sigma$~peak of less than $30$~MeV/$c^2$ in the $\Xi\pi\pi$~spectrum, based on just 18~events, though. The $\Xi^-\pi^+\pi^-$~signal is shown in Fig.~\ref{Figure:2250-2370} (left side). The observation was not confirmed in a subsequent experiment at Argonne~\cite{Hassall:1981fs} in the early 1980s. 

Additional confirmation was published in the 1980s from the group at BNL~\cite{Jenkins:1983pm} in $K^- \,p$~interactions at 5~GeV/$c$. The $\Xi(2250)$~state was observed in the missing mass in the reaction $K^-\,p\to K^+\,X$ and the resonance mass was determined to be $M = (2214\pm 5)$~MeV/$c^2$. Figure~\ref{Figure:2030} (right side) shows the missing-mass peak observed at BNL associated with the $\Xi(2250)^-$. The additional observation of the $\Xi(2250)$ by the group at CERN~\cite{Biagi:1986zj} in $\Xi^-\,Be$~interactions remains very speculative.

In summary, several groups have reported some evidence for this structure and thus, a two-star assignment seems justified. However, little is known about the decay modes and the RPP mass estimate of about 2250~MeV/$c^2$, giving the state its name, is based on a wide range of mass measurements. The reported width ranges from as small as less than 30~MeV/$c^2$ to possibly 150~MeV/$c^2$. If real, different $\Xi^\ast$~states might have been observed. 

\subsubsection*{\mbox{The $\Xi(2370)\,\ast\ast$ Resonance}} The evidence for the existence of this state is less controversial and the reported mass measurements are more consistent. The RPP mass estimate is about 2370~MeV/$c^2$ giving the state its name. The best evidence comes from the 1983 observation of a clear signal in $K^-\,p$~interactions by the group at BNL~\cite{Jenkins:1983pm}. The observed peak can be seen in Fig.~\ref{Figure:2030} (right side) at $M^2\approx 5.6$~(GeV/$c^2)^2$. The authors themselves speculated that the signal could be associated with the observed $6\sigma$~peak in the $(\Lambda/\Sigma)\overline{K}\pi$~spectrum originating from $K^-\,p$~interactions at 8.25 GeV/$c$ (CERN experiment) discussed in the 1980 Ref.~\cite{Birmingham-CERN-Glasgow-MichiganState-Paris:1979oxu}. Figure~\ref{Figure:2250-2370} (right side) shows the combined 
$(Y\overline{K}\pi)^-$ and $(Y\overline{K}\pi)^0$~spectrum. A clear signal for the $\Xi(2370)$ is visible. Somewhat weaker evidence for this state, also observed in the $(\Lambda/\Sigma)\overline{K}\pi$~spectrum, was presented in a $K^-\,p$~experiment at 6.5~GeV/$c$ (Argonne experiment) and discussed in the 1981 Ref.~\cite{Hassall:1981fs}. 

The authors of the BNL publication~\cite{Jenkins:1983pm} made an interesting point about the production of the $\Xi(2370)$. According to Ref.~\cite{Jenkins:1983pm}, an indication of a true $\Xi$~resonance is that its cross section varies in a reasonable manner as a function of incident $K^-$~momentum $p_{\rm \,lab}$, and that its decay modes are consistent at a variety of $p_{\rm \,lab}$~values. An empirical relationship for the
variation of the two-body production in the reaction $K^-\,p\to \Xi^{\ast\,-}\,K^+$ is
\begin{eqnarray}
  \sigma = A\,p_{\rm \,lab}^{\,\alpha}\,,
\end{eqnarray}
where $\alpha$ ranges from about 3.0 to 3.5~\cite{DiBianca:1975ey}. The authors of Ref.~\cite{Jenkins:1983pm} claimed good agreement of this empirical relationship with their measurements for all the well-established $\Xi^\star$~states. On the basis of the 8.25-GeV/$c$ data from the CERN experiment, it was expected that a cross-section value of at least $(6.3\pm 1.7)$~pb be observed in the 5~GeV/$c$ BNL experiment. However, a significantly lower cross section of only $(0.9\pm 0.3)$~pb was observed. The authors concluded that "in view of this discrepancy,
it appears that the $\Xi(2370)$ reported earlier is not a normal $\Xi^\ast$~resonance and does not seem to be produced via normal baryon exchange." 

\subsection{The status of $\Omega$ baryons}
Only five triply strange $\Omega$~baryons are currently listed by the Particle Data Group in the latest edition of the RPP~\cite{ParticleDataGroup:2022pth} and only three appear in the Summary Table. The ground-state decuplet $\Omega^-$~baryon is the only listed $\ast\ast\ast\,\ast$~resonance with only two additional $\ast\ast\ast$~states, $\Omega(2012)$ and $\Omega(2250)$. The remaining two resonances, $\Omega(2380)$ and $\Omega(2470)$ have a $\ast\,\ast$~status. 

Despite the discovery of the $\Omega^-$ in 1964, nothing was reported before the 1980s on excited $\Omega^-$~states. The reason is clearly that these searches are challenging because the $K^-\,p$~inclusive cross sections are very small and the experimental setups needed the required sensitivity and capability to reconstruct the complex decay chains.

\subsubsection{The decuplet $\Omega$ ground state}
The $\Omega^-$~baryon is the heaviest among all the ground-state baryons with a mass of ($1672.43\pm 0.32)$~MeV/$c^2$ (RPP average)~\cite{ParticleDataGroup:2022pth}. Its discovery in 1964 was a great success for the classification of hadrons based on the SU(3)$_{\rm flavour}$~symmetry ("Eightfold Way") because it was searched for and found after its existence, mass, and decay modes had been predicted by the American physicist Murray Gell-Mann~\cite{Gell-Mann:1961omu} and, independently, by the Israeli physicist Yuval Ne'eman~\cite{Neeman:1961jhl}. At the time, according to the rules of the eightfold way, the $J^P = \frac{3}{2}^+$~multiplet containing $4~\Delta$, $3~\Sigma$, and $2~\Xi$~resonances could be a {\bf 10} or a {\bf 27}. However, the {\bf 27} would involve baryons with positive strangeness, which had not been found. For this reason, Gell-Mann declared the multiplet a {\bf 10} at the 1962 Rochester Conference and made the prediction that the missing member had to be an $S=-3$, $I=0$, $J^P = \frac{3}{2}^+$~state with a mass of $M_\Omega - M_\Xi(1530) \approx M_\Xi(1530) - M_\Sigma(1385) \approx M_\Sigma(1385) - M_\Delta \approx 150$~MeV/$c^2$ and thus, $M_\Omega\approx 1680$~MeV/$c^2$. The state had to decay weakly since the lightest available $S = -3$~system, $\Lambda\,\overline{K}^{\,0} K^-$, has a mass of more than 2100~MeV/$c^2$. The classification scheme was not based on any underlying theory of fundamental structure. It simply provided a concise representation that exhibited symmetry and order and, additionally, predictive power. Now an established baryon, the triply strange $\Omega^-$ decays only via the weak interaction and has therefore a relatively long lifetime of $(0.821\pm 0.011)\times 10^{-10}$~s (RPP~average)~\cite{ParticleDataGroup:2022pth}. Furthermore, all $\Omega$~baryons including the $\Omega_c^0\,|ssc\rangle$, the charmed partner of the $|\Omega^-|sss\rangle$,
%which was discovered in 1985 by the WA62 Collaboration at CERN~\cite{Biagi:1984mu}, 
have isospin $I=0$ since they do not contain any up or down quarks.

The $\Omega^-$~baryon was discovered at BNL using the 80-inch bubble chamber in the reaction $K^-\, p\to \Omega^- K^+ K^0$~\cite{Barnes:1964pd}. The photograph in which the $\Omega^-$ was identified is shown in Fig.~\ref{Figure:OmegaDiscovery}. The interaction conserves strangeness and proceeds by the strong interaction. The $\Omega^-$~baryon has now been observed in various production mechanisms, e.g. in $K^-\, p$~interactions and using hyperon beams, but also in the decay of charmed baryons, e.g. in $\Xi_c^0\to \Omega^- K^+$ and $\Omega_c^0\to \Omega^- K^+$ at BaBar~\cite{BaBar:2006omx}. The decay proceeds dominantly via $\Lambda K^-$ with a branching fraction of $(67.8\pm 0.7)$\,\%~\cite{ParticleDataGroup:2022pth} or with a $\Xi$~hyperon in the final state in decays such as $\Xi^0\pi^-$ $(23.6\pm 0.7)\,\%$~\cite{ParticleDataGroup:2022pth} or $\Xi^-\pi^0$ $(8.6\pm 0.4)\,\%$~\cite{ParticleDataGroup:2022pth}. We refer to the RPP for other decay modes. The radiative decay into $\Xi^-\gamma$ is only poorly known and currently listed as $< 4.6\times 10^{-4}$~\cite{ParticleDataGroup:2022pth}. The magnetic moment was measured at Fermilab~\cite{Wallace:1995pf,Diehl:1991pp} at the FNAL hyperon facility and is listed as $(-2.02\pm 0.05)~\mu_N$~\cite{ParticleDataGroup:2022pth} (RPP average).

\begin{figure}[t]
\caption{\label{Figure:OmegaSpin} Spin measurement of the $\Omega$~baryon. Shown are the efficiency-corrected cos$\,\theta_h(\Lambda)$ distribution for $\Xi_c^0\to \Omega^- K^+$ from BaBar~\cite{BaBar:2006omx}. Left: The dashed curve shows the $J = \frac{3}{2}$~fit, in which the fit function allows for a possible non-zero asymmetry as a consequence of parity violation in the charmed baryon and the $\Omega^-$~weak decay. The solid curve represents the corresponding fit without the assumption of such an asymmetry. Right: The solid line represents the expected
distribution for $J = \frac{1}{2}$, while the dashed curve corresponds to $J = \frac{5}{2}$ without assuming any asymmetry. Reprinted figure with permission from~\cite{BaBar:2006omx}, Copyright (2006) by the American Physical Society.}
\vspace{2mm}
\begin{minipage}{1.0\textwidth}
\begin{tabular}{cc}
\includegraphics[width=0.49\textwidth,height=0.25\textheight]{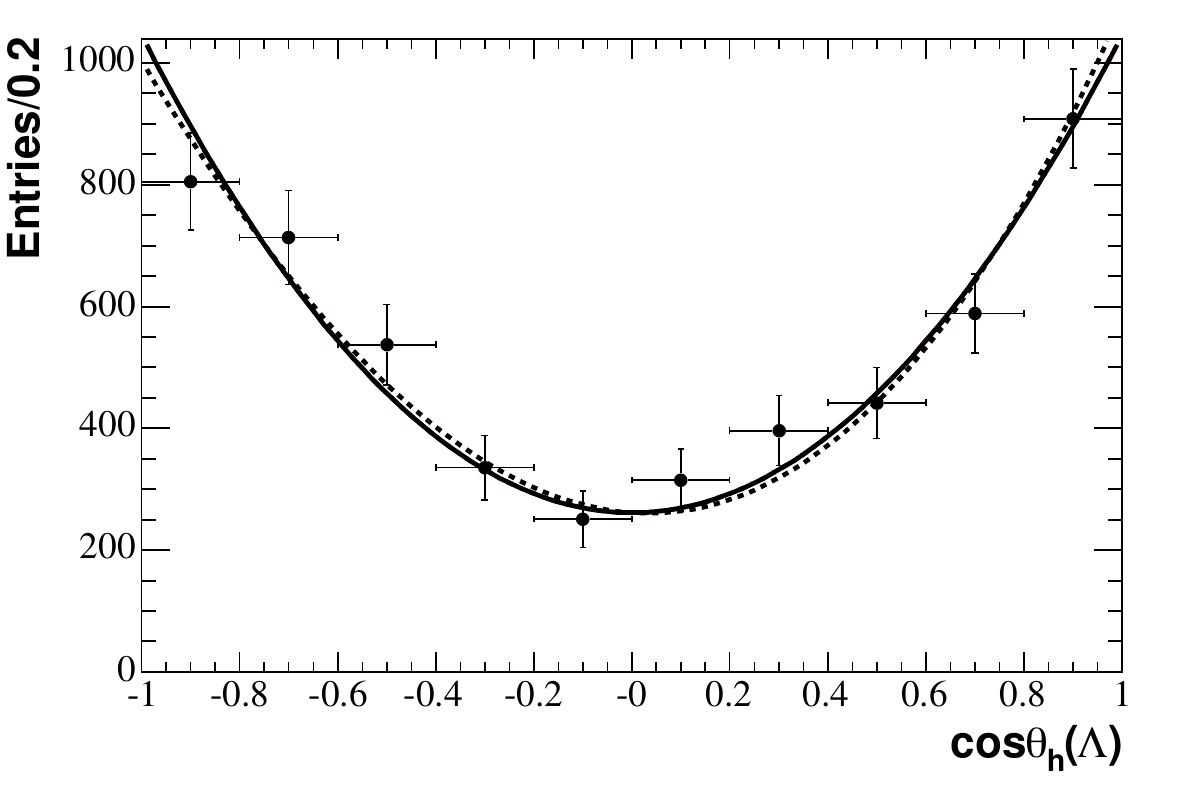} &
\includegraphics[width=0.49\textwidth,height=0.25\textheight]{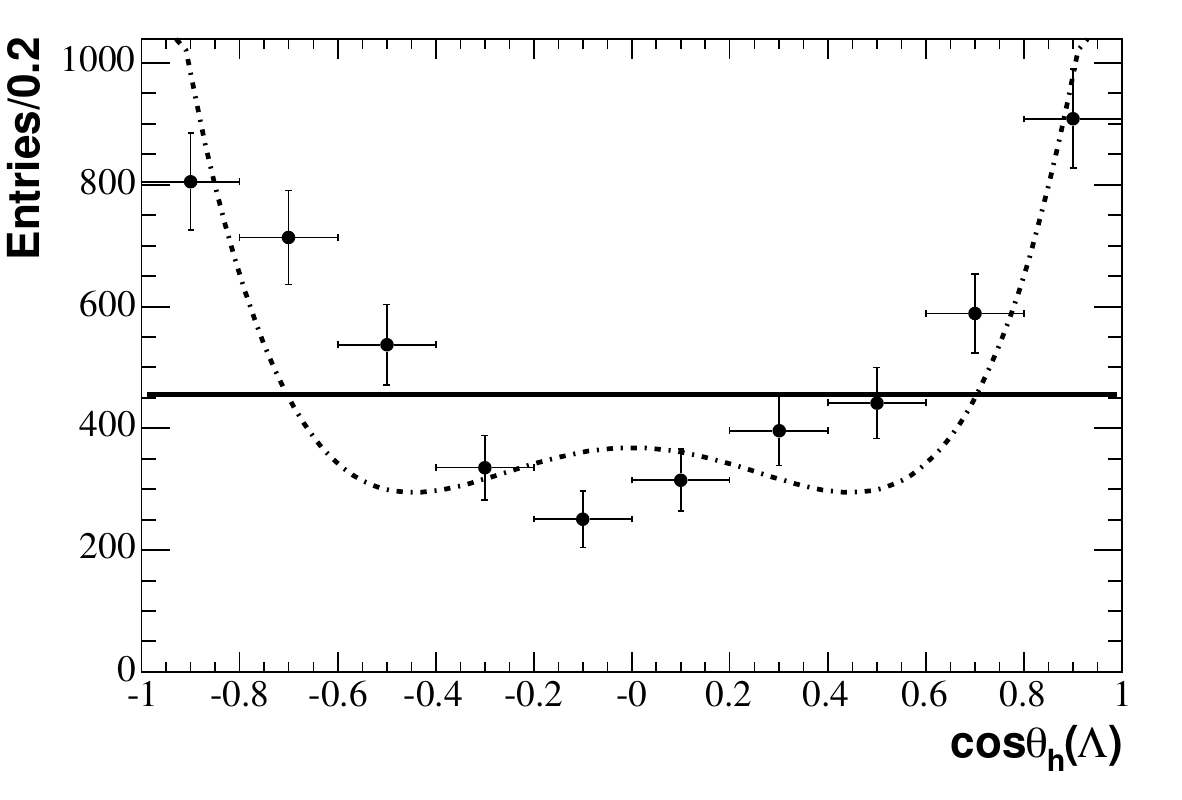}
\end{tabular}
\end{minipage}
\end{figure}

The $J^P$~quantum numbers of the $\Omega^-$ follow from the assignment to the ground-state decuplet. But attempts to measure the spin have been made. In the late 1970s, $J=\frac{1}{2}$ was ruled out and consistency with $J=\frac{3}{2}$ was claimed by the Aachen-Berlin-CERN-Innsbruck-London-Vienna Collaboration~\cite{Aachen-Berlin-CERN-Innsbruck-London-Vienna:1977ojz} and the Birmingham-CERN-Glasgow-MichiganState-Paris Collaboration at CERN~\cite{Birmingham-CERN-Glasgow-MichiganState-Paris:1978nrw}. In 2006, the BaBar Collaboration reported on a spin measurement based on $\Xi_c^0$ and $\Omega_c^0$~decays and found $J=\frac{3}{2}$, assuming that the spin for the charmed baryons is $J=\frac{1}{2}$, though. BaBar studied the helicity angle $\theta_h$ of the decay~$\Lambda$ defined as the angle between the direction of the $\Lambda$ in the rest-frame of the $\Omega^-$ and the quantisation axis (helicity frame).
The efficiency-corrected angular distribution of the $\Lambda$ from BaBar originating from the $\Omega^-$~decay is shown in Fig.~\ref{Figure:OmegaSpin}. The left side shows the $J_\Omega = \frac{3}{2}$~fit and the right side shows the $J_\Omega = \frac{1}{2},\,\frac{5}{2}$~fits. Clear consistency with $J = \frac{3}{2}$ is observed, see Ref.~\cite{BaBar:2006omx} for more details.

\subsubsection{The 2000--2300~MeV mass region}
Two triply strange $\Omega^\ast$~baryons are currently known in this mass range and these are listed in the RPP as $\Omega(2012)^-$ and $\Omega(2250)^-$. The lower-mass state has just recently been added in the 2019 edition of the RPP~\cite{ParticleDataGroup:2018ovx} and assigned a 3-star status from the beginning by the Particle Data Group. The spin of this resonance remains unknown but the parity is listed as negative. The higher-mass state was added in the 1988 edition and was also assigned a 3-star status right from the beginning. The $J^P$~quantum numbers of this resonance are entirely unknown, though.

Excited $\Omega$~resonances are generally expected to decay via the strong interaction. However, these triply strange states have isospin $I=0$ and thus, the decay mode into $\Omega^-\pi^0$ with the largest phase space is highly suppressed. For the lowest-lying excited states, the decay proceeds therefore most likely via $\Xi^- K^0_S$ or $\Xi^0 K^-$. A more complex decay chain for higher-lying states involving excited $\Xi$~states, e.g. $\Xi(1530)$, or involving an excited Kaon, e.g. $K^\ast(892)$, which both result in $(\Xi\pi K)^-$~final states, is also a possibility. 
These are the most promising final states to search for $\Omega^\ast$~resonances.

\subsubsection*{\mbox{The $\Omega(2012)^-\,\ast\ast\,\ast$ Resonance}} 
\begin{figure}[t]
\caption{\label{Figure:Omega2012-Omega2250} (Colour online) Left: Best (and sole) evidence to date for the $\Omega(2012)$ from the Belle Collaboration. The (a) $\Xi^0 K^-$ and (b) $\Xi^- K_S^0$ invariant 
mass distributions are shown in data taken at the $\Upsilon(1S)$, $\Upsilon(2S)$, and 
$\Upsilon(3S)$ resonance energies~\cite{Belle:2018mqs}. The curves show a simultaneous fit to the two distributions with a common mass and width. Reprinted figure with permission from~\cite{Belle:2018mqs}, Copyright (2018) by the American Physical Society. Right: Best evidence to date for the $\Omega(2250)$ from the LASS Collaboration at SLAC~\cite{Aston:1987bb}. Shown is the $\Xi(1530)^0 K^-$ mass distribution presented in 0.1~GeV/$c^2$ bins except in the region of the peak at $\sim 2.25$~GeV/$c^2$, where the bin size is 0.05~GeV/$^2$. The curve represents the result of a fit to the data using a single Breit-Wigner function and a linear background in (a) and using two Breit-Wigner functions in (b). Listed in the RPP is the average of these two fit results. The second peak is not listed in the RPP as an $\Omega^\ast$~candidate. Reprinted from~\cite{Aston:1987bb}, Copyright (1987), with permission from Elsevier.} 
\vspace{2mm}
\begin{minipage}{1.0\textwidth}
\begin{tabular}{cc}
\includegraphics[height=0.35\textheight]{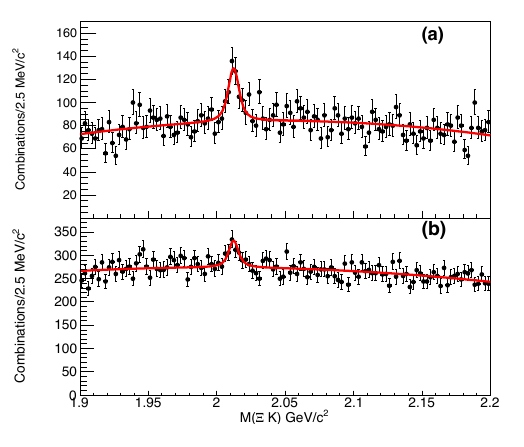} &
\includegraphics[height=0.35\textheight]{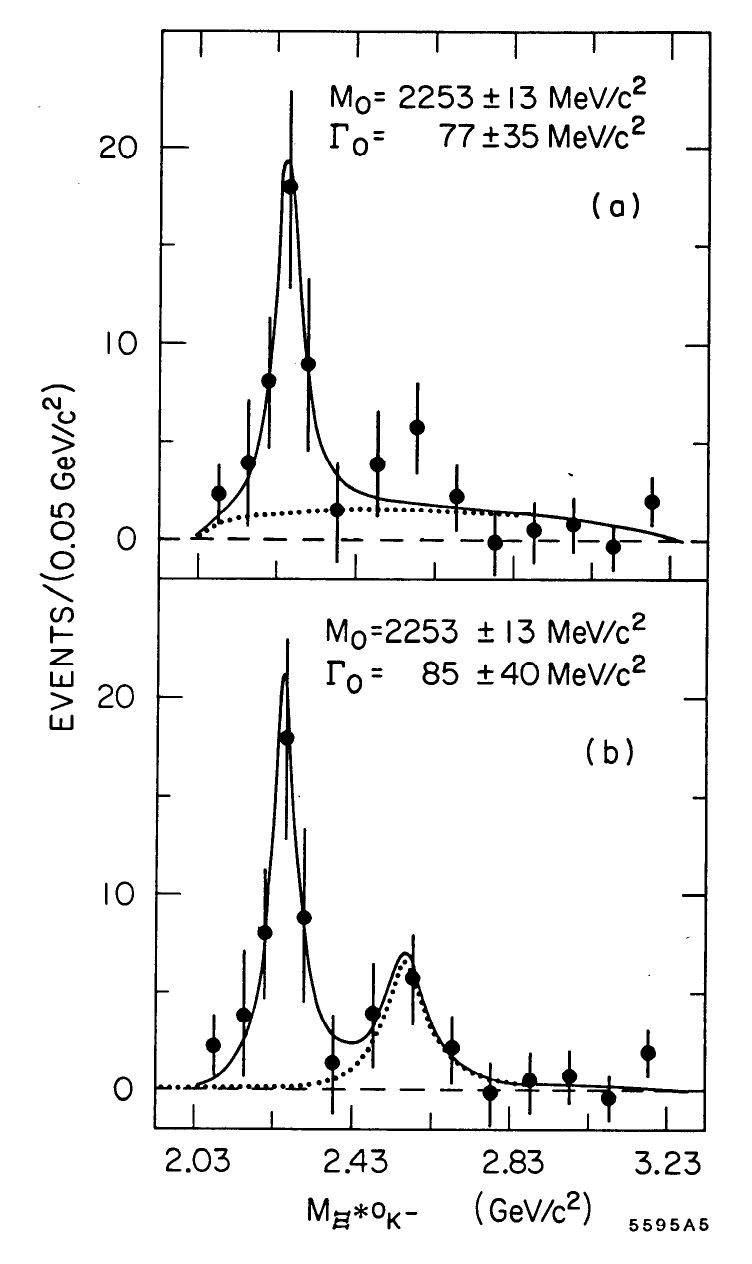}
\end{tabular}
\end{minipage}
\end{figure}

In 2018, the Belle Collaboration reported on the observation of this until-then-unknown excited $\Omega$~state in $e^+ e^-$~collisions~\cite{Belle:2018mqs}. Not surprisingly, the new resonance was observed in the invariant $\Xi^0 K^-$ and $\Xi^- K_S^0$~mass distributions. The Belle signals are shown in Fig.~\ref{Figure:Omega2012-Omega2250} for data taken at the $\Upsilon(1S)$, $\Upsilon(2S)$, and $\Upsilon(3S)$ resonance energies, where the latter two often decay
to $\Upsilon(1S)$, which is known to be a good factory for both baryons 
and strangeness. A simultaneous fit to the $\Xi K$~mass distributions gives a mass of $M = [2012.4\pm 0.7\,({\rm stat.})\pm 0.6\,({\rm syst.})]$~MeV/$c^2$ and a fairly narrow width of $\Gamma = [6.4^{+2.5}_{-2.0}\,({\rm stat.})\pm 0.6\,({\rm syst.})]$~MeV/$c^2$. Based on various mass predictions~\cite{Faustov:2015eba}, the authors of Ref.~\cite{Belle:2018mqs} speculate that the spin of the resonance is $J^{(P)} = \frac{3}{2}^{(-)}$ as an $\Omega^{\ast-}$ with this spin-parity would proceed via D-wave in the decay into $\Xi K$, whereas the decay of a state with $J^P = \frac{1}{2}^-$ would proceed via S-wave and thus be wide. The resonance has been confirmed by the same experiment, albeit with low statistics, in the substructure of $\Omega_c$ decays~\cite{Belle:2022mrg}.
Further investigation by the Belle Collaboration indicated that no signal was present for
$\Omega(2012)^-\to\Xi(1530)^0 K^-$, a decay on the edge of permissable phase 
space~\cite{Belle:2019zco}. However, a later (unpublished) analysis on the same data, using a more sophisticated modeling of the shape of both the parent $\Omega(2012)^-$ 
and daughter $\Xi(1530)^0$, 
implies a considerable branching fraction into this mode~\cite{Belle:2022mrg}.
This has been taken by some authors as an indication that the state should not be taken as a standard three-quark state~\cite{Ikeno:2023wyh}. However, if confirmed, its real value is in giving further evidence that the $\Omega(2012)^-$ has $J^P=\frac{3}{2}^-$. Actual calculations of the branching fractions are extremely difficult when the decay depends so critically on the width of the parent particle.
It is finally worth noting that the dominant systematic uncertainty of the mass measurement stems from the uncertainties in the ($\Xi^0,\Xi^-$)~masses, which enter almost directly into the calculation of the $\Omega^{\ast -}$~mass.

\subsubsection*{\mbox{The $\Omega(2250)^-\,\ast\ast\,\ast$ Resonance}} The best evidence for the $\Omega(2250)$~resonance was reported in the mid-1980s by the LASS Collaboration at SLAC based on the $K^-\,p$~program at \mbox{5--16~GeV/$c$}. The experiment and analysis are described in more details in Ref.~\cite{Aston:1987bb}. The state was observed in the $\Xi(1530)\overline{K}$~decay mode with a mass of 
\mbox{$M = (2253\pm 13)$~MeV/$c^2$} and a width of $\Gamma = (81\pm 38)$~MeV/$c^2$. The structure observed at LASS associated with the $\Omega(2250)^-$ is shown in Fig.~\ref{Figure:Omega2012-Omega2250} (right side).

Weaker evidence for this resonance was already published a year earlier based on data from the CERN SPS charged hyperon beam~\cite{Biagi:1985rn}. The resonance was observed in initial $\Xi^-$\,-nucleon interactions based on $(78\pm 23)$~events in the decay into $\Xi^-\pi^+ K^-$. The mass and width were extracted from a fit using a polynomial background and two Breit-Wigner functions. A mass of $M = (2251\pm 12)$~MeV/$c^2$ and a width of $\Gamma = (48\pm 20)$~MeV/$c^2$ were determined. This value is smaller than the width of $\Gamma = (81\pm 38)$~MeV/$c^2$ observed later at SLAC, which itself was based on just $(44\pm 11)$~events (4$\sigma$~significance).

\subsubsection{The 2300-2500~MeV mass region} Two excited $\Omega$~states in this mass range are listed in the latest version of the RPP~\cite{ParticleDataGroup:2022pth} and known as $\Omega(2380)$ and $\Omega(2470)$. Both states have a 2-star classification by the Particle Data Group and are omitted from the Summary Table. The existence of these states is plausible but nothing is known about the properties beyond the dominant decay mode in which they were observed. 

\subsubsection*{\mbox{The $\Omega(2380)^-\,\ast\ast$ Resonance}} 
\begin{figure}[t]
\caption{\label{Figure:Omega2380-Omega2470} Best evidence for the $\Omega(2380)$ and $\Omega(2470)$. Left: Invariant $\Xi^- \pi^+ K^-$ mass distribution from an experiment at the CERN SPS charged hyperon beam using incident $\Xi^-$~hyperons~\cite{Biagi:1985rn}. The fit on the left describes the mass spectrum with a third-order polynomial and a single Breit-Wigner function, whereas on the right, a polynomial background and two Breit-Wigner functions are used. The results of the latter fit are listed in the RPP. Reproduced from~\cite{Biagi:1985rn}, with permission from Springer Nature. Right: Invariant $\Omega\,\pi^+\pi^-$ mass distribution from the LASS Collaboration at SLAC~\cite{Aston:1988yn}. The fit uses a Breit-Wigner line shape and a polynomial background. Reprinted from~\cite{Aston:1988yn}, Copyright (1988), with permission from Elsevier.}
\vspace{2mm}
\begin{minipage}{1.0\textwidth}
\begin{tabular}{cc}
\includegraphics[height=0.3\textheight]{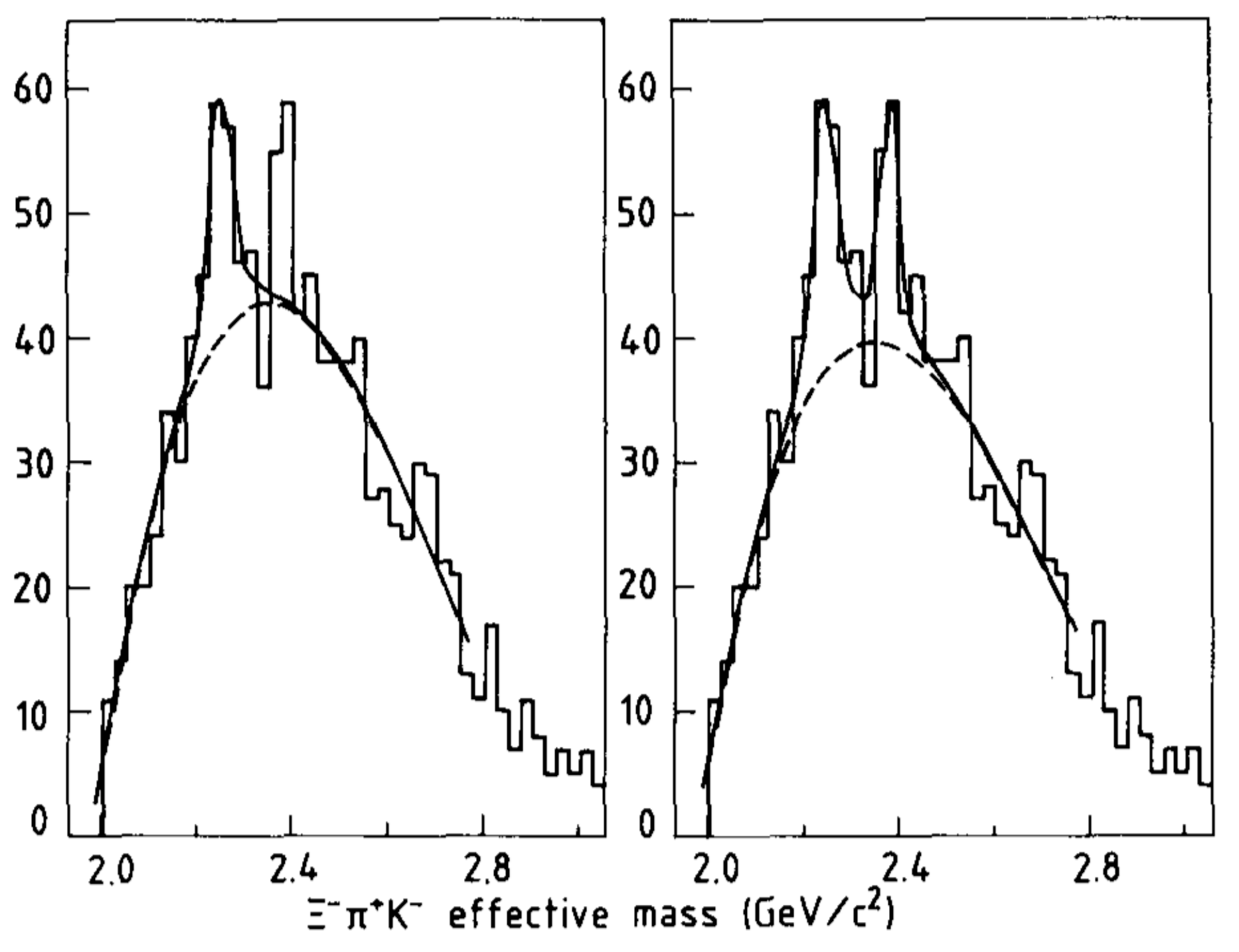} &
\includegraphics[height=0.3\textheight]{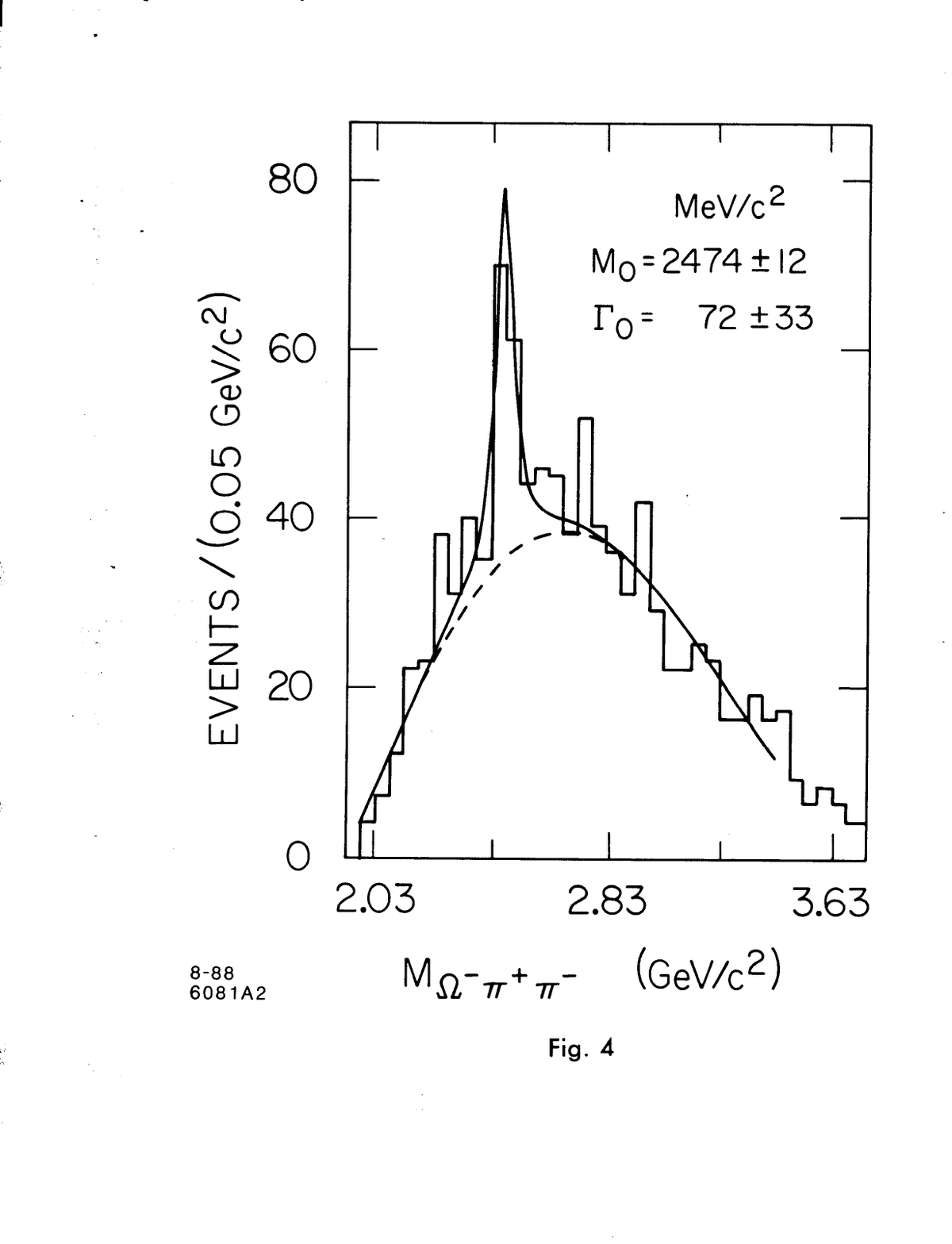}
\end{tabular}
\end{minipage}
\end{figure}

The sole evidence for this resonance stems from the same experiment at the CERN SPS that also reported on the $\Omega(2250)^-$ in $\Xi^-$\,-nucleon interactions. In the same 1986 publication~\cite{Biagi:1985rn}, a structure was presented slightly below 2400~MeV/$c^2$ in the $\Xi^-\pi^+ K^-$~spectrum based on $(45\pm 10)$~events (Fig.~\ref{Figure:Omega2380-Omega2470}, left side). The best estimate for the $\Omega(2380)^-$~signal parameters were determined from the same fit providing the parameters for the $\Omega(2250)^-$ and reported as $M = (2384\pm 9)$~MeV/$c^2$ and $\Gamma = (26\pm 23)$~MeV/$c^2$. Some weak branching ratio estimates relative to the $\Xi\pi^+ K^-$~decay mode were also reported for the branching fractions into $\Xi(1530)^0 K^-$ and $\Xi^- \,\overline{K}^{\,\ast}(892)^0$. However, nothing else is known about this excited $\Omega$~resonance and in retrospect, a 2-star assignment seems surprising in light of the evidence for other doubly strange states.

\subsubsection*{\mbox{The $\Omega(2470)^-\,\ast\ast$ Resonance}} This high-lying excited $\Omega$~state was the second signal observed by the LASS Collaboration at SLAC in $K^- \,p$~interactions at 11~GeV/$c$. The details of this analysis can be found in Ref.~\cite{Aston:1988yn}. A signal significance of at least $5.5\sigma$ was claimed based on the $\Omega^-\pi^+\pi^-$~decay mode and no further evidence was found for a direct decay into $\Xi^- K^- \pi^+$. The signal observed at LASS is shown in Fig.~\ref{Figure:Omega2380-Omega2470} (right side). A fit using a Breit-Wigner line shape gave a mass and width of $M = (2474\pm 12)$~MeV/$c^2$ and $\Gamma = (72\pm 33)$~MeV/$c^2$. No other evidence has been reported from a different experiment or production mechanism.

%% file: CharmedBaryons.tex
In the 2023 updated edition of the RPP, the PDG has listed 26~singly charmed, one doubly charmed, 
and 25~beauty baryons in its Baryon Summary Table~\cite{ParticleDataGroup:2022pth}. Only three heavy baryons, the $\Lambda_c^+$, $\Sigma_c(2455)$, and $\Xi_c^0$ have a 4-star assignment. Quite remarkably, the number of beauty baryons has more than quadrupled in the last ten years (up from six in the 2012 edition) and three additional charmed baryons are now listed. The number of beauty baryons in the Summary Table has even increased by six more states from the 2022 edition indicating the rapid progress in heavy baryon spectroscopy. For most of the beauty baryons, with
the exception of the $\Lambda_b^0$, the $(\Xi_b^-,\Xi_b^0)$~doublet, and the $\Omega_b^-$, only one decay mode has been observed and the status is merely given as {\it seen}. Many of the $I$,~$J$, or $P$ quantum numbers have not been measured, particularly parities, but are merely based on quark model expectations. However, the BESIII Collaboration has recently measured the $\Lambda_c^+$~ground-state spin to be $J = \frac{1}{2}$~\cite{BESIII:2020kap}, and several other studies are suggestive of the spin of higher-mass states based on decay angular distributions.

\subsection{The $\Lambda_c^+$ states}
There are four ground-state charmed baryon resonances, $\Lambda_c^+, \Xi_c^0, \Xi_c^+$, and $\Omega_c^0$, that can only decay weakly (note that the $\Sigma_c$ decays strongly), and the lowest-mass state of these, the $\Lambda_c^+$, 
comprises $cud$ quarks in an isospin-0 $(I=0)$ configuration. Not surprisingly, 
it was the first of the four $\frac{1}{2}^+$~ground-state resonances to be discovered and remains the most studied. The $\Lambda_c^+$~state was first reported in 1976~\cite{Knapp:1976qw}, which was a short time after the discovery of charm itself. 
The spin of the $\Lambda_c^+$ has now been directly measured by BESIII~\cite{BESIII:2020kap} to be $J=\frac{1}{2}$, an assignment which had previously been assumed due to quark model predictions.
There are now around 100 different decay modes measured. In general, the decay rates are extracted relative to that of
the decay $\Lambda_c\to pK^-\pi^+$. This mode was chosen because it has a comparatively high branching fraction and is experimentally
easy to detect because its final state consists of only stable charged particles. However, as precision has improved, the choice can be seen as
unfortunate because this three-body decay has a very complicated resonant substructure~\cite{LHCb:2022sck}, 
including interference effects between
the intermediate states, which means that finding its detection efficiency precisely using Monte Carlo simulations becomes difficult.
For years, results for the absolute branching fraction of any mode depended on a ``guesstimate'' of 
${\cal{B}}(\Lambda_c^+ \to pK^-\pi^+)$, but now there are two independent and rather direct measurements. Belle measured 
the absolute branching fraction using the ratio of fully reconstructed $e^+e^-$ events including the 
$\Lambda_c^+ \to p K^- \pi^+ $ decay to those where 
the existence of the $\Lambda_c^+$ is inferred by the missing mass~\cite{Belle:2013jfq}, and found
${\cal{B}}(\Lambda_c^+\to pK^-\pi^+) = (6.84\pm 0.24^{+0.21}_{-0.27})$\,\%.
BESIII used a more straightforward method by running at the $\Lambda_c^+\bar{\Lambda}_c^-$ 
threshold, reconstructing one ``side'' of the event, and measuring how often the other ``side'' is reconstructed~\cite{BESIII:2015bjk}. 
Their measurement of $B(\Lambda_c^+\to pK^-\pi^+)\ =\ (5.84\pm0.27\pm0.23)\%$ is in slight tension with the Belle result. 
Making reasonable assumptions for unmeasured decays (particularly
those with multiple neutrals), the truth seems to exist in between the two numbers. No evidence exists
that decays are missing from the lists, and any undiscovered decays must be rare.

By far the most precise measurement of the $\Lambda_c^+$ mass comes from the BaBar Collaboration in their 2005 study of low-$Q^2$ decays into 
$\Lambda K^0_S K^+$ and $\Sigma^0K^0_SK^+$, and is $M_{\Lambda_c^+}=(2286.4 \pm 0.14)$~MeV/$c^2$~\cite{BaBar:2005wur}. 

\subsubsection{The $\Lambda_c(2595)^+$ and $\Lambda_c(2625)^+$ resonances}
It took 17 years between the discovery of the $\Lambda_c^+$~resonance and the first observation of an excited state. The report was finally made by the ARGUS Collaboration,
looking at the decay mode $\Lambda_c^{*+}\to \Lambda_c^+\pi^+\pi^-$ and finding evidence~\cite{ARGUS:1993vtm} of a narrow state we now know as the $\Lambda_c(2625)^+$. 
This announcement was followed shortly afterwards by a report from the CLEO Collaboration, who at the time had more data than ARGUS, 
confirming this state and even adding one at a lower mass
we now know as the $\Lambda_c(2595)^+$~\cite{CLEO:1994oxm}.
Other experiments soon confirmed the existence of both these two particles, and their major decay modes. 
For the higher-mass state, the three-body decay proceeds mostly non-resonant, and for the lower-mass state, the decay resonates through an intermediary $\Sigma_c$.
Their masses
and decay characteristics led to the classification of the $(\Lambda_c(2595)^+, \Lambda_c(2625)^+)$ doublet as first orbital excitations of the $\Lambda_c^+$ with $J^P\ =\ \frac{1}{2}^-,\frac{3}{2}^-$, respectively. 
It is interesting to note that the most referenced studies at the time indicated that these states were expected to 
be $\approx 10$~MeV/$c^2$ apart~\cite{Capstick:1986ter}, compared with 
the experimentally found greater value of $\approx 30$~MeV/$c^2$. The latter mass difference set the scale for this type of hyperfine splitting, 
which we will see can then be used to 
predict the analagous splitting in higher mass charmed baryons. 

\begin{figure}[b]
\caption{\label{Figure:Lambda2595-2625} Evidence for the $\Lambda_c(2595)^+$ and $\Lambda_c(2625)^+$~resonances. Left: The Dalitz plot, $M^2(\pi^+\pi^-)$ versus $M^2(\Lambda_c^+\pi^+)$, for $\Lambda_c(2625)\to\Lambda_c^+\pi^+\pi^-$ decays from Belle~\cite{Belle:2022voy} showing clear bands
from the $\Sigma_c^{++}$ and the reflection of the $\Sigma_c^0$. Reprinted figure with permission from~\cite{Belle:2022voy}, Copyright (2023) by the American Physical Society. Right: The mass difference $M(\Lambda_c^+\pi^+\pi^-)-M(\Lambda_c^+)$ in semi-leptonic $\Lambda_b$ decays from LHCb~\cite{LHCb:2017vhq},
clearly displaying the shape of the $\Lambda_c(2595)$. Reprinted figure with permission from~\cite{LHCb:2017vhq}, Copyright (2017) by the American Physical Society.}
\begin{minipage}{1.0\textwidth}
\vspace{1mm}
\begin{tabular}{cc}
\includegraphics[height=0.21\textheight]{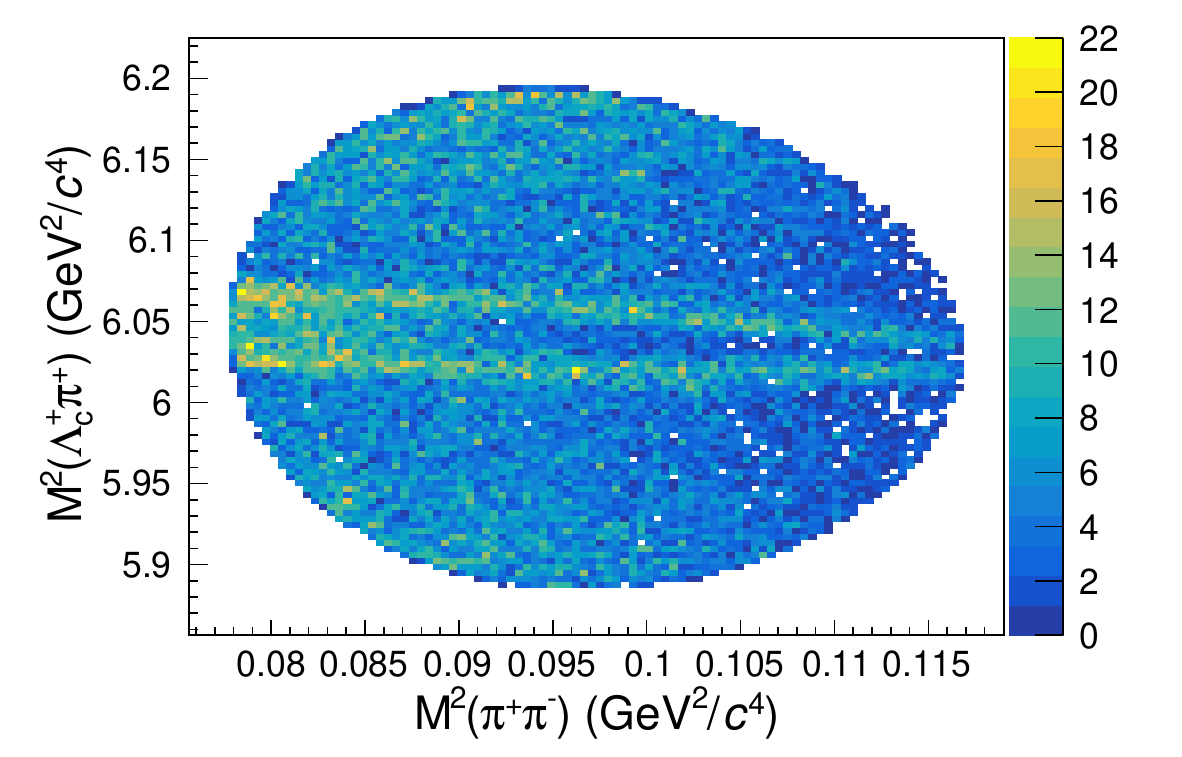} &
\includegraphics[height=0.22\textheight]{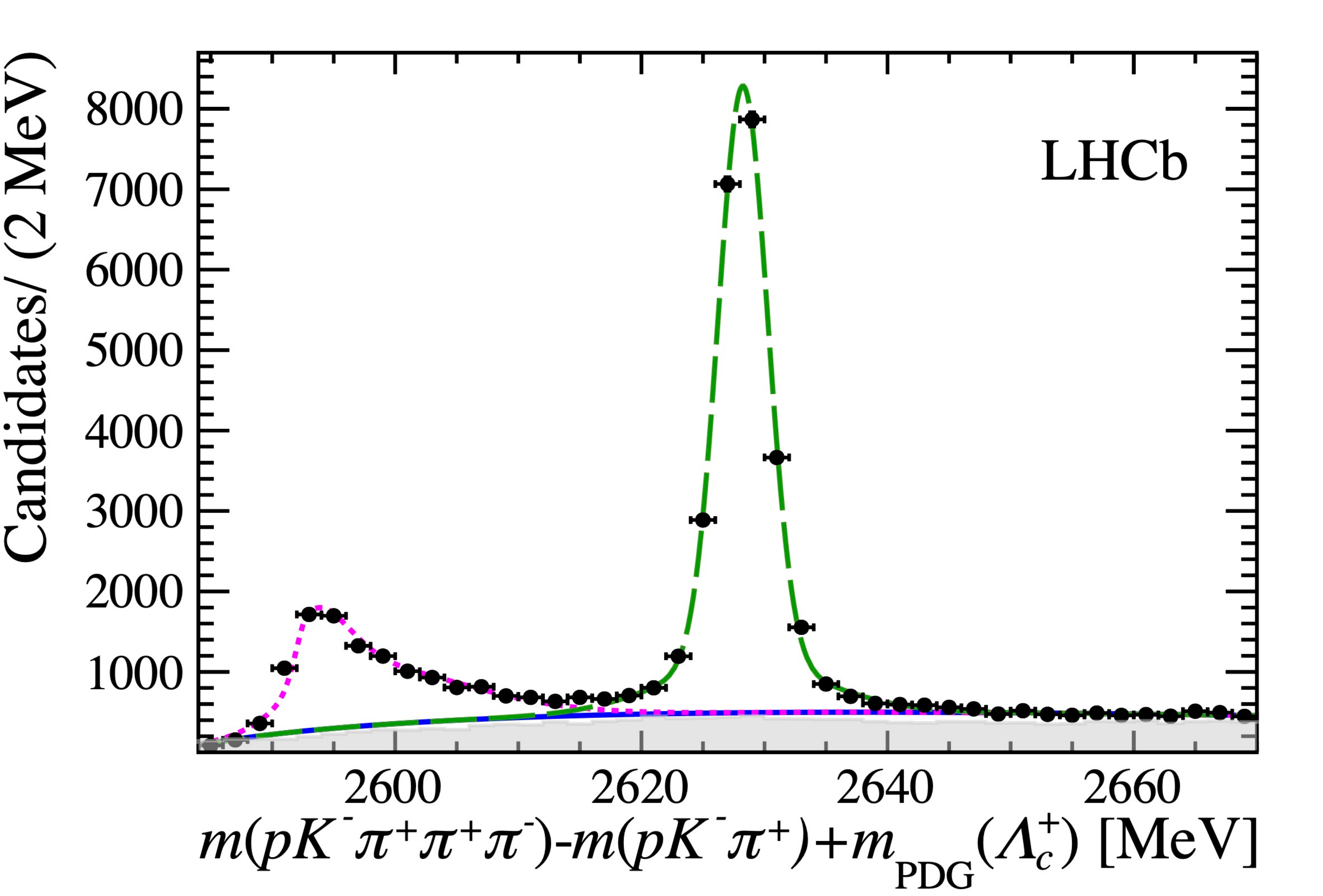}
\end{tabular}
\end{minipage}
\end{figure}

The most detailed study of the $\Lambda_c(2625)^+$ has been performed by Belle~\cite{Belle:2022voy}, 
who have done a Dalitz analysis of the three-body 
decay (Fig.~\ref{Figure:Lambda2595-2625}, left side). The results are consistent with the $J=\frac{3}{2}^-$ identification, with a small fraction decaying via $\Sigma_c(2455)$, some
through $\Sigma_c(2520)$ (off-shell), some via three-body decays, 
and of course interference between these different decay mechanisms. 
Unfortunately, although knowing the fraction decaying
into $\Sigma_c$ (which is a D-wave decay) is interesting, finding the partial width would be much more informative if it could be combined with a measurement of the full 
width. However, the full width is too narrow for the Belle detector to measure given the uncertainties in the resolution.    

Experimentally and theoretically, the $\Lambda_c(2595)^+$ is much harder to study than its $\frac{3}{2}^-$ partner. 
Its pole mass appears to be below $M(\Sigma_c^{++/0})+M(\pi^{\pm})$
but above $M(\Sigma_c^+)+M(\pi^0)$. 
This means the decays to the first of these proceed because of a combination of the finite width
of the parent state and that of the intermediary $\Sigma_c$, and are therefore 
suppressed compared with the $\pi^0$ transitions which have an advantage due to the isospin splitting in the $\pi$ system. However, experimentally finding the two low-momentum transition 
$\pi^0$~mesons in the decay chain $\Lambda_c(2595)^+\to\Sigma_c^+\pi^0\to \Lambda_c^+\pi^0\pi^0$ is very difficult, and all studies so far have concentrated on the charged transitions. 
It was first pointed out by Blechman {\it et al.}~\cite{Blechman:2003mq} that
this threshold effect meant that the first measurements of the $\Lambda_c(2595)^+$ were too high by around 2~MeV/$c^2$. 
The Blechman {\it et al.} formulation 
of the phase space terms was then used by the CDF Collaboration in their analysis of the state~\cite{CDF:2011zbc}. 
Although the CDF data is clearly inferior in statistics
and signal-to-noise ratio than that of current experiments, this analysis has presented the most precise measurements published.
The CDF~results agree with the Blechman {\it et al.}
statement~\cite{Blechman:2003mq} that the masses of the earlier measurements were too high. 
The asymmetric peak has also been shown in data with high statistics and low background by
the LHCb Collaboration~\cite{LHCb:2017vhq} and is displayed in Fig.~\ref{Figure:Lambda2595-2625} (right side).
It is interesting to note that the CDF analysis explicitly assumed that all the decays
resonate through an intermediate $\Sigma_c$, whereas some earlier results indicated that this might not be true.
However, those earlier results
did not take into account these phase space complications and so the CDF assumption seems reasonable. 
The long high-mass tail of the $\Lambda_c(2595)$ extends beyond even the $\Lambda_c(2625)$, 
complicating the analysis of the latter particle. 

When discovered, the $\Lambda_c(2625/2595)^+$ doublet was immediately 
identified as a pair of orbitally excited $\Lambda_c^+$ baryons~\cite{CLEO:1994oxm}. 
This identification was made mostly because of the decay chain --
the narrow width of the higher-mass state was due to the fact that it was below the threshold for
its otherwise preferred decay via $\Sigma_c(2520)$ -- and this indicated a spin-parity of $J^P\ = \frac{3}{2}^-$,  
whereas the $\Lambda_c(2595)$ could $just$ decay into its preferred $\Sigma_c(2455)$ intermediate state, indicating $J^P\ = \frac{1}{2}^-$. 
In a standard heavy-quark, light-diquark model,
this splitting is to first order inversely proportional to the heavy quark mass. 
Thus, these measurements, made back in 2002, 
immediately indicated that there should be a similar pair in the $\Xi_c$ system with a similar splitting, 
as well doublets in the $\Lambda_b$ and $\Xi_b$ systems
with smaller splittings. As we shall see, these predictions have proved successful, giving extra credibility to this
interpretation of the quark structure of these states. 
Some theoretical treatments went further~\cite{Rosner:1995dr},
and stated that, as the $\Lambda_c^+$ spectrum was so well understood, 
we could now extrapolate down in mass to the spectrum of strange baryons and improve our understanding of that system, too. 
This latter 
opinion, which appeared in the RPP review as late as 2010~\cite{ParticleDataGroup:2010dbb}, proved more controversial. 
Recently, 
many papers have looked again at the $\Lambda_c(2595)$ in view of its proximity to the $\Sigma_c\pi$ thresholds, and concluded
that it may be a molecular state or a dynamically generated state, 
rather than a simple heavy-quark, light-diquark system with orbital angular momentum~\cite{Nieves:2019nol}. 
However, there has been 
little in the way of experimental tests proposed that can be performed to differentiate these models. 
It is possible to use the heavy-quark, light-diquark
model as a guide to the mass spectra, and still believe that those combinations with a mass close to the sum of a baryon and 
meson mass have their properties affected by the corresponding threshold. 

\subsubsection{Higher mass $\Lambda_c^+$ states}
\begin{figure}[t]
\caption{\label{Figure:Lambda2765-2880} Evidence for the $\Lambda_c(2765)^+$ and $\Lambda_c(2880)^+$~resonances. Left: The $M(\Lambda_c\pi^+\pi^-)$ mass distributions from Belle~\cite{Belle:2019bab} (a) without, and (b) with a cut on an intermediate $\Sigma_c(2455)$~resonance. Right: The  $M(\Sigma_c\pi)$ mass distributions from Belle~\cite{Belle:2019bab} for (a) doubly charged, and (b) neutral $\Sigma_c$ baryons. No signals are found, see text for more details of the analysis. Reprinted figures with permission from~\cite{Belle:2019bab}, Copyright\,@\,2020, World Scientific.}
\begin{minipage}{1.0\textwidth}
%\vspace{1mm}
\begin{tabular}{cc}
\includegraphics[width=0.58\textwidth]{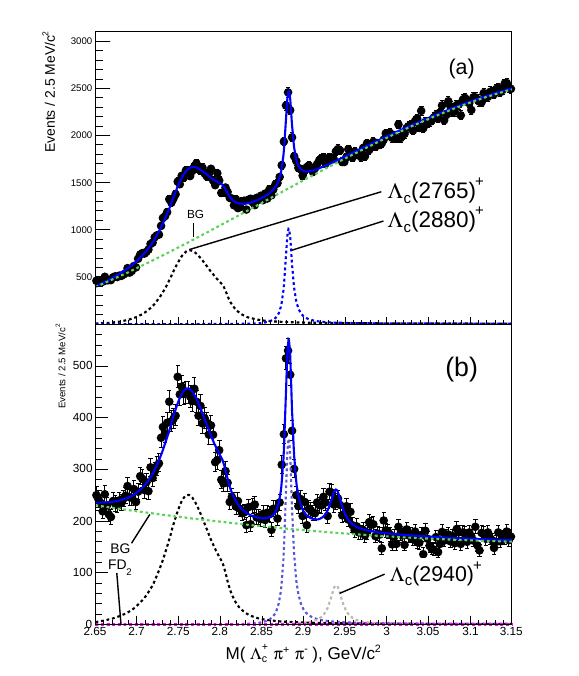} &
\begin{minipage}{.5\textwidth}
    \vspace{-9.8cm}
    \includegraphics[height=0.206\textheight]{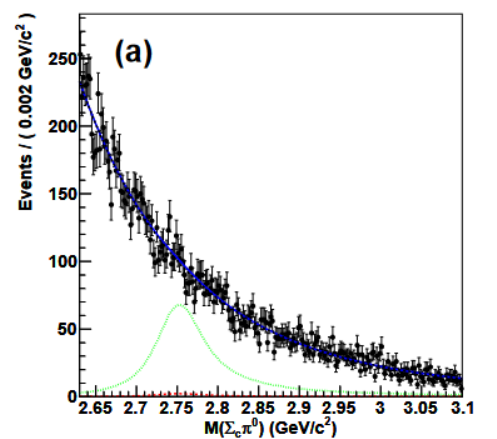}\\[1.2ex]
    \includegraphics[height=0.206\textheight]{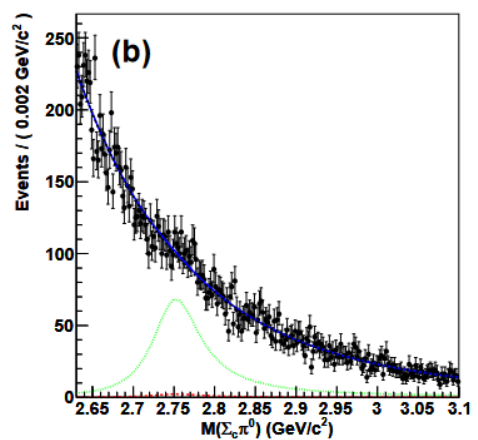}\\
\end{minipage}
\end{tabular}
\end{minipage}
\end{figure}

In the $\Lambda_c^+\pi^+\pi^-$ spectrum at masses
higher than that of the $\Lambda_c(2625)^+$, 
several resonances have been seen starting with the CLEO observation~\cite{CLEO:2000mbh} of a wide state, $\Lambda_c(2765)^+$,
and a narrow state, $\Lambda_c(2880)^+$. 
The $\Lambda_c(2765)^+$ has proven difficult to investigate in detail, and indeed it only has 
attained ``one-star'' status in the RPP~\cite{ParticleDataGroup:2022pth},
as the PDG only  references the publication that claimed its discovery.
However, in our opinion there is little doubt the state exists due to the statistical quality of the experimental data detailed below from both the Belle and LHCb Collaborations.
The first observation by CLEO did not claim any knowledge of whether it was a $\Lambda_c$ or a $\Sigma_c$~resonance, although the latter always seemed unlikely. 
The status was clarified in a conference paper by Belle~\cite{Belle:2019bab} which showed no excess in 
the $(\Lambda_c^+\pi^{\pm})_{\Sigma_c}\pi^0$ decay mode (Fig.~\ref{Figure:Lambda2765-2880}, right side)
but a large signal in $\Lambda_c^+\pi^{+}\pi^-$ (Fig.~\ref{Figure:Lambda2765-2880}, left side). 
The Belle analysis of
the $\Lambda_c(2880)\to\Sigma_c^{++/0}\pi$~decay~\cite{Belle:2006xni} also showed a large excess corresponding to the $\Lambda_c(2765)$, 
and even LHCb showed a large excess~\cite{LHCb:2017vhq}
in the region, even though they made no explicit measurements of it. 
The reason hard data on the $\Lambda_c(2765)$ properties is so hard to come by is that it is clearly a wide resonance, 
which then decays via a resonant substructure. There are
no convenient sidebands to make a sideband subtraction, and the resonant substructure into all of
($\Sigma_c(2455)^0,\Sigma_c(2520)^0,\Sigma_c(2455)^{++},\Sigma_c(2520)^{++}$) produces many peaks and reflections of peaks in the $\Lambda_c\pi$ mass
distributions. 
However, there is a consensus that the most likely spin-parity of the state is $J^P=\frac{1}{2}^+$, making 
it an analogue to a ``Roper-like'' radial excitation. 
Circumstantial evidence supporting this comes from its mass of $\approx 500$~MeV/$c^2$ above the ground state, 
the decay patterns, the similarity to the $\Xi_c(2970)$, whose 
spin-parity has been measured as described below, and one other feature that is not generally mentioned -- its production cross section. 
The cross-section for the production of the $\Lambda_c(2765)^+$ in $e^+e^-$ annihilation, 
though not a published measurement, is clearly big. 
For an interpretation as a radial excitation, the particle must not be part of a pair of particles (in the manner of a $J^P=\frac{1}{2}^-,\frac{3}{2}^-$ 
pair described above), 
which have similar cross sections. We do not see signs of a second state and  
a radial excitation is expected to be a single peak.
It is also interesting to note that an unpublished analysis of the Belle data in a thesis~\cite{Joo:2015ese} already concluded that the
$\Lambda_c(2765)$ has spin-parity of $J^P=\frac{1}{2}^+$.

The CLEO paper of 2002~\cite{CLEO:2000mbh} which first showed the wide state ($\Lambda_c(2765)^+$)  
also showed a narrow peak decaying into $\Lambda_c^+\pi^+\pi^-$, clearly resonating approximately one third of the time
through $\Sigma_c(2455)\pi$, but not showing any $\Sigma_c(2520)\pi$ substructure. 
To find a narrow resonance, now known as the $\Lambda_c(2880)$, at so high a mass was a surprise at the time. 
With much more data available, 
Belle~\cite{Belle:2006xni} demonstrated that the greatly preferred spin assignment of the state was $J = \frac{5}{2}$. 
They also found a small but finite $\Sigma_c(2520)\pi$ component,
and concluded that the state was likely to have $J^P=\frac{5}{2}^+$ and 
comprised a heavy quark with two units of orbital angular momentum relative to the light diquark, 
i.e. $l_{\,\lambda} = 2$ (see also Table~\ref{Table:CharmedBaryonClassification}). 
They argued that the low ratio between the $\Sigma_c(2520)$ and $\Sigma_c(2455)$ was consistent with this assignment, 
with both proceeding 
via ``F-wave'' decays. However, the missing piece in this argument is that the preferred ``P-wave''
decay of a $J^P=\frac{5}{2}^+$ particle would be to $\Sigma_c(2520)\pi$.
More recently LHCb, in their partial-wave analysis of $\Lambda_b\to pD^0 X$~decays~\cite{LHCb:2017jym}, 
have confirmed this $J^P=\frac{5}{2}$ 
assignment by looking at the particle in the $pD^0$ decay mode.
This means that the reason for the
suppression of the $\Lambda_c(2880)\to\Sigma_c(2520)\pi$ partial width remains a mystery. 

When BaBar analysed the spectrum of $pD^0$ combinations in $e^+e^-$ annihilation~\cite{BaBar:2006itc}, 
they found not only the $\Lambda_c(2880)^+$, but a further resonance,
the $\Lambda_c(2940)^+$. This observation was quickly confirmed by Belle in the $\Sigma_c(2455)^{0/++}\,\pi^\pm$ spectrum~\cite{Belle:2006xni}. 
The mass and, in particular, width measurements of the
two experiments are in reasonable agreement indicating the likelihood these are the same state. 
More experimental data was then lacking until LHCb, as part of their analysis of $\Lambda_b$
decays, measured a preference for the particle to be $J^P=\frac{3}{2}^-$~\cite{LHCb:2017jym}. 
This particle is definitely in need of more experimental investigation. Because it is just a few
MeV/$c^2$ below the $pD^{*0}$ threshold, 
some authors have suggested that this is a baryon-meson molecular state~\cite{Wang:2020dhf} or maybe a 2P state
affected by the threshold~\cite{Luo:2019qkm}. 

\begin{figure}[t]
\caption{\label{Figure:Lambda2910-2940} Evidence for the $\Lambda_c(2910)^+$ and $\Lambda_c(2860)^+$~resonances. Left: The $M(\Sigma_c^{++/0}\,\pi^\pm)$ mass distribution from $B$~meson decays based on data from Belle~\cite{Belle:2022hnm}. Reprinted figure with permission from~\cite{Belle:2022hnm}, Copyright (2023) by the American Physical Society. Right: The $M(D^0p)$ mass distribution in $\Lambda_b^0$ decays from LHCb~\cite{LHCb:2017jym}. The fit shows
the $\Lambda_c(2860)$, $\Lambda_c(2880)$ and $\Lambda_c(2940)$. Reproduced from~\cite{LHCb:2017jym}, with permission from Springer Nature.}
\begin{minipage}{1.0\textwidth}
%\vspace{1mm}
\begin{tabular}{ccc}
\includegraphics[height=0.21\textheight]{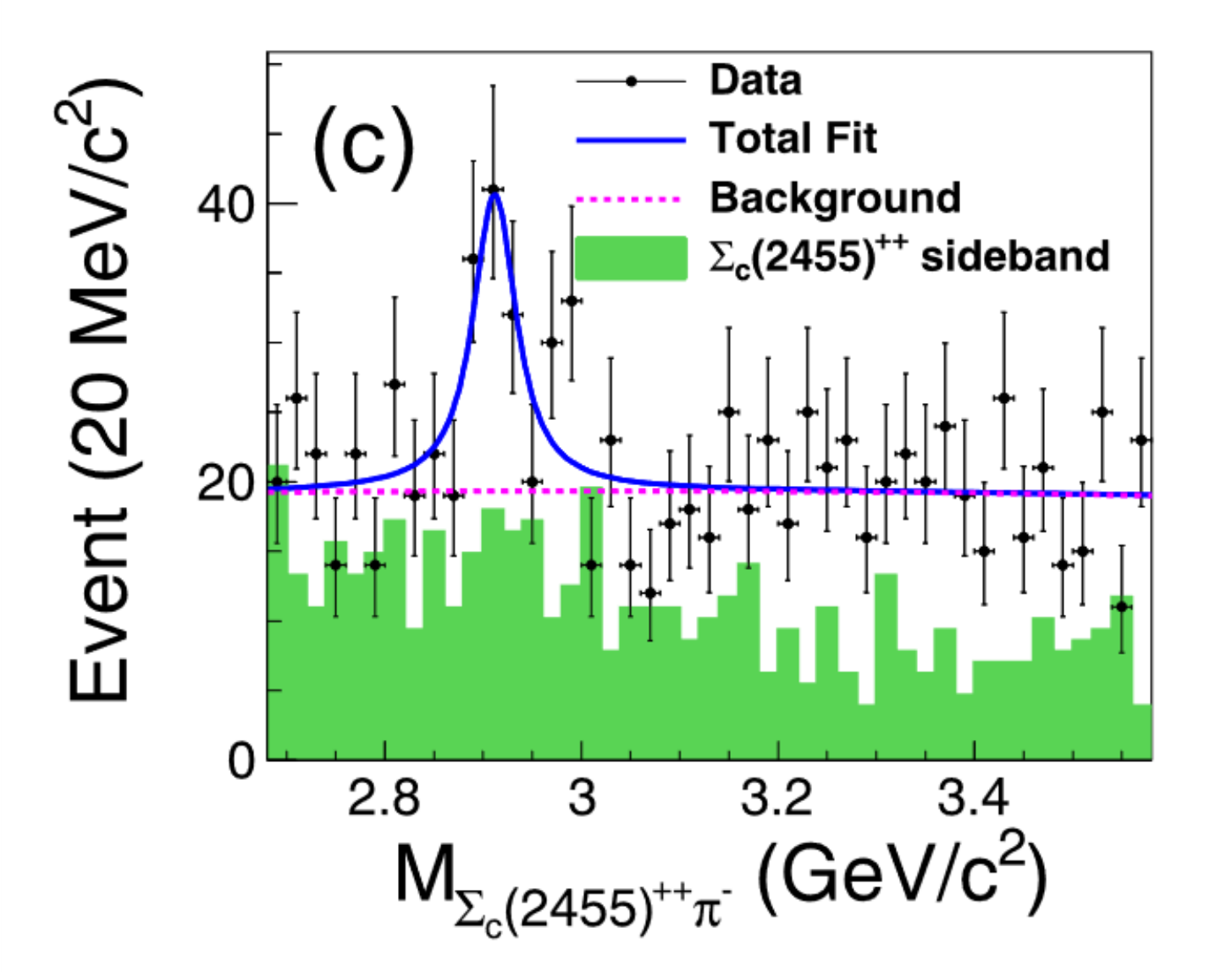} & &
\includegraphics[height=0.21\textheight,width=0.49\textwidth]{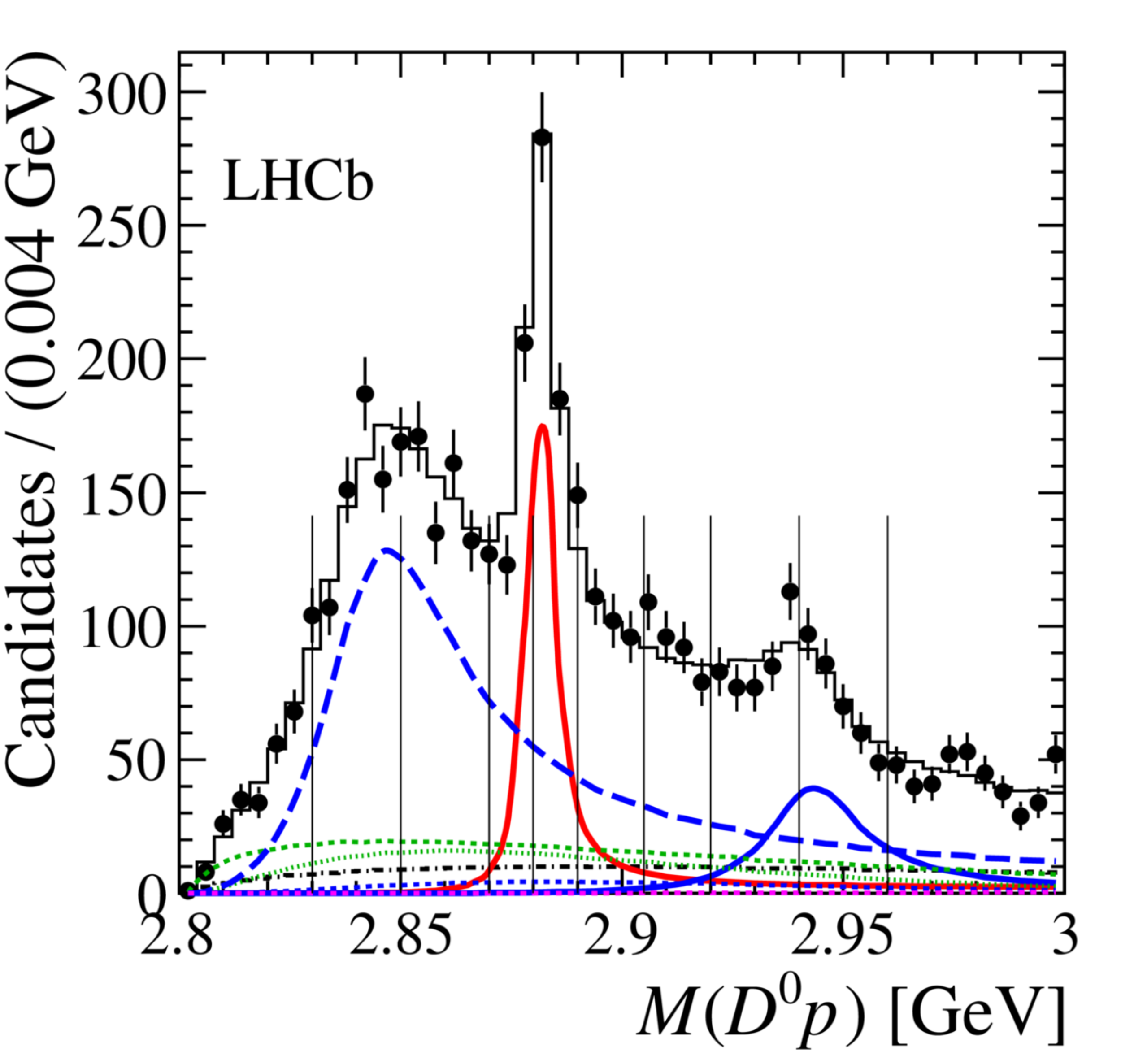}
\end{tabular}
\end{minipage}
\end{figure}

The Belle analysis that measured the $\Lambda_c(2880)$ spin~\cite{Belle:2006xni}
also showed a statistically significant excess in the $\Sigma_c\pi$ mass distribution 
between the $\Lambda_c(2880)$ and $\Lambda_c(2940)$ states.
This complicated their analyses, but the observation was insufficient to claim as evidence of a new particle. 
Further information was lacking until 2022 when Belle analysed the decay 
$B\to\Sigma_c\pi p$~\cite{Belle:2022hnm}. Here, in the $\Sigma_c\pi$ mass spectrum, no signals for the 
$\Lambda_c(2880)$ or $\Lambda_c(2940)$~resonances were observed, 
but a large, wide peak spanning the region and centered
at a mass of 2910~MeV/$c^2$ (Fig.~\ref{Figure:Lambda2910-2940}, left side). This could explain the excess found by Belle in the continuum data. 
They conjecture that the new peak could be due to a $J^P\ =\ \frac{1}{2}^-,\ 2P$ state. However, we must also be wary of 
assuming a wide excess of this nature is due to a single particle, 
and not the overlap of two or more particles as has been seen in $\Xi_c$ production from $B$ decays. 

Furthermore, the LHCb Collaboration, in their partial-wave analysis of the decay 
$\Lambda_b\to pD^0 X$~\cite{LHCb:2017jym}, 
found, as noted above, peaks due to the $\Lambda_c(2880)^+$ 
and $\Lambda_c(2940)^+$. However, they could only explain their data (Fig.~\ref{Figure:Lambda2910-2940}, right side) by introducing a new spin $\frac{3}{2}$ particle, 
which they call the $\Lambda_c(2860)^+$.

\subsection{The $\Sigma_c$ states}
The naming convention that differentiates between the charmed $\Lambda$ and $\Sigma$ states carries over from the strange sector, 
where it was invented long before 
the advent of charm. The $\Sigma_c$ states, by definition, have isospin $I=1$, and thus, exist as $\Sigma_c^{++}$, $\Sigma_c^+$, and $\Sigma_c^0$~particles.
Mass measurements are always made in terms of mass differences with 
respect to the ground-state~$\Lambda_c^+$. Following the guide of the non-charmed states, 
it was always expected that we would find one $J^P\ =\ \frac{1}{2}^+$ and one
$J^P\ =\ \frac{3}{2}^+$ iso-triplet.
The first signal claimed for a $\Sigma_c$ was at the same time as that of a $\Lambda_c$. 
This is in part because the kinematics for a low-$Q^2$ decay,
such as $\Sigma_c\to\Lambda_c\pi$, is experimentally convenient, i.e. the resolution is good and thus, the background is low.

From a theory point of view, the singly charged state is similar to its isospin partners, but
from an experimental point of view, it is 
much more difficult to detect and measure. The decay proceeds via a $\pi^0$~meson, which typically has worse resolution, purity and detection efficiency than a charged particle.
The Belle Collaboration has measured the $\Sigma_c^{++}$ and $\Sigma_c^0$ masses and widths with good precision~\cite{Belle:2014fde}, 
and the singly charged states with reasonable 
precision~\cite{Belle:2021qip}. 
One complication that arises once we reach the level of precise measurements is that the decay of the $\Lambda_c(2595)^+$, which as discussed above 
is on the edge of the phase space for a $\Sigma_c\pi$ combination, can bias the resultant $\Sigma_c$ to the low-mass end of its Breit-Wigner lineshape, and this needs
to be accounted for in any measurement of its pole mass.
The discovery of the $J^P\ =\ \frac{3}{2}^+$~state we now know as the $\Sigma_c(2520)$~resonance took much longer, largely because of its greater width and less favourable 
kinematics. It is also subject to the complication from feed-down, this time from the $\Lambda_c(2625)$ which gives an irregularity to the background shape inconveniently
close to the pole mass. 

The recent measurements of the $\Sigma_c(2455)$ and $\Sigma_c(2520)$ iso-triplets do allow for some investigation of the isospin mass splittings due
to electromagnetic effects. According to the model first proposed
by Franklin~\cite{Franklin:1975yu}, the value of the mass relationship
$M(\Sigma_c^{++})+M(\Sigma_c^0)-2\times M(\Sigma_c^+)$ should be the same for the $\Sigma_c(2455)$
and $\Sigma_c(2520)$ iso-triplets.
Combining recent measurements with those of the Particle Data Group~\cite{ParticleDataGroup:2010dbb}
for the others, produces values of $2.46^{+0.17}_{-0.34}$~MeV/$c^2$ and $2.2^{+1.0}_{-1.4}$~MeV/$c^2$ for the two systems,
respectively, consistent with the model of Ref.~\cite{Franklin:1975yu}.

Clearly, at higher masses, we expect orbital excitations of $\Sigma_c$ baryons. 
In particular, we note that one unit of orbital angular momentum between the heavy
quark and light diquark $(l_\lambda=1)$ will combine with the spin-1 diquark to give a quintuplet of 
iso-triplets with $J^P=\frac{1}{2}^-,\frac{1}{2}^-,\frac{3}{2}^-,\frac{3}{2}^-,\frac{5}{2}^-$, respectively. All these particles will have allowed single-pion transitions
to the ground-state~$\Lambda_c^+$, but with varying widths. 
There has been only one report of an excited $\Sigma_c$ iso-triplet
in the correct mass range for these particles~\cite{Belle:2004zjl}.
Unambiguous peaks have been observed for each of the three charged states, but it is unclear why only one peak should be visible in each distribution. 
A likely explanation is that the observed structure
is actually based on two or more overlapping states. For instance, one model~\cite{WangWangKaiLei:2021kdd} identifies the seen peak as the overlap of a
$J^P\ =\ \frac{3}{2}^-$ and $\frac{5}{2}^-$, with two of the other states decaying via an intermediate $\Sigma_c$ where they could be easily hidden, at least in
decays involving charged pions, under the $\Lambda_c(2765)^+$ peak. However that begs the question of where the fifth state (with $J^P\ =\ \frac{1}{2}^-$) is hidden, 
as the model claims it should be comparatively narrow.
The BaBar Collaboration has shown a peak in the neutral channel~\cite{BaBar:2008get} by
looking in $B$ decays, but their
mass is only marginally consistent with that of Belle. Clearly this is a topic worthy of further investigation either by 
Belle~II or LHCb.

\subsection{The $\Xi_c$ states}
The $\Xi_c$~states comprise $csu$ and $csd$ quark combinations. Although now we have three quarks of distinctly different masses, the heavy-quark, light-diquark model still works 
rather well, with the strange quark relegated to the role of one of the light quarks. The discovery of the 
charged state was first reported in fixed target experiments~\cite{Biagi:1983en}, but the uncertainties of the mass measurements
were so large that it is not clear if new information was gained. However, by 1990, relatively precise and accurate measurements had been made of the masses of both 
ground states by $e^+e^-$ experiments~\cite{CLEO:1989msq} as well as fixed target experiments~\cite{ACCMOR:1989xbf}.

The first higher-mass versions have the two light quarks in a spin-1 configuration,
analagous to the $\Sigma_c$. However, here the mass difference between these and the ground states is lower because of the increased mass of the diquark. 
This results in the $J^P=\frac{1}{2}^+\ \Xi_c^{\prime}$ being
below the mass threshold for pion decays, and thus, it decays electromagnetically. 
From an experimental standpoint, this makes the detection difficult because of the large
number of photons from $\pi^0$ decays which typically occupy the same region of the phase space. 
An advantage of the lower mass differences is that for the 
$J^P=\frac{3}{2}^+\ \Xi_c(2645)$, the decay is more restricted than the analogous one of the $\Sigma_c(2520)$, and this leads to a narrower width and a better signal-to-noise ratio.
Because of the difference in the detection techniques in terms of charged pions and photons, which are more difficult to detect, 
the $J^P=\frac{3}{2}^+$~states were found first~\cite{CLEO:1995amh,CLEO:1996zcj}, 
followed a few years later by the ``prime'' states~\cite{CLEO:1998wvk}.
At the time, there were already many theoretical predictions of the
mass difference with respect to the ground states, and the observations fit in nicely with these predictions. 
For instance, one of the first predictions, made 
in 1980~\cite{Maltman:1980er} had
the prime states at 95 MeV/$c^2$ above the ground states and the $J^P=\frac{3}{2}^+$~states a further 70 MeV/$c^2$ higher, whereas recent experimental 
values of these splittings are 109~MeV/$c^2$ and 77~MeV/$c^2$, respectively~\cite{ParticleDataGroup:2022pth}.
 
The most precise measurement of these two doublets, now known as the $\Xi_c^{\prime}$ and $\Xi_c(2645)$, 
has been performed by the Belle experiment~\cite{Belle:2016lhy}. This includes the measurement of the intrinsic widths of the
$J^P=\frac{3}{2}^+$ doublet particles. 
Note that the measurement of the latter is further complicated by the feed-down from higher states. Decays of the $\Xi_c(2815)\to\Xi_c\pi\pi$ give
satellite peaks near the true particles. It is particularly disconcerting to see peaks arise in $\Xi_c^+\pi^+$ mass plots where no standard particle can exist, but
such peaks are just due to the limited phase space of the decays of higher mass particles. 

Once the $J^P=\frac{1}{2}^+,\frac{3}{2}^+$~doublet had been discovered, the next logical search was 
for the $\Xi_c$ analogues of the $\Lambda_c(2595,2625)^+$. 
Based on the simple model, we would expect the same excitation energy above the ground state of $\approx 300$~MeV/$c^2$ (to first order the same for 
all heavy-quark, light-diquark baryons), and also a similar splitting between states of $\approx 30$~MeV/$c^2$, i.e. inversely proportional to the heavy quark mass, which is charm in both cases. 
Furthermore, predictions could then be made that the favoured decay of the $J^P=\frac{1}{2}^-$~state would be to $\Xi_c^{\prime}\pi$, and of the $J^P=\frac{3}{2}^-$ state
would be to $\Xi_c(2645)\pi$. As these decay products had been found in $e^+e^-$ continuum events, 
it followed that this new pair of iso-doublets should be searched
for with these characteristics and in this same environment. 
At the time, CLEO was the experiment running in the right energy range to perform these searches and indeed, they found all four particles, the $\Xi_c(2790)$ and $\Xi_c(2815)$ iso-doublets, once their data set was sufficiently large. 
Moreover, they found them with masses and decay characteristics very close to those 
expected~\cite{CLEO:2000ibb,CLEO:1999msf}. 
The most precise measurements of the properties of these particles are again due to Belle~\cite{Belle:2016lhy} operating in a similar accelerator environment but with 
much greater statistics.

One notable measurement of the properties of the $\Xi_c(2815)$ and $\Xi_c(2790)$ is the observation by the Belle Collaboration~\cite{Belle:2020ozq} 
of their photon transitions to their ground state at a rate competing with their 
strong decays, but only for the neutral partners. 
Such a difference between isospin partners had been predicted by a conventional constituent quark model~\cite{Wang:2017kfr}, and it adds credibility
to the many other predictions for radiative decays made by the authors.

\subsubsection{The $\Xi_c(2970)$}
\begin{figure}[t]
\caption{
$\Xi_c^0\,\pi^+\pi^-$ mass distribution from Belle~\cite{Belle:2020tom} showing the $\Xi_c(2970)$~peak (left), and the decay angle
distribution (right) showing that 
$J=\frac{1}{2}$ (black line) fits the data well. Reprinted figure with permission from~\cite{Belle:2020tom}, Copyright (2021) by the American Physical Society.}
\begin{minipage}{1.0\textwidth}
\begin{tabular}{cc}
\includegraphics[width=0.49\textwidth]{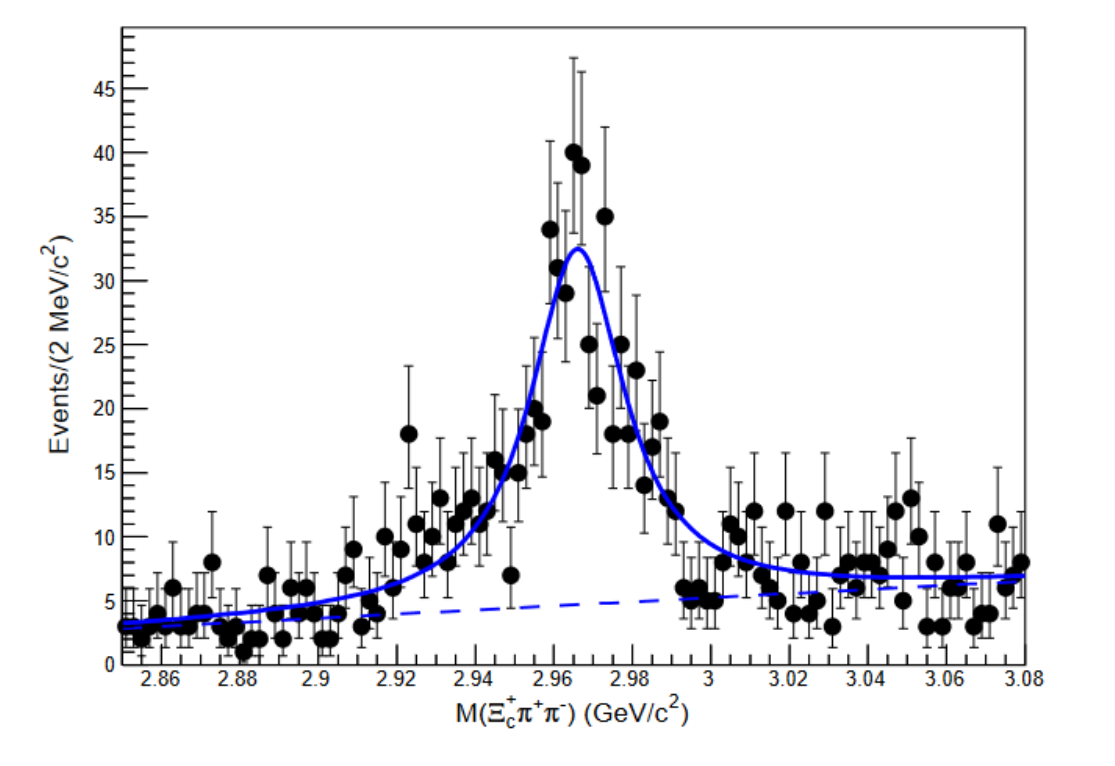}
\includegraphics[width=0.49\textwidth]{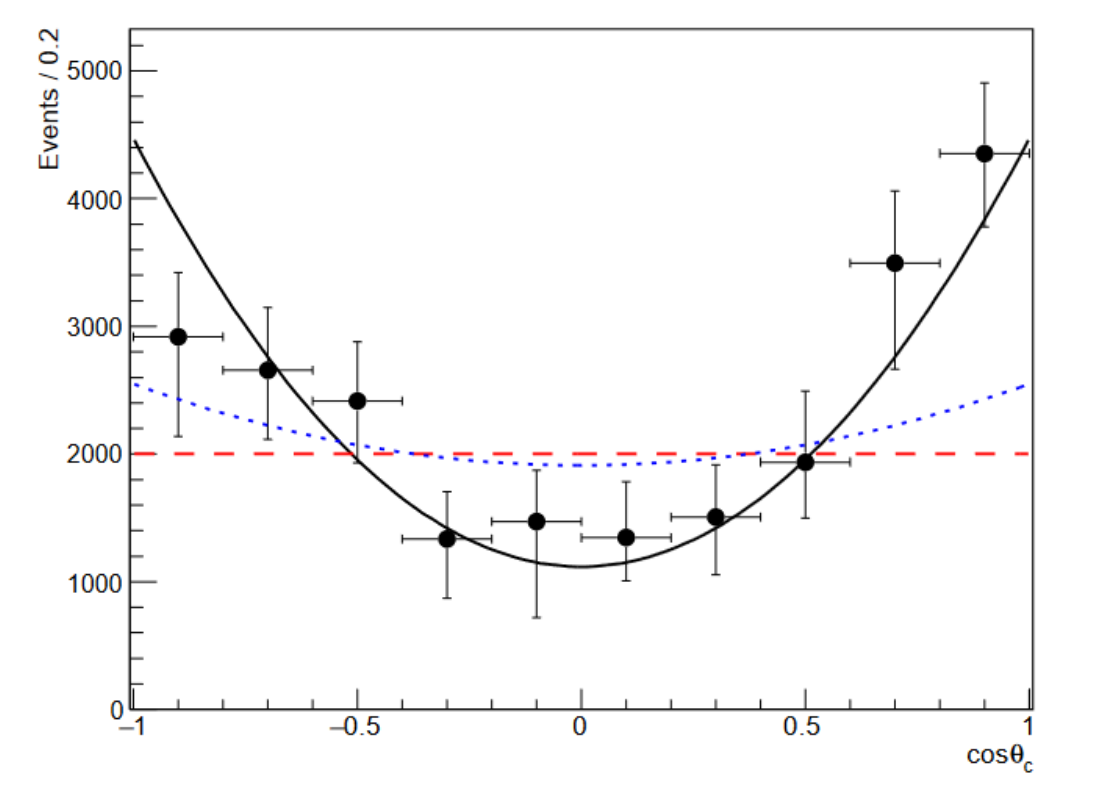}
\end{tabular}
\end{minipage}
\label{fig:2970}
\end{figure}

Chronologically, the next excited state to be found was the $\Xi_c(2980/2970)$. Here, the picture took much longer to gain focus. 
Confusingly, the PDG's rounding of the mass used
in its name changed from 2980~MeV/$c^2$ to 2970~MeV/$c^2$ as more experiments found evidence for it and the measured mass decreased. 
There is still considerable uncertainty as to how many states exist in this mass range.  
The original discovery was reported by Belle~\cite{Belle:2006edu} which showed a broad peak in the $\Lambda_c^+K^-\pi^+$~spectrum at a mass of 2978.5~MeV/$c^2$, with a hint of 
an isospin partner at a similar mass. 
These observations were then confirmed by BaBar~\cite{BaBar:2007zjt}, albeit with slightly lower masses and widths, and the neutral state still suffering 
from lack of statistics. 
Next, Belle~\cite{Belle:2008yxs} showed firm evidence of both isospin states using their decay into $\Xi_c(2645)\pi$, and their widths were measured to be less than 20 MeV/$c^2$.  
The same experiment followed up with a reanalyis of the $\Lambda_c^+K^-\pi^+$ 
spectrum~\cite{Belle:2013htj}, now using their entire data set, and determined values for mass and width in good agreement with their results in $\Xi_c(2645)\pi$. 
Further analysis by Belle~\cite{Belle:2016lhy}, using more decay modes
of the $\Xi_c$ ground states, pushed the width average wider again, without being in actual disagreement with the above measurements. They also showed evidence 
of decays to $\Xi_c^{\prime}\pi$. With more data than previous measurements, the statistical uncertainties on the masses were much reduced. Lastly, LHCb looked in a 
new decay mode, $\Xi_c^*\to\Lambda_c^+ K^-$, and found a very substantial peak in this mass region~\cite{LHCb:2020iby}. 
However, their mass measurement of $M=(2964.88 \pm 0.26 \pm 0.20)$~MeV/$c^2$ is in
distinct tension with that of the Belle Collaboration, $(2970.8 \pm 0.7 \pm 0.2)$~MeV/$c^2$, indicating the likelihood that these are not the same particles. This is discussed 
further below. 
One welcome note of clarity is the Belle measurement of the spin-parity quantum numbers~\cite{Belle:2020tom} based on the decay of the $\Xi_c^+(2790)$ to $\Xi_c(2645)^0$ and $\Xi_c(2645)^{\prime +}$~(see Fig.~\ref{fig:2970}). 
The authors concluded that the spin-parity assignment of $J^P=\frac{1}{2}^+$ is the most likely scenario,  
with the light-quark degrees of freedom being $s_{qq} = 0$, i.e. a ``Roper-like'' radial excitation (see also Table~\ref{Table:CharmedBaryonClassification}). This agrees with previous expectations for this particle
which, from its properties, appears to be the charmed-strange
analogue of the $\Lambda_c(2765)^+$~resonance. 
Its copious production in $e^+e^-$ annihilation is further circumstantial evidence that this assignment is correct. 

\subsubsection{The $\Xi_c$ spectrum at high masses}
The 2006 Belle paper~\cite{Belle:2006edu} presented more than just evidence of the wide $\Xi_c(2970)^+$ (as it is now called) in the $\Lambda_c^+K^-\pi^+$
decay mode. It also showed a relatively narrow peak known as the $\Xi_c(3080)^+$ in the same mass distribution, 
with its isospin partner also visible (using neutral Kaons 
which are intrinsically less efficient to find). 
These were then both confirmed, with a similarly sized data set, by a BaBar analysis~\cite{BaBar:2007zjt} 
who went further and showed
that the $\Xi_c(3080)^+$ resonates through both the $\Sigma_c(2455)$ and $\Sigma_c(2520)$, and the 
$\Xi_c(2970)$ resonates through the $\Sigma_c(2455)$. Furthermore, they
showed evidence for two more signals: the first, $\Xi_c(3055)^+$, resonating through $\Sigma_c(2520)^{++}K^-$, and the second, $\Xi_c(3123)^+$, resonating through $\Sigma_c(2520)$.
The latter has not been confirmed, and its significance was always marginal. Once again the neutral partners of these excited $\Xi_c$ baryons are harder to find, but
the paper did show evidence of the analogous decays of the $\Xi_c(2980)^0$ as noted above, as well as the $\Xi_c(3080)^0$. 
The Belle Collaboration~\cite{Belle:2013htj} then confirmed the BaBar results
for the charged $\Xi_c(3055)$ and $\Xi_c(3080)$ in an analysis geared towards searching for doubly charmed baryons. Lastly for this group, Belle looked in the decay
mode $\Lambda D$, serving as the $\Xi_c$~analogue of the $pD^0$ decay mode, which had proven fruitful in searches for the decays of excited $\Lambda_c$ baryons.  
Furthermore, both isospin partners were investigated easily by using decays to both $D^0$ and $D^+$. 
This analysis found a strong signal for the $\Xi_c(3055)^{+/0}$, 
a clear signal for the $\Xi_c(3080)^+$, but at best a weak signal for the $\Xi_c(3080)^0$. 
The identification of the spin structure of the $\Xi_c(3055/3080)^{\pm}$ pair is so far just hand-waving. However, it is clear that if we follow the idea that
the $\Xi_c$~spectrum is simply the $\Lambda_c^+$~spectrum with an increase in mass of $\approx 185$~MeV/$c^2$, the $\Xi_c(3055)$ and $\Xi_c(3080)$ can be thought
of as the charmed-strange analogues of the $\Lambda_c^+(2860)$ and $\Lambda_c^+(2880)$. And as we have seen, these are often identified as the $J^P = (\frac{3}{2}^+,\frac{5}{2}^+$)~pair with $l_\lambda = 2$ (Table~\ref{Table:CharmedBaryonClassification}). 

\subsubsection{Excited $\Xi_c$ decays into $\Lambda_c^+K^-$}
One energy range so far not discussed is around 2930 MeV/$c^2$. Here, the BaBar Collaboration reported on a broad structure in the $\Lambda_c^+K^-$ 
substructure of $B$ decays~\cite{BaBar:2007yvx}.
It was many years before this was confirmed, but in 2018, Belle~\cite{Belle:2017jrt} showed evidence for a resonance in the 
region at the level of more than 5$\sigma$. They followed this 
up with finding the expected isospin partner in the $K^0_S\Lambda_c^+$ substructure of B decays~\cite{Belle:2018yob}. 
However, in 2020~\cite{LHCb:2020iby} and 2022~\cite{LHCb:2022vns}, 
LHCb showed convincingly that this structure comprised two distinct particles. 
Although neither the BaBar nor the Belle observations were in any way incorrect, their resolution 
and, more importantly, low statistics, were not enough to resolve the peaks. 
The internal structure of the particles decaying into $\Lambda_c^+K^-$ will be discussed in the 
following section on the excited $\Omega_c$~spectrum.

%% file: CharmedOmegaBaryons.tex
The early history of the $\Omega_c^0$ is very convoluted. Early sightings include an observation of the resonance with a mass of $(2747\pm 10)$~MeV/$c^2$~\cite{Biagi:1984mu} and an observation
in $e^+e^-$~annihilations with a cross section clearly ruled
out by another experiment~\cite{ARGUS:1992mwl}. 
One experiment purported to measure the lifetime even though the masses in the various decay modes did not agree~\cite{WA89:1995lbz}, 
and another lifetime measurement
which was in a decay mode that has not been confirmed by any other experiment despite the much larger data sets~\cite{E687:1995cvt}. 
Although CLEO did find a reasonable signal by adding together
many decay modes~\cite{CLEO:2000dhf}, 
and BaBar showed several modes without publishing the mass~\cite{BaBar:2007jdg},
the situation was not satisfactory until Belle showed a large peak in the $\Omega^-\pi^+$~spectrum~\cite{Solovieva:2008fw}. 
This decay mode still dominates studies as it is both copious and easy to detect.
 
\subsubsection{The excited $\Omega_c^0$ spectrum}
When discussing the excited $\Omega_c^0$~spectrum, we must always remember that we are dealing with a diquark in a spin-1 configuration, and thus the $\Omega_c^0$ is the 
analogue of the $\Sigma_c$ and the $\Xi_c^{\prime}$, and not the analogue of $\Lambda_c^+$ or the ground-state $\Xi_c$. Thus, it immediately becomes apparent that we expect the
first excited state to be at an excitation energy of $M(\Sigma(2520))-M(\Sigma(2455))=M(\Xi_c(2645))-M(\Xi_c^{\prime})\approx 70$~MeV/$c^2$. This of course would reveal itself
in a photon transition. Despite the difficulties in finding a low energy photon in a $\pi^0$-rich environment, it was
no surprise that first BaBar~\cite{BaBar:2006pve} and then Belle~\cite{Solovieva:2008fw} found this particle, and the average value of the 
mass difference is $70.6^{+0.8}_{-0.9}\ {\rm MeV}/c^2$.

Orbital excitations were much harder to find. As discussed above in the section on $\Sigma_c$ excitations, we can expected first a family of five orbital excitations, with
the spin of the light quarks combining with the $l_\lambda$-excitation to give $J^P=\frac{1}{2}^-,\frac{1}{2}^-,\frac{3}{2}^-,\frac{3}{2}^-,\frac{5}{2}^-$. However, unlike the
$\Sigma_c$ particles, strong decays via one-pion emission would violate isospin and thus, would be heavily suppressed. 
Possible decay modes would thus be $\Xi_c^+K^-$ and $\Xi_c^0K^0_S$ 
and the latter 
would be experimentally much more difficult to find because of the lower efficiency and higher background due to the ambiguous strangeness of the $K^0_S$. The decay into
$\Omega_c\pi\pi$ would also be possible,
but we note the general paucity of non-resonant
three-body decays in other excited charmed baryon systems. Lastly, electromagnetic decays to the ground state $\Omega_c^0$ might be
competitive, as has been shown in the case for the $\Xi_c(2815)^0$~\cite{Belle:2020ozq}, but there has been little theoretical work on this. It is interesting to note that the difficulty
in finding ground-state $\Omega_c^0$ decays is partly due to the fact that many of their excited states decay into $\Xi_c$~ground states. This is in contrast
to the $\Lambda_c$ and $\Sigma_c$ spectra, where most of the excited states tend to funnel down to ground state $\Lambda_c^+$ particles. 

Given the above, it was the logical next step for the LHCb Collaboration to search for excited $\Omega_c$~baryons in the $\Xi_c^+K^-$ invariant mass spectrum. The results were spectacular, as they found five well-separated peaks,
and all of them were narrow (some with no measurable width)~\cite{LHCb:2017uwr}. The Belle Collaboration, with effectively 1/30 the data set size, reported results on the same mass
distribution and confirmed the existence of four of the states~\cite{Belle:2017ext}. This was old data from Belle which had not been published because it 
did not stand up by itself as evidence of several particles without the input
from LHCb of the intrinsic widths of the peaks. Given that a quintuplet of five states was predicted and five states
were indeed found, and that higher spin typically produces higher mass, the most naive interpretation of the spectrum was that the five states should be $J^P=\frac{1}{2}^-,\frac{1}{2}^-,\frac{3}{2}^-,\frac{3}{2}^-,\frac{5}{2}^-$ (in that order). However, it is not a necessity that higher spin produces
higher mass. Many models claim to be able to explain some of the particles but not others, implying their internal quark configurations are quite different. However,
such theories need to explain the similarity of the cross sections of the particles. 

One notable feature of the spectrum was that the highest mass of the five, the $\Omega_c(3120)$, also had the smallest signal, and in the (low statistics) 
Belle data, was missing completely. In general, for particles within one multiplet, the production cross section rises with the spin of the particle. This has been 
tested over many states in continuum data by Belle~\cite{Belle:2017caf}. Even allowing for a mass suppression and the possibility of decays to $\Xi_c^{\prime +}$, we should
not expect the highest-mass signal to be so clearly the smallest. The plot then thickened when LHCb analysed the substructure of 
$\Lambda_b$ decays to $\Xi_c^+K^-\pi^+$. Here, the constrained
system allows the possibility to find the spin-parity of the states, and there are no complications from $\Xi_c^{\prime +}$ production. Once again, the highest-mass
state had gone missing. 
This low statistics of the second LHCb measurement is not sufficient to produce unambiguous results for the spin-parity of the states. For the $\Omega_c(3000)^0$
and $\Omega_c(3090)^0$, there is effectively no discrimination between $J=\frac{1}{2},\frac{3}{2}$ and $\frac{5}{2}$, whereas for the 
middle two, $J^P=\frac{3}{2}$ is clearly preferred. Those models favouring a family of five states then have several questions to address: 
\begin{enumerate}
    \item Firstly, what is the nature of the clearly
evident narrow state, the $\Omega_c(3120)$? 
    \item Secondly, if the $\Omega_c(3120)$ is excluded from the family of five, where is the missing fifth state?
\end{enumerate}
\noindent
For the former question, one explanation would be that the next
level up of states with one unit of orbital excitations are the ``$\rho=1$'' states, where the excitation is between the light quarks.
In the $\Omega_c$ spectrum, the mass difference between $\lambda$ and 
$\rho$ excitations is clearly going to be smaller than it is for the 
$\Lambda_c$ and $\Xi_c$ spectra. A $J^P=\frac{3}{2}^-, l_\rho = 1$ state below the $\Xi_c^*K$ threshold might well be narrow, whereas a 
$J^P=\frac{1}{2}^-$ state above $\Xi_c^{\prime}K$ threshold may well be wide and in any case, may not be seen in the $\Xi_c K$ mass spectrum. An alternative explanation of the
$\Omega_c(3120)$ is that it is a molecular state, and in this model, the charmed analogue of the $\Omega(2012)$~\cite{Ikeno:2023uzz}.
For the missing state, there have been various conjectures. In both LHCb papers, it is noted that there are events very close 
to the kinematic threshold. In the first analysis, this was thought to be feed-down from $\Xi_c^{\prime +}$ decays. In the second, there is no such background
and they claim there is a $4.2\sigma$ enhancement in that region. 
Alternative models imply that one of the observed peaks is in fact two overlapping particles. 
Other possibilities are that the state is completely below threshold, so will be seen in other decay modes. 
A fourth possibility is that it is so wide that it does not
appear in any plot as a recognizable signal. This is entirely reasonable as the
state with spin-zero $(s_{qq}=0)$ light quarks should be able to decay via an S-wave.

The LHCb Collaboration then repeated their original analysis with more data~\cite{LHCb:2023sxp}, 
and reported two more particles. Already hinted at in their first paper and by Belle, a first wide 
($\Gamma \sim 50$~MeV/$c^2$) signal at a mass of 3185~MeV/$c^2$, 
and more surprisingly a comparatively narrow peak up at a mass of 3327~MeV/$c^2$.
Immediately, various models have been developed to explain these particles including radial excitations, and molecular $(\Xi D$).
Certainly, the lower one has the hallmarks of the series of particles, $\Lambda_c(2765), \Xi_c(2970)$, generally associated with the first
radial excitation. However, despite the large number of models arising from the first paper, including pentaquark states including an $s\bar{s}$ quark pair,~\cite{An:2017lwg}, there were no previous $predictions$ of a separated narrow 
state as high as 3327~MeV/$c^2$. 

\begin{figure}[t]
\caption{\label{Figure:Primes} (Colour online) A comparison of charmed-baryon spectra. Left: The mass spectra for resonances thought to be $\Sigma$-like states. The masses are plotted with respect to the spin-weighted average of the lowest two states -- the lowest state in the $\Xi_c$~column is the $\Xi_c^{\,\prime}$. Right: The mass spectra for the $\Lambda$-like resonances, i.e. light-quark spin zero. The masses are plotted with respect to the ground state. The similarity of the spectra helps guide the identification of the states.}
\vspace{2mm}
\begin{minipage}{1.0\textwidth}
\vspace{1mm}
\begin{tabular}{cc}
\includegraphics[height=0.28\textheight]{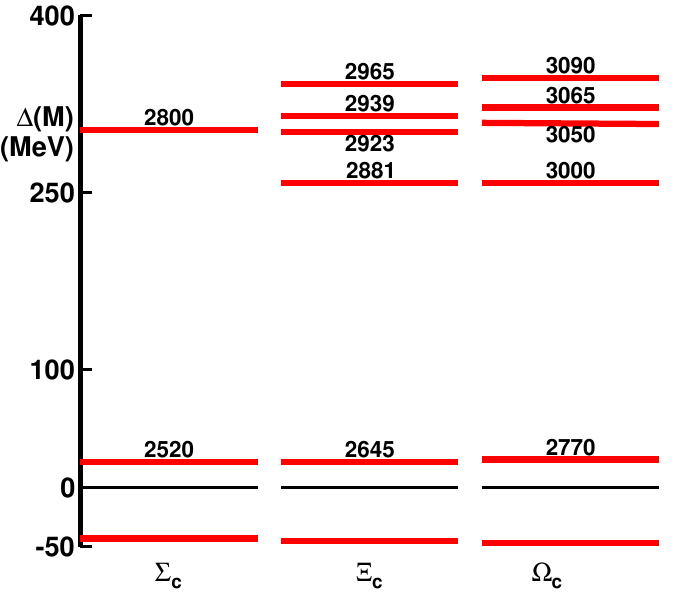} &
\includegraphics[height=0.28\textheight]{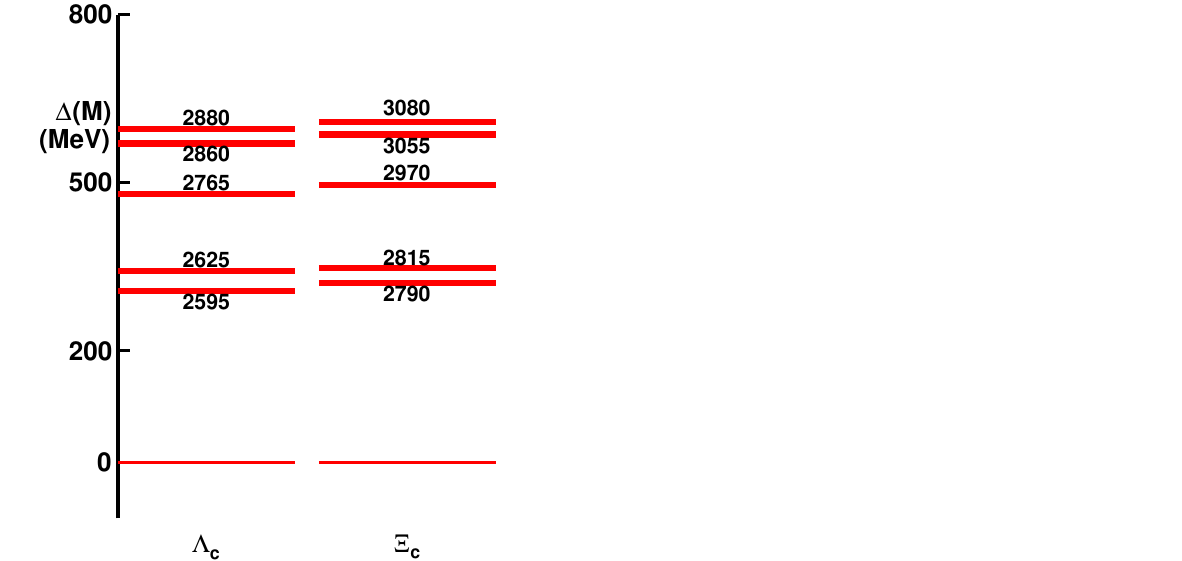}
\end{tabular}
\end{minipage}
\end{figure}

\subsubsection{Comparison of $\Xi_c^*\to\Lambda_c^+K^-$ and excited $\Omega_c^0$ spectra}
\label{Subsubsection:CharmedXiOmega}
We now loop back to the LHCb investigation of the $\Lambda_c^+K^-$ spectrum. Following the lead of the $\Omega_c^{0*}\to\Xi_c^+K^-$ spectrum, this 
has now been done in two ways:
first inclusively~\cite{LHCb:2020iby}, and then in a low statistics but very pure $B$ meson decay mode~\cite{LHCb:2022vns}. 
The first paper showed three clear peaks, with masses of 2923~MeV/$c^2$, 2939~MeV/$c^2$, and 2965~MeV/$c^2$. Furthermore, there is 
another large enhancement at lower mass which overlaps with expected feed-down lumps. In their second
paper, the 2923~MeV/$c^2$ and 2939~MeV/$c^2$ mass peaks are clearly defined, the 2965~MeV/$c^2$ state is too high in mass to be produced, but the peak at around 2882~MeV/$c^2$, though statistically not overwhelming,
is clear enough for a mass and width measurement to be made. The immediately visible pattern,
noticed by the collaboration, was how the mass splittings mirrored those of the excited $\Omega_c$ baryons. Given that the hyperfine splitting should be 
dependent on the heavy quark mass but not the light quark masses, this clearly indicates the likelihood that the spin-parity of the states are similarly aligned.
This is shown in Fig.~\ref{Figure:Primes}.
The 2923~MeV/$c^2$ and 2939~MeV/$c^2$ states, overlapping because of their natural widths, 
make it rather clear that the signals reported by BaBar and Belle, and named the $\Xi_c(2930)$, 
are in fact these two states unresolved.   
If we take the four resolved excited $\Xi_c$~states as corresponding to the four (out of five) $l_\lambda=1$ excitations of the $\Xi_c^{\prime}$~resonance, we once again have the question
of where the fifth state lies. Here, we cannot look to the state being below kinematic threshold, but certainly having a wide state is a possibility. 
For instance, one detailed theoretical analysis of the spectrum~\cite{Wang:2020gkn} shows the 2880~MeV/$c^2$ region comprising the two $J^P = \frac{1}{2}^-$~states
mixing into one comparatively narrow state, and one probably too wide to observe. It should be noted that one distinct difference between this excited 
$\Xi_c$~spectrum and the analogous $\Omega_c$~spectrum is that the orbitally excited $\Xi_c$~particles can decay to a pion plus a ground-state~$\Xi_c$,
$\Xi_c^{\prime}$ or $\Xi_c(2645)$. Unfortunately, the mass spectra of all of these are crowded with overlapping resonances and satellite peaks from kinematic reflections of decays.

\begin{figure}[t]
\caption{\label{Figure:Decays}(Colour online) The mass spectra and decays of all the known excited charmed baryons.}
\includegraphics[height=0.35\textheight,width=0.99\textwidth]{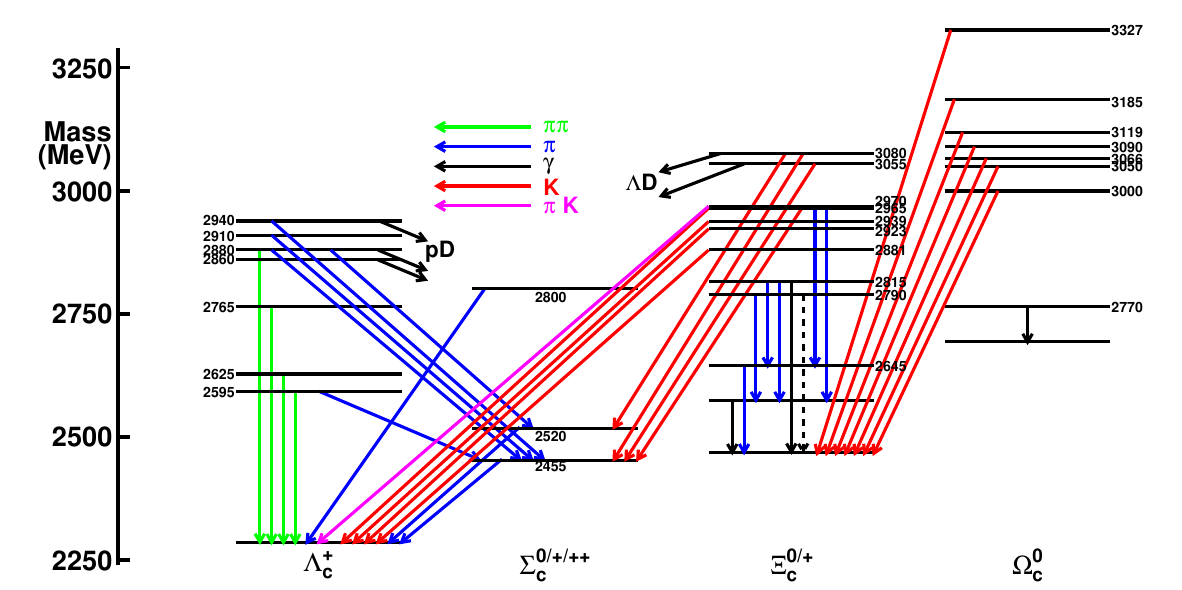}
\end{figure}

Finally, Figure~\ref{Figure:Decays} shows the mass spectra of all the excited charmed baryons and the detected transitions between them. 

\subsection{The Doubly Charmed $\Xi_{cc}^{++}$ baryon}
The quark model clearly predicts a spectrum of doubly charmed baryons. When the two charm quarks are accompanied by a light quark, they are denoted according to the standard naming system as $\Xi_{cc}$ baryons.

The first reports claiming the discovery of a doubly charmed baryon, the 
$\Xi_{cc}^{+}$~state, was made by the SELEX Collaboration~\cite{SELEX:2002wqn,SELEX:2004lln} in a $\Sigma^-$ beam, fixed-target experiment. However, these results were immediately controversial. Although the statistical significance of the signals was superficially sufficient to claim an observation of a new particle, the production rate relative to that of the singly charmed $\Lambda_c^+$~baryon was much larger than any production model could accommodate and the experiment showed no signals for other singly charmed baryons, which would be expected to be much easier to detect than doubly charmed baryons. Furthermore the lifetime was shorter than expected for this weakly decaying state, and the observed mass was at the low end of the quark model predictions.

The LHCb experiment at CERN is ideal for searching for doubly charmed baryons. The cross-section in high-energy proton collisions should be high and the experiment is designed around tagging on the lifetime of weakly decaying heavy particles.
In 2017, the LHCb Collaboration reported unambiguous evidence of a doubly charged, doubly charmed baryon in the invariant $\Lambda_c^+ K^- \pi^+\pi^+$ mass distribution, where
the $\Lambda_c^+$~baryon is reconstructed in its decay into $p K^- \pi^+$~\cite{LHCb:2017iph}. After further studies and combining decay modes, LHCb quoted $(3621 \pm 0.23\pm 0.30)\ {\rm MeV}/c^2$ as the most precise mass measurement of this state~\cite{LHCb:2019epo}. 
The references in the first of these LHCb publications~\cite{LHCb:2017iph} is an excellent source of the mass predictions which ranged from 3500--3700~MeV/$c^2$.
Several predictions, both from the standard quark model~\cite{Ebert:2002ig, Karliner:2014gca} and from lattice gauge theory (see, for instance, ~\cite{Perez-Rubio:2015zqb} and references therein), proved to be in excellent agreement with the measurement. These predictions, made despite the disconcerting SELEX 
claims, add credibility to both theoretical approaches. The measured lifetime, $\tau(\Xi_{cc}^{++}) = (0.256^{+0.024}_{-0.022} \pm 0.014)$~ps, is at the low end of predictions.

The masses of the $\Xi^{++}_{cc}$ and $\Xi^{+}_{cc}$~states are expected to differ by only a few MeV/$c^2$, due to the approximate isospin symmetry~\cite{Hwang:2008dj,Brodsky:2011zs,Karliner:2017gml}. This makes the SELEX result even less believable. LHCb have searched for the $\Xi_{cc}^+$ and have a hint of a signal near the mass of their $\Xi_{cc}^{++}$~\cite{LHCb:2021eaf}. Most predictions indicate that the
lifetime of the singly charged state will be lower than that of the doubly charged state which is a disadvantage in separating signal from combinatorial background. Clearly in the next few years, we can expect more discoveries in this sector and there are many predictions of the properties of excited $\Xi_{cc}$ states and other doubly heavy baryons that will be tested.

%% file: BottomBaryons.tex
\subsection{The ground states}
The first bottom baryon found was, not surprisingly, the ground state $\Lambda_b^0$. There were several rough measurements of its mass from CERN as well as evidence of its semi-leptonic decays, 
and then these efforts were followed by rather more convincing evidence from CDF at Fermilab~\cite{CDF:1996rvy}.
However, the first reasonably precise mass measurement was not performed until CDF's second
measurement in 2006~\cite{CDF:2005qop}, and this report can be said to have marked the beginning of $B$~baryon
spectroscopy. 

The first exclusive reconstruction of $\Xi_b$~states was performed by the Fermilab collider experiments, though
evidence of the semi-leptonic decays of the $\Xi_b^-$, and indeed a life time measurement,
was already presented by experiments at the LEP collider.
We shall see that the rather large isospin-splitting of the $(\Xi_b^0,\,\Xi_b^-)$~pair has ramifications in the discovery of the excited states.

The first report of the $\Omega_b$~mass in 2008 by the D0 experiment at Fermilab~\cite{D0:2008sbw} was $\sim 100$~MeV/$c^2$ higher than standard potential
model predictions. The model of Ref.~\cite{Karliner:2008sv} predicted 6052.1~MeV/$c^2$, for instance.
It was thus a success of the models that subsequent measurements, first by the CDF~\cite{CDF:2009sbo} Collaboration, and more recently
by the LHCb Collaboration~\cite{LHCb:2013wmn}, not only agreed with each other, but also disagreed with the first report, and are now in good agreement with models. The present world average
is $(6045.2\pm 1.2)$~MeV/$c^2$~\cite{ParticleDataGroup:2022pth}.

\subsection{$\Sigma_b$ and Excited $\Lambda_b$ Baryons}
It is interesting to conjecture what
we would expect for bottom baryons based on the charmed baryon spectrum and compare that with what has actually been found. This
gives a direct check on whether the understanding of the charm spectrum is complete.

\subsubsection{The $\Sigma_b$ spectrum and decays}
\begin{figure}[tb]
\caption{\label{fig:g2} (Colour online) Data of the coupling constant $g_2$ which characterizes the $L=1$ strong decays of $\Sigma$-like baryons.}
\centerline{\includegraphics[width=0.46\textwidth]{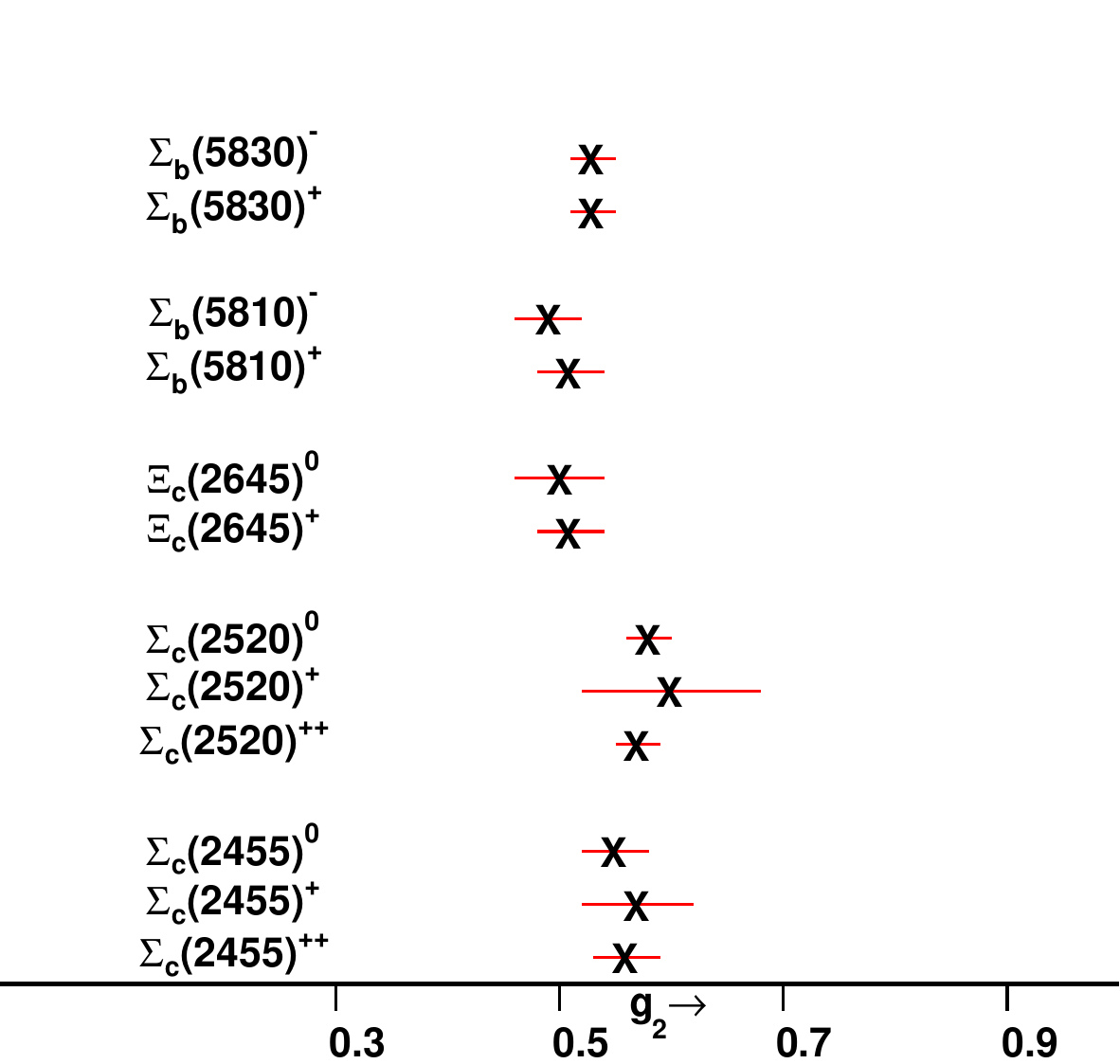}}
\end{figure}

According to the standard quark model, the excitation energy of the spin-weighted $\Sigma_b$ and $\Lambda_b$~ground states
should be independent of the heavy quark mass. In addition, the hyperfine splitting between the $\frac{1}{2}^+$ and 
$\frac{3}{2}^+$ states should be inversely proportional to the heavy quark mass. For the latter, 
we can take the constituent 
quark masses for charm and bottom which are given in the ratio of~0.31. 
This means that we expect a $\Sigma_b/\Sigma_b^*$~doublet with masses around 5817~MeV/$c^2$ and 5837~MeV/$c^2$. 
These particles have since been found, first by the CDF Collaboration~\cite{CDF:2007oeq} and then by the LHCb Collaboration~\cite{LHCb:2018haf}.
The mass splitting is in good agreement with expectations, but the masses themselves are 
approximately 5~MeV/$c^2$ lower. We can go further and look at the widths of these particles, as the decays 
of all the $\Sigma$ particles
should be directly related to their masses, using the formalism of Pirjol and Yan~\cite{Pirjol:1997nh} to describe the P-wave decay:
\begin{equation}
  \Gamma(\Sigma_h)=\frac{g_2^2}{2\pi f^2_{\pi}}\frac{M_{\Lambda_h}}{M_{\Sigma_h}}|p^{\pi}|^3\,,
\end{equation}
where $h$ refers to the heavy quark, and the value of the coupling $g_2$ should be the same for all resonances. Figure~\ref{fig:g2} shows these values for the $\Sigma$-like states (including $\Xi_c$), 
and indeed, the agreement is impressive. Note that the shown uncertainties are approximate and no attempt should be made 
to average these values
without understanding the correlations between measurements.  

\subsubsection{The excited $\Lambda_b$ and $\Sigma_b$ spectra}
Moving to the orbital excitations, the hyperfine splitting should obey the rule that the mass splitting of the ($\frac{1}{2}^-,\,\frac{3}{2}^-$)~doublet is expected to be about 3.2~times bigger in the charm sector than it is for the bottom analogues. 
The one unit of energy associated with the orbital excitation should also be the same to zeroth order. 
However, there was always an expectation that this energy decreases with 
higher heavy-quark mass. 
Specific quark model predictions were made by Karliner, Keren-Zur, 
Lipkin and Rosner~\cite{Karliner:2008sv}, who predicted masses of 
5929~MeV/$c^2$ and 5940~MeV/$c^2$ for the $\frac{1}{2}^-$ and $\frac{3}{2}^-~\Lambda_b^0$ states, respectively. We now know 
that the masses are 5912~MeV/$c^2$ and 5920~MeV/$c^2$, and the agreement seems reasonable.
The hyperfine splitting of 8~${\rm MeV}/c^2$ agrees rather better with an
estimate based on the
$\Xi_c$ splitting of 25 ${\rm MeV}/c^2$ than with the latest value for the $\Lambda_c$~splitting
of 36~${\rm MeV}/c^2$. This might be evidence that the latter splitting is enhanced by
the interaction of the $\Lambda_c(2595)$ with the $\Sigma_c\pi$ threshold.
We note that there are the calculations of Capstick and Isgur~\cite{Capstick:1986ter} who predicted the masses of this doublet to within 1~MeV/$c^2$ of their experimental values - however this must be a coincidence.
 
In the $\Lambda_c^+\pi^+\pi^-$~spectrum, the next higher-mass state is the poorly understood $\Lambda_c^+(2765)$.
An analogous state is  observed in the $\Lambda_b\pi^+\pi^-$ invariant mass distribution, the 
wide peak known as the $\Lambda_b(6070)$ seen by the LHCb Collaboration~\cite{LHCb:2020lzx} and with lower statistics, by the CMS Collaboration~\cite{CMS:2020zzv}.
Like the $\Lambda_c^+(2765)$, 
this particle is copiously produced but wide, and it cannot be ruled out that it is in fact two or more states
overlapping. The natural explanation is its interpretation as the 2S radial excitation, since its mass and properties match that of the $\Lambda_c(2765)$ and $\Xi_c(2970)$~resonances. 

The next doublet up in mass would be the equivalent of the $\Lambda_c(2860,2880)$ doublet. If we assume that these
have been correctly identified as a $J^P=\bigl(\frac{3}{2}^+,\,\frac{5}{2}^+\bigr)$ pair, we would expect a similar pair 
in the $\Lambda_b$~spectrum, but this time separated by only about $20/3\approx 6$~MeV/$c^2$. 
The LHCb Collaboration~\cite{LHCb:2019soc} have indeed found such a pair, the $\Lambda_b(6146,6152)$ doublet. 
At first view, the peak in the $\Lambda_b\pi^+\pi^-$ invariant mass plots appears to be one
particle, however, by looking at the resonant substructure, they showed that the higher-mass state decays preferentially into
$\Sigma_b$, whereas the lower-mass state decays into $\Sigma_b^*$. 
The fact that the upper state decays so little into
$\Sigma_b^*$ would actually make its identification as a $J^P=\frac{5}{2}^+$ state problematic, but it is
very reminiscent of the $\Lambda_c(2880)$ decays discussed above. In other words, it is
a problem, but only the same one problem shared by the two systems, charm and bottom. The lower-mass state has a much narrower width
than the corresponding charmed state, presumably (at least in part) because it is below threshold for the decay into $pB$ which would
be the decay analogous to that of $\Lambda_c(2860)\to p D^0$.

We note that there has been no observation of a bottom equivalent of the $\Lambda_c(2940)$~resonance (see Tables~\ref{Table:CharmedBaryonClassification} and~\ref{Table:BottomBaryonClassification}). If one is found, it
will shed new light on its underlying quark structure. 

One more $\Sigma_b$ candidate has been observed. The $\Sigma_b(6097)$~resonance has been observed in two charge states, and it is copiously produced and wide. The situation is very similar to the $\Sigma_c(2800)$ in that it is in the region
where the quark model would predict five states, but we do not know if this is one or many of them, or some other quark combination. 

\subsection{Excited $\Xi_b$ baryons} 
We can expect the excited $\Xi_b$ spectrum to look similar to the $\Xi_c$ spectrum, with the usual narrowing of
the gap between spin partners. There is the complication that the isospin mass splitting is now of similar
size to the spin splitting. By accident, the first excited negatively charged $\Xi_b$ (the ``prime'' state)
is just above the pion threshold
and the state was discovered, along with its $J^P=\frac{3}{2}^+$ partner, by the LHCb Collaboration~\cite{LHCb:2014nae} in decays to $\Xi_b^0\,\pi^-$. On the other hand, the neutral prime state, which is presumably below threshold for
pion decays, remains to be discovered. The neutral $J^P=\frac{3}{2}^+$ partner state 
was first reported by the CMS Collaboration~\cite{CMS:2012frl} and confirmed
by LHCb~\cite{LHCb:2016mrc}. However, we note that its identification as a $J^P=\frac{3}{2}^+$~state rather than a $J^P=\frac{1}{2}^+$~state is simply due to its mass.

Progress in the search for higher-spin states was first made with the discovery of the $\Xi_b(6100)^-$, reported by the CMS Collaboration~\cite{CMS:2021rvl} in the decays to
$\Xi_b^-\pi^+\pi^-$. Reconstructing as many as four decay vertices in this one decay chain is a great achievement for CMS which
was not designed for heavy baryon spectroscopy. 
The authors interpreted their results as evidence of the 
$J^P=\frac{3}{2}^-$~state equivalent of the $\Xi_c(2815)$. Their result was 
confirmed by LHCb~\cite{LHCb:2023zpu} who found this same state, measured its intrinsic width as narrow, and also found its neutral partner, ($\Xi_b(6095)$.
In addition, they also found the neutral $J^P=\frac{1}{2}^-$ state 8 MeV/$c^2$ lower than the $J^P=\frac{3}{2}^-$ state. The LHCb Collaboration has also shown circumstantial evidence of the charged $J^P=\frac{1}{2}^-$ state, but its detection is complicated by the fact that presumably all its decays involve neutral particles. Thus, we have an almost complete set of the first orbital excitations of $\Lambda_c,\Xi_c, \Lambda_b$ and $\Xi_b$. The very narrow widths of these particles is not surprising considering the very limited phase space available, but calculations are complicated by the isospin splittings. We note that the mass splitting of the two neutral states, named by LHCb the $\Xi_b(6095)$ and $\Xi_b(6087)$), is very similar to the mass splitting of the $\Lambda_b(5920)$ and $\Lambda_b(5912)$ as expected, and even the isospin mass splitting of the $\Xi_b(6100)^-$ and $\Xi_b(6095)^0$ follows the pattern found in the analogous charm sector.

Lastly in this system, is the $\Xi_b(6227)$, found in both its charge states by the LHCb Collaboration~\cite{LHCb:2018vuc,LHCb:2020xpu}.
Their measured natural widths are $\approx 20$~MeV/$c^2$, which is sufficiently large that there could be more than one overlapping particles. Although there are many possible interpretations of these state, it is noteworthy
how similar its properties are to the $\Lambda_c(2765)$, $\Xi_c(2970)$ and $\Lambda_b(6070)$.

\subsection{Excited $\Omega_b$ Baryons}
Progress in the excited $\Omega_b$ sector has been fast -- 
no doubt given momentum by the excited $\Omega_c$~spectrum. 
The first excited
state ($J^P=\frac{3}{2}^+$) has not been found, but this is not surprising as it will involve a low-energy photon transition that will be
very difficult to detect. However, the LHCb Collaboration~\cite{LHCb:2020tqd} has reported the discovery of (probably four) excited states that closely echo the corresponding states in the excited $\Omega_c$~spectrum. The states are narrow and well separated. Of course, as we have seen, identification of the excited $\Omega_c$ states
is still controversial, and the $\Omega_b$ spectrum does not add much clarity except that any explanation for the $\Omega_c$ spectrum must also
work for $\Omega_b$ with the simple substitution of the heavy quark mass. It should be noted that the statistical significance of two of
the signals
is below the significance threshold that is usually a prerequisite for claiming the existence of a new particle. 
However, it would be surprising, given the symmetry
with the $\Omega_c$~signals, if they did not increase in significance with more data.

It can be noted that the fifth of the observed narrow $\Omega_c$ states, 
the $\Omega_c(3120)$~resonance, is not seen in this excited $\Omega_b$ spectrum.
This might, of course, be because of the lack of statistics. 
However, it can also be noted that if the $\Omega_c(3120)$~resonance was a $\rho$~excitation, 
the equivalent $\Omega_b$~resonance would be more separated from the $\lambda$~excitations (at a mass of approximately 6445~MeV/$c^2$) than is the case in the charm sector.

%\section{Summary of Charm-Bottom Comparison}
%In Fig.~\ref{fig:cb} we demonstrate the symmetry found between the charm and bottom sectors. We note that the hyperfine splitting being inversely proportional to the heavy quark mass, is obeyed in a large number of different states. In addition, the expected rather small decrease in the excitation energy associated of, for instance, a $\lambda$-excitation is also apparent. These patterns appear exist even for particles for which the precise spin-parity is not known - and thus any identification of a state must look at both spectra simultaneously. As shown in Fig.\ref{fig:cb}, the masses of no fewer than 16 excited bottom baryons could have been well-predicted using these simple rules just by extrapolating from the charmed baryons. The bottom spectrum has understandably been slower to fill up than the charm one, and we will have to wait to see if there is a bottom analog of, for instance, the $\Lambda_c(2940)$.

%\begin{figure}[htb]
%\includegraphics[height=8cm]{figures/charmbottom2.pdf}
%\caption\noindent{Comparison of the charmed baryons and bottom baryons spectra. Note that isospin splitting is ignored, and the base line is given as the weakly decaying ground-state for the $\Lambda$ and $\Xi$, and the spin-weighted average of the lowest two states for the $\Xi_c^{\prime}$ and $\Omega$. The position of the first excited state of the $\Omega_b$ (shown by the dashed line) is assumed.}
%\label{fig:cb}
%\end{figure}

%% file: Discussion.tex
\section{Discussion and Open Questions}
\label{Section:Discussion}
The proton has been known for more than a century but some of its basic properties, e.g. spin, mass, intrinsic structure, size, are still poorly understood. The valence quark contribution to the nucleon spin is only about half of the total spin~\cite{Aidala:2012mv}. Moreover, only a small fraction of the nucleon mass is generated by the Higgs mechanism. The biggest share in the mass is dynamically generated from the underlying strong interaction. The composite nature of the nucleon manifests itself in the existence of a rich spectrum of excited states and one of the important questions to be addressed in spectroscopy is that of the relevant degrees of freedom in a baryon. Does a baryon consist of three symmetric quark degrees of freedom, a dynamical quark-diquark structure, or does it even have a very different, more complex structure? A better understanding of the properties of the known resonances and an improved mapping of the baryon spectra will help elucidate this question.

One approach to better understand the nucleon is nicely outlined in Ref.~\cite{Granados:2017cib}. To explore the intrinsic structure of an object,
\begin{itemize}
    \item excite it,
    \item scatter from it,
    \item replace some of its building blocks with other, similar ones.
\end{itemize}
Recent photoproduction experiments have addressed the excitation spectrum of the nucleon as part of the global $N^\ast$~program. Several new resonances have been identified, but their confirmation remains challenging since any discovery is not inferred from a direct observation~\cite{Gross:2022hyw,Ireland:2019uwn,Klempt:2009pi,Crede:2013kia,Thiel:2022xtb}. Replacing one or more of the light $u$ or $d$~quarks in the nucleon with a strange, charm, or bottom quark results in a hyperon. To this effect, the very strange $\Xi$ and $\Omega$~hyperons provide the link between the light-flavour sector and the heavy-flavour sector, which have been reviewed in this contribution. Historically, the doubly and triply strange hyperons were instrumental in revealing the quarks as the building blocks of the nucleons and other hadrons. As a consequence, the intrinsic structures of hyperons and nucleons must be intimately related. On the structure side, hyperon electromagnetic and transition form factors contain complementary information to the nucleon and $\Delta$~form factors. Their knowledge would provide crucial tests for our current picture of the nucleon structure. However, the experimental information on
hyperon form factors is rather limited. Magnetic moments have been measured for the octet hyperons~\cite{ParticleDataGroup:2022pth} and the $\Omega$~decuplet ground state. But the decuplet-octet transitions are only poorly known and a program at the upcoming facilities on radiative and Dalitz decays of hyperons in transitions such as $\Sigma^0\to \Lambda e^+ e^-$ and $\Xi^0\to \Lambda e^+ e^-$ would be very interesting. The latter decay was studied by the NA48 Collaboration at CERN~\cite{NA48:2007smd} and the branching fraction was determined with a fairly large uncertainty~\cite{ParticleDataGroup:2022pth}. 

In the following, we give a brief summary of the status of hyperons; we will also discuss possible quantum numbers and multiplet assignments for the multi-strange resonances. A section on the {\it current and planned experiments} will outline relevant open questions and aspects of a spectroscopy program on multi-strange resonances at the current and upcoming facilities, e.g. at Jefferson Lab ($K_L$\,-beam experiments at GlueX), J-PARC ($K^-$\,-beam experiments), and FAIR ($p\overline{p}$~experiments at PANDA) on the nuclear side, and KEKB ($e^+ e^-$~interactions at Belle) and CERN ($pp$~interactions at LHCb) on the high-energy side. We conclude with some relevant and open questions regarding charmed baryons. 

\subsection{A brief summary of the status of the very strange resonances}
\subsubsection*{The doubly strange $\Xi^\ast$~states}
The lowest-mass $\Xi$~resonances, $\Xi(1320)$ and $\Xi(1530)$, have been well established ever since their discovery more than 60 years ago. Different phenomenological models can successfully explain their properties, but predictions for the excitations
are very different and partially conflicting. A spread of 100 to 200~MeV/$c^2$ in the mass of any particular state is not uncommon among the predictions
of these various treatments, particularly for excited states. These two lowest-mass states are assigned to the octet and decuplet for the ground-state baryons, respectively. While there is no reason to question these assignments, we note that only the spin of the $\Xi(1530)$ has been experimentally addressed. The BaBar Collaboration examined the $\Xi(1530)$ in the reaction $\Lambda_c^+ \to \Xi^- \pi^+ K^+$ to determine its quantum numbers, and assuming that $\Lambda_c^+$ has $J^P = \frac{1}{2}^+$, they find conclusively that the spin of the $\Xi(1530)$ is $J=\frac{3}{2}$ and that positive parity is favoured. The details of this study can be found in Ref.~\cite{BaBar:2008myc}. The $J^P$~quantum numbers of the $\Xi(1320)$ are merely quark model assignments.

\begin{figure}[t]
\caption{\label{Figure:Photoproduction} (Colour online) Left: Preliminary invariant $\Lambda\pi^-$ mass from GlueX showing the doubly strange octet ground-state
  $\Xi(1320)$ resonance in the reaction $\gamma p\to K^+\,((\Lambda\pi^-)_{\Xi(1320)}\,K^+)$. A signature for the $\Sigma(1385)^-$ is also visible due to some background originating from a $\pi^+$ in the final state misidentified as a $K^+$. Right:
  Preliminary invariant $\Lambda K^-$~mass showing signatures of the $\Xi(1690)^-$ and $\Xi(1820)^-$~resonances
  in the reaction $\gamma p\to K^+\, ((\Lambda K^-)_{\Xi^\ast}\,K^+)$.}
\vspace{2mm}
\begin{minipage}{1.0\textwidth}
\begin{tabular}{cc}
\includegraphics[width=0.47\textwidth,height=0.25\textheight]{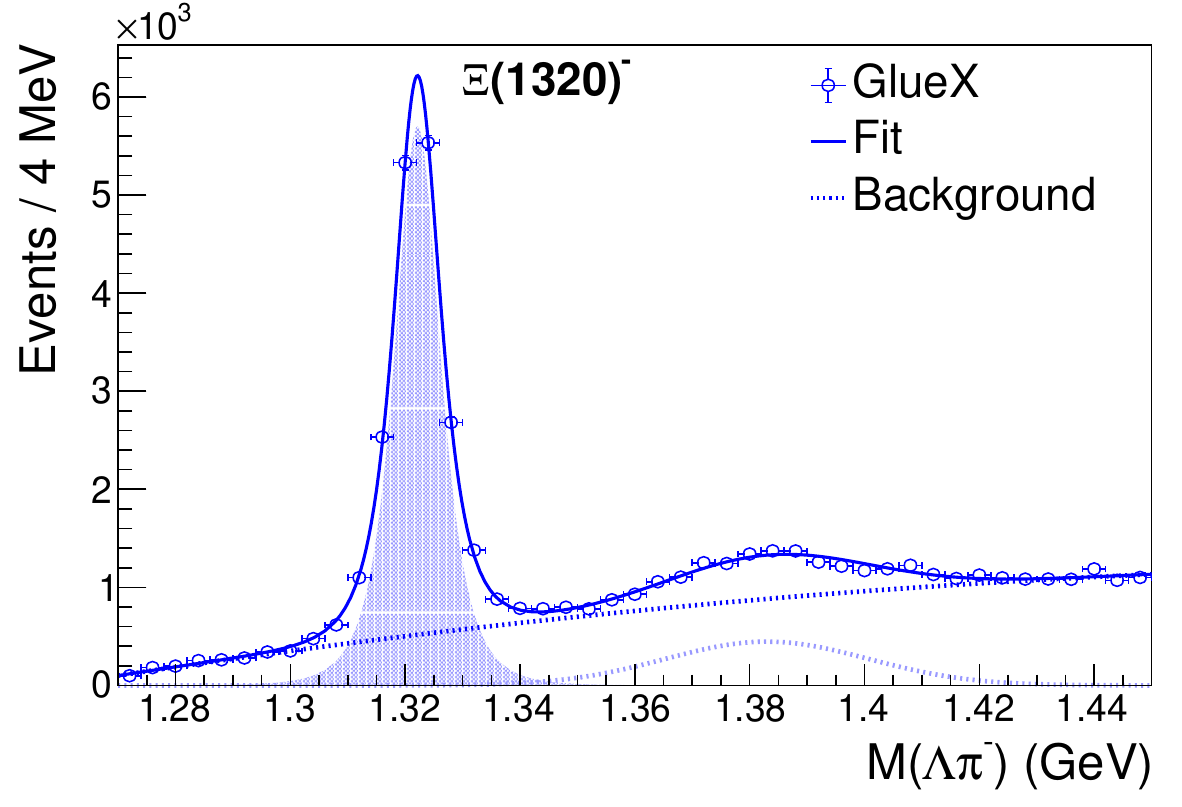} &
\includegraphics[width=0.47\textwidth,height=0.25\textheight]{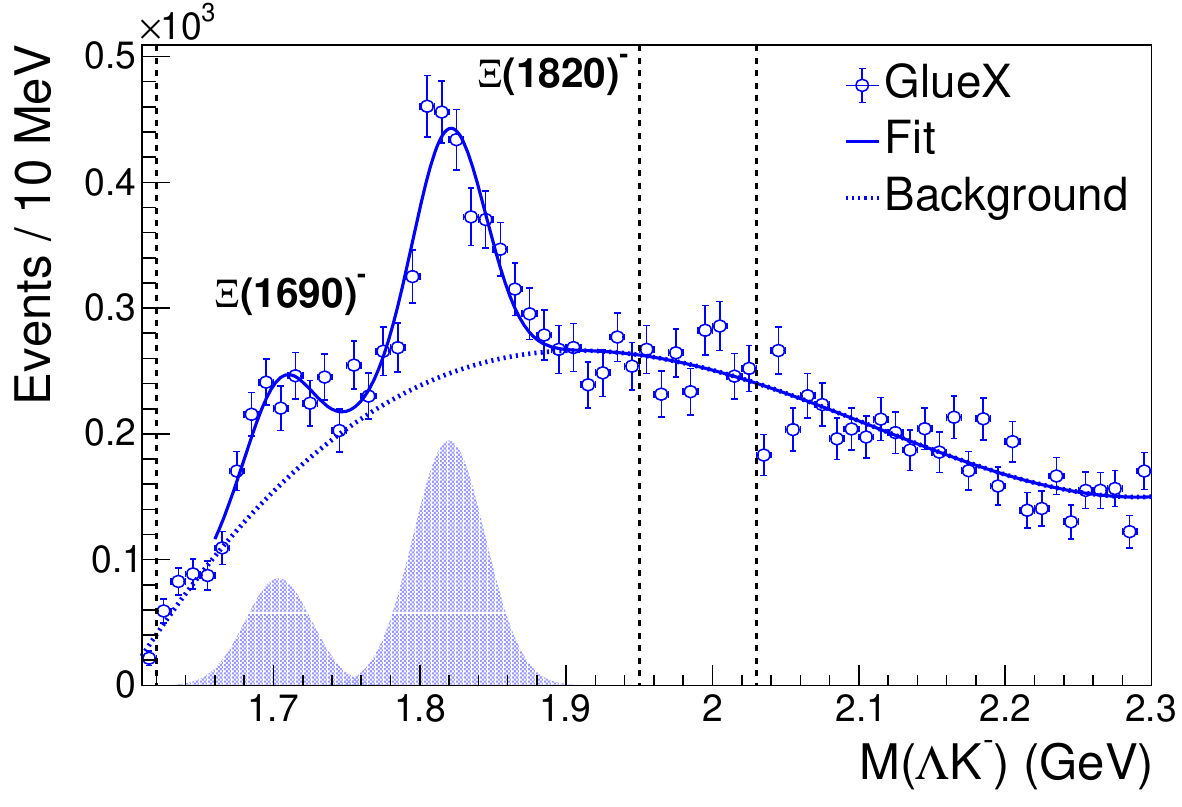}
\end{tabular}
\end{minipage}
\end{figure}

The $\Xi$~spectrum beyond the $\Xi(1530)$ is more controversial and not well established. Based on the accumulated experimental evidence, the three states, $\Xi(1690)$, $\Xi(1820)$, and $\Xi(2030)$, deserve their 3-star status as genuine doubly strange resonances. A clear signal for the $\Xi(1820)$ has recently also been observed in photoproduction at GlueX~\cite{Crede:2023ncq} (Fig.~\ref{Figure:Photoproduction}, right side) and in the reaction $e^+ e^- \to (K^-\Lambda)_{\Xi^\ast} \overline{\Xi}^+$~\cite{BESIII:2023mlv} (at BESIII). Since this state can consistently been described in almost all theoretical approaches, we believe an upgrade to a well-established 4-star state is warranted in future editions of the RPP. 

In addition to the $\Xi(1690)$, we also look at the structure around 1620~MeV/$c^2$ as an established doubly strange signal that has now been observed in various production mechanisms, e.g. in $K^-\,p$ interactions and the decay of the charmed $\Xi^+_c$. An upgrade to at least a 2-star assignment \b{should be considered}. However, the nature of this state as a genuine resonance is questionable and given its comparatively large width, the structure is most likely a superposition of two or more $\Xi$~states. The reported width of $\sim 60$~MeV/$c^2$ is about two-three times larger than the width of $\sim 25$~MeV/$c^2$ reported for the $\Xi(1690)$, $\Xi(1820)$, and $\Xi(2030)$~resonances. The $\Xi(1620)$ exhibits a dominant $\Xi\pi$~decay mode, see for example the results of the Belle Collaboration~\cite{Belle:2018lws}. But very weak hints for additional decay modes, e.g. into $\Lambda K^-$, now come from photoproduction experiments, see Fig.~\ref{Figure:Photoproduction} (right side). We note that the phase space for $\Xi(1620)\to \Lambda K$ is greatly limited and the decay into $\Sigma K$ even kinematically forbidden. More statistics is needed in the future to properly establish the different decay modes of the $\Xi(1620)$~structure and to perform a lineshape measurement aimed at disentangling different resonant contributions.

The 1900--2000~MeV mass range, which contains the $\Xi(1950)$~resonance, also needs further investigation. Many excited $\Xi$~resonances are expected in this region and the experimental results for the mass and width of the $\Xi(1950)$ are more conflicting than for any other established $\Xi$~resonance. In our opinion, the current experimental evidence for the $\ast~\Xi(1620)$ is stronger than for the $\ast\ast\ast~\Xi(1950)$. 
%\sout{but they are currently [...]}

\subsubsection*{The triply strange $\Omega^\ast$ resonances} The $\Omega^-$~ground state has fairly well known properties with solid evidence for its existence. Excited states have proven difficult to find, though. Out of only four known excited states, only the two lowest-lying excited states, $\Omega(2012)$ and $\Omega(2250)$, have a 3-star classification in the RPP~\cite{ParticleDataGroup:2022pth}, i.e. their {\it existence is likely to certain, but further confirmation is desirable.} However, we note that clear experimental evidence for the $\Omega(2012)$~resonance has so far only been reported by the Belle Collaboration~\cite{Belle:2018mqs}, whereas the initial claim for the $\Omega(2250)$~resonance at CERN~\cite{Biagi:1985rn} was confirmed later by the LASS Collaboration at SLAC~\cite{Aston:1987bb}. Usually, confirmation from a second experiment is expected before a peak is elevated to the Summary Table as emphasized in the RPP introductory remarks for the $\Omega(2470)$. Nevertheless, a 3-star classification for both resonances is justified and acceptable. The remaining two $\Omega^\ast$~states, $\Omega(2380)$ and $\Omega(2470)$, have a 2-star classification, i.e. {\it evidence of existence is only fair.} We agree with the RPP statement for the $\Omega(2470)$ that {\it there is no reason to seriously doubt the existence of this state,} which is also fair to say about the $\Omega(2380)$. 
%\sout{In light [...], the 2-star classification may seem slightly inconsistent.}

\subsection{Possible quantum numbers and multiplet assignments}
\subsubsection*{The doubly strange $\Xi^\ast$~resonances}
Seven excited $\Xi$~states with negative parity are expected in the mass region up to about 2100~MeV/$c^2$ based on SU(3)~$\otimes$~O(3)~symmetry. These first orbitally excited states are members of the $({\bf 70},\,1_1^-)$~supermultiplet which decomposes into a decuplet with spin~$\frac{1}{2}$, and two octets with spin~$\frac{1}{2}$ and spin~$\frac{3}{2}$, see Eq.~(\ref{Equation:70plet}). Combining these with one unit of orbital angular momentum yields three states with $J^P = \frac{1}{2}^-$, three states with $J^P = \frac{3}{2}^-$, and one state with $J^P = \frac{5}{2}^-$. These states are predicted by all quark models with varying masses, but the numbers for each partial wave are also consistent with predictions from various lattice efforts~\cite{Edwards:2012fx}. Furthermore, the mass of the first radial excitation of the $\Xi(1320)$ is expected to fall into this low-mass range. Based on the masses of other radial excitations, $N(939) \leftrightarrow N(1440)$, $\Lambda(1116) \leftrightarrow \Lambda(1600)$, $\Sigma(1193) \leftrightarrow \Sigma(1660)$, we would not expect the $\Xi^{\,\ast}\,\frac{1}{2}^+$ at masses far below 1900~MeV/$c^2$. The calculations in a relativized quark model described in Ref.~\cite{Capstick:1986ter} predict that the first radial excitation of the ground-state~$\Xi$ should have a mass of about 1840~MeV/$c^2$ and the constituent quark model studies of Ref.~\cite{Valcarce:2005rr} place the state at a mass slightly below 1800~MeV/$c^2$. However, the lattice-QCD efforts of the Hadron Spectrum Collaboration~\cite{Edwards:2012fx} have predicted the mass of the $\Xi\,\frac{1}{2}^+$ to be greater than the mass of the negative-parity states. Finally, dynamically generated $\Xi$~states based on final-state meson-baryon degrees of freedom have been predicted in this mass region, see for example Ref.~\cite{Ramos:2002xh,Miyahara:2016yyh}.

Among the established $\Xi$~resonances, the $\Xi(1820)$ is the least controversial with a $J^P = \frac{3}{2}^-$ classification that is rarely questioned. Almost all theoretical approaches predict a $\frac{3}{2}^-$~state that closely matches the experimental $\Xi(1820)$~state. Since this resonance almost exclusively decays into $Y\overline{K}$ and seems to largely decouple from $\pi$~decays, assigning this state to the lowest-lying $\frac{3}{2}^-$~wave is consistent with the arguments presented in Section~\ref{Subsection:LightBaryonSpectroscopy}. These predict the lowest-mass excitations in the $\frac{1}{2}^-$\,- and $\frac{3}{2}^-$\,-\,waves to be dominantly based on excitations between the two strange quarks. As a result, they would decouple from the decay into $\Xi\pi$ and thus be narrow. This is the observation for the $\Xi(1820)$, likely making this state an octet member and the partner of $N(1520)$, $\Lambda(1690)$, and $\Sigma(1670)$ with $J^P = \frac{3}{2}^-$. Likewise, the $\Xi(1690)$~state is frequently discussed as a $\Xi\,\frac{1}{2}^-$~resonance and the low mass almost rules out the interpretation as a radial excitation. Belle clearly observes the $Y\overline{K}$~decay mode to dominate over $\Xi\pi$. For this reason, a dominant component in the structure of this resonance seems to be an excited $|ss\rangle$~diquark. The state could thus be the lowest-lying $\Xi\,\frac{1}{2}^-$~resonance and the $^2\bf{8}_{\frac{1}{2}^-}$~octet partner of $N(1535)$, $\Lambda(1670)$, and $\Sigma(1620)$ with $J^P = \frac{1}{2}^-$. Or alternatively, the $^4\bf{8}_{\frac{1}{2}^-}$~octet partner of $N(1650)$, $\Lambda(1800)$, and $\Sigma(1750)$ with $J^P = \frac{1}{2}^-$. Given the clear observation of the lower-mass $\Xi(1620)$~structure, we tentatively assign $\Xi(1690)$ to the $^4{\bf 8}_{J^-}$~triplet. We note that some weak experimental evidence for the $J^P = \frac{1}{2}^-$ classification was reported in 2008 by BaBar~\cite{BaBar:2008myc}, see also Section~\ref{Subsubsection:1690}. 

The classification of the $\Xi(1690)$ and $\Xi(1820)$ as $\frac{1}{2}^-$ and $\frac{3}{2}^-$~states is supported by the majority of theoretical approaches, e.g. by calculations in a chiral quark model~\cite{Xiao:2013xi} and by the calculations in the constituent quark models of Refs.~\cite{Valcarce:2005rr,Pervin:2007wa}. A study by means of two-point QCD sum rules considering the $\Xi(1690)$ either an orbitally excited state with $J^P = \frac{1}{2}^-$ or the radial excitation with $J^P = \frac{1}{2}^+$ clearly favoured the negative-parity assignment based on results for the ratio of its decay into $\Sigma K$ and $\Lambda K$~\cite{Aliev:2018hre}. The $\frac{1}{2}^-$~classification of the $\Xi(1690)$ is also favoured in calculations from a Skyrme model~\cite{Oh:2007cr} and in the chiral unitary approach of Ref.~\cite{Miyahara:2016yyh}. For the latter, under the assumption of $J^P = \frac{1}{2}^-$ for the $\Xi(1690)$, a clear peak for the dynamically generated $\Xi(1690)$ is observed in the $\Xi\pi$ and $\Lambda\overline{K}$~distributions originating from the decay $\Xi_c\to \pi^+\,MB$. Earlier theoretical efforts have also explained the $\Xi(1690)$~resonance as a dynamically generated state~\cite{Kolomeitsev:2003kt,Garcia-Recio:2003ejq,Sarkar:2004jh}. 
For completeness, we also list references for calculations that favor a $\Xi\,\frac{1}{2}^+$~classification for the $\Xi(1690)$, e.g. the QCD sum rules of Ref.~\cite{Lee:2002jb} or the quark model calculations of Refs.~\cite{Melde:2008yr}.

If a superposition of more than one state, the lowest-lying $\Xi\,\frac{1}{2}^-$~wave containing an excited $|ss\rangle$~diquark could also be part of the $\Xi(1620)$~structure. However, the $Y\overline{K}$~decay mode will be challenging to study owing to the limited kinematic phase space. The $\frac{1}{2}^-$~classification is likely, though. The assignment is supported by calculations in a Skyrme model~\cite{Oh:2007cr} and the chiral unitary approach of Refs.~\cite{Ramos:2002xh,Miyahara:2016yyh,Garcia-Recio:2003ejq}. The authors of Ref.~\cite{Ramos:2002xh} argue that the addition of $\Xi(1620)\,\frac{1}{2}^-$ to $\Lambda(1670)$, $\Sigma(1620)$, and $N(1535)$ completes the octet of $J^P = \frac{1}{2}^-$~resonances dynamically generated in the chiral unitary approach through multiple scattering of meson-baryon pairs. We note that the composition of this octet of dynamically generated states is identical to the proposed octet based on the traditional quark model.
The $\Xi(1620)$ has frequently been considered an unconventional state and the doubly strange analogue to the $\Lambda(1405)$~hyperon, which is now listed in the RPP as two different states. We tentatively assign $\Xi(1620)$ and $\Xi(1820)$ to the $^2 {\bf 8}_{J^-}$~doublet. However, the $(M_{\frac{3}{2}^-} - M_{\frac{1}{2}^-})$~mass difference of about 200~MeV/$c^2$ is unusually large in comparison with the corresponding mass differences of the multiplet partners, which is $\leq 50$~MeV/$c^2$. The study of the 1600--1700~MeV/$c^2$ mass range in $K N$~interactions should be a priority in upcoming experiments.

The only additional established $\Xi$~resonance with weak evidence for its spin is the $\Xi(2030)$ that is listed with $J^P \geq \frac{5}{2}^?$~\cite{ParticleDataGroup:2022pth}. If true, the state could be the expected $\frac{5}{2}^-$~state and likely be an octet member with partners $N(1675)$, $\Lambda(1830)$, and $\Sigma(1775)$. However, the mass would be surprisingly high and moreover, according to the calculations discussed in Ref.~\cite{Chao:1980em}, the $^4\bf{8}$~state is predicted to dominantly decay into $\Xi\pi$. The experimental state is exclusively observed to decay into $Y\overline{K}$, though. According to the SU(3)~symmetry
of hadrons, predictions were made already in the 1970s that the $\Xi(2030)$ could be the partner of $N(1680)$, $\Lambda(1820)$ and $\Sigma(1915)$ with $J^P = \frac{5}{2}^+$~\cite{Samios:1974tw}. However, the more recent
strong-decay analysis of Ref.~\cite{PavonValderrama:2011gp} did not support this classification. We tentatively assign $\Xi(2030)$ to the $^4 {\bf 8}_{J^-}$~triplet with $J^P = \frac{5}{2}^-$. Table~\ref{Table:BaryonOctets} shows our suggestions for assignments of the ground state and known negative-parity light baryons to the lowest-lying quark model spin-flavour SU(6) $\otimes$ O(3) singlets and octets.

\begin{table}[t]
\begin{center}
\caption{\label{Table:BaryonOctets}Tentative assignments of the ground state and known negative-parity light baryons to the lowest-lying SU(6)\,$\otimes$\,O(3) singlets and octets. States marked with $^\dagger$ are merely educated guesses because the evidence for their existence is poor or they can be assigned to other multiplets. A hyphen indicates that the state does not exist, an empty box that it is missing.}
%\begin{indented}
%\item[] 
\begin{tabular}{l|l|l|c|l|l|l|l||c}
\br
$N$ & $(D,\,L^P_N)$ & $S$ & $J^P$ & \multicolumn{4}{c||}{Octet Members} & Singlets\\
\mr
0 & $(56,\,0_0^+)$ & $\frac{1}{2}$ & $\frac{1}{2}^+$ & $N(939)$ & $\Lambda(1116)$ & $\Sigma(1193)$ & $\Xi(1318)$ & $-$\\
\mr
1 & $(70,\,1_1^-)$ & $\frac{1}{2}$ & $\frac{1}{2}^-$ & $N(1535)$ & $\Lambda(1670)$ & $\Sigma(1620)$ & \bl{$\Xi(1620)^\dagger$} & $\Lambda(1405)$\\
   &                        &                      & $\frac{3}{2}^-$ & $N(1520)$ & $\Lambda(1690)$ & $\Sigma(1670)$ & $\Xi(1820)$ & $\Lambda(1520)$\\
   &                        & $\frac{3}{2}$ & $\frac{1}{2}^-$ & $N(1650)$ & $\Lambda(1800)$ & $\Sigma(1750)$ & \bl{$\Xi(1690)^\dagger$} & $-$\\
   &                        &                      & $\frac{3}{2}^-$ & $N(1700)$ & & & & $-$\\
   &                        &                      & $\frac{5}{2}^-$ & $N(1675)$ & $\Lambda(1830)$ & $\Sigma(1775)$ & $\Xi(2030)^\dagger$ & $-$\\
\mr
\br
\end{tabular}
\end{center}
%\end{indented}
\end{table}

Finally, the $\Xi(1950)$ is likely a superposition of more than one state. A structure in this mass region is clearly observed, but the evidence for a genuine resonance is weaker owing to the reported conflicting experimental results. Unlike the $\Xi(1690)$, $\Xi(1820)$, and $\Xi(2030)$ states, the $\Xi(1950)$ is significantly broader. We expect the two remaining $\Xi\,\frac{3}{2}^-$~states, possibly another $\Xi\,\frac{1}{2}^-$~state, and the radial excitation in this mass region. These states are considered missing. The authors of Ref.~\cite{PavonValderrama:2011gp} also supported the idea of several $\Xi(1950)$~resonances. In particular, they proposed the existence of three states: one of these states would be part of a spin-parity $\frac{1}{2}^-$~decuplet and the other two probably could belong to the $\frac{5}{2}^+$ and $\frac{5}{2}^-$~octets. Based on their theoretical arguments, they have proposed to search for the negative-parity $\frac{5}{2}$~state as a broader structure in the $\Xi\pi$~spectrum, whereas the positive-parity state would be observed as a narrow structure in the $\Lambda \overline{K}$~spectrum. The new upcoming data could put these predictions to the test. High-statistics baryon samples can be directly produced on the nucleon by the process $\overline{K} N \to \Xi^\ast K$ with large production cross sections.

For the higher-lying $\Xi$~states, the decays into $\Xi\pi\pi$ and $Y\overline{K}\pi$ become increasingly important. Since all known states have been observed as isolated peaks in various mass spectra, it is likely that higher-lying states will decay by populating intermediate excited Kaons or hyperons. The results from the BNL experiment~\cite{Jenkins:1983pm} using the multi-particle spectrometer nicely illustrate that peak hunting may be an option, see Fig.~\ref{Figure:1950} (right side). For this reason, it is likely that many $\Xi$~resonances have not been observed, yet, because high-quality data and the statistics for studies of decays into $\Xi(1530)\pi\to \Xi\pi\pi$, $Y\,\overline{K}^{\,\ast}\to Y\,\overline{K}\pi$, or $\Sigma(1385)\overline{K}$ have not been available to date. The expected high-statistics data samples from the $K_L$~facility at JLab and J-PARC will change this.

\subsubsection*{The triply strange $\Omega^\ast$~resonances} Two excited $\Omega$~states with negative parity 
are expected in the first excitation band with masses up to about 2050~GeV/$c^2$. These originate from the decuplet with spin~$\frac{1}{2}$ which, combined with one unit of orbital angular momentum, yields an excited $\frac{1}{2}^-$ and a $\frac{3}{2}^-$~state. According to the early quark model calculations of Ref.~\cite{Chao:1980em}, these two negative-parity $(\frac{1}{2}^-,\frac{3}{2}^-)$~states have masses slightly above 2000~MeV/$c^2$, are mass degenerate and decay entirely via $\Xi\,\overline{K}$ in an $(S,D)$-wave since the decay into $\Xi(1530)\,\overline{K}$ in a $(D,S)$-wave is phase-space suppressed. Only the first radial excitation is expected in the same mass region and also decays dominantly via $\Xi\,\overline{K}$~\cite{Chao:1980em}. However, the excitation energy of the radial excitation in this model is unusually low. All known Roper-like states have a mass about 400--500~MeV/$c^2$ higher than the corresponding ground state. The $\Omega$~ground state has $J=\frac{3}{2}^+$ but naively, we would still expect the $\Omega^{\,\ast}\,\frac{3}{2}^+$~radial excitation close to 2100~MeV/$c^2$. Improved calculations within a {\it relativized} approach~\cite{Capstick:1986ter} predict a mass for the radial excitation well above 2100~MeV/$c^2$. More recently, the conventional quark model calculations of Refs.~\cite{Loring:2001ky,Faustov:2015eba,Pervin:2007wa,Arifi:2022ntc} and the Skyrme model calculations of Ref.~\cite{Oh:2007cr} have also predicted that the first radial excitation has a mass greater than 2100~MeV/$c^2$. All these models expect masses for the $J^P = (\frac{1}{2}^-,\frac{3}{2}^-)$~pair of orbital excitations in the 2000~MeV/$c^2$ region with a lower mass for the $\Omega\,\frac{1}{2}^-$ below 2000~MeV/$c^2$. 

\begin{table}[t]
\begin{center}
\caption{\label{Table:BaryonDecuplets}(Colour online) Tentative assignments of the ground state and known negative-parity light baryons to the lowest-lying SU(6)\,$\otimes$\,O(3) decuplets. States marked with $^\dagger$ are merely educated guesses because the evidence for their existence is poor or they can be assigned to other multiplets. A hyphen indicates that the state does not exist, an empty box that it is missing.}
%\begin{indented}
%\item[] 
\begin{tabular}{l|l|c|c|l|l|l|l}
\br
$N$ & $(D,\,L^P_N)$ & Spin, $S$ & $J^P$ & \multicolumn{4}{c}{Decuplet Members}\\
\mr
0 & $(56,\,0_0^+)$ & $\frac{3}{2}$ & $\frac{3}{2}^+$ & $\Delta(1232)$ & $\Sigma(1385)$ & $\Xi(1530)$ & $\Omega(1672)$\\
\mr
1 & $(70,\,1_1^-)$ & $\frac{1}{2}$ & $\frac{1}{2}^-$ & $\Delta(1620)$ & & & \\
   &                        &                      & $\frac{3}{2}^-$ & $\Delta(1700)$ & & & $\Omega(2012)^\dagger$\\
\mr
\br
\end{tabular}
\end{center}
%\end{indented}
\end{table}

Lattice-QCD results of the Hadron Spectrum Collaboration~\cite{Edwards:2012fx} and of the BGR [Bern-Graz-Regensburg] Collaboration~\cite{Engel:2013ig} predict masses for the positive-parity states that are all greater than the masses of the two negative-parity states, consistent with the model calculations of Refs.~\cite{Loring:2001ky,Faustov:2015eba,Pervin:2007wa,Oh:2007cr,Arifi:2022ntc}.

Table~\ref{Table:BaryonDecuplets} shows our suggestions for assignments of the ground state and known negative-parity light baryons to the lowest-lying quark model spin-flavour SU(6)~$\otimes$~O(3)~decuplets. The $\Omega(2012)$ likely has spin-parity $J^P = \frac{3}{2}^-$ due to its fairly low mass and rather narrow width which was initially observed in the decay into $\Xi K$~\cite{Belle:2018mqs}. If phase-space allowed, the preferred decay of an excited $\Omega^\ast$~state with $J^P = \frac{3}{2}^-$ is expected to proceed via S-wave to the $\Xi(1530)\,\overline{K}$~final state, which has a minimum mass of about 2025~MeV/$c^2$. However, the lowest-lying excited state, $\Omega(2012)$, has a mass slightly below that of the $\Xi(1530)\,\overline{K}$~system. Therefore, the decay to $\Xi(1530)\,\overline{K}$ corresponds to a decay from the high-mass end of the mother's natural width to the low-mass end of the daughter's natural width, distorting both mass spectra. 
Belle looked for such a signal~\cite{Belle:2019zco} and reported no excess. However, on further analysis using a more sophisticated event selection and fitting function, results have been presented in an unpublished preprint which imply a branching fraction into $\Xi(1520)\,\overline{K}$ comparable to that into $\Xi\,\overline{K}$~\cite{Belle:2022mrg}. We assign the $\Omega(2012)$ to the $^2{\bf 10}_{J^-}$~doublet with $J=\frac{3}{2}$.
Even if the state is identified as having spin-parity of $J^P=\frac{3}{2}^-$, there will still be differing theoretical views of its internal structure. Ikeno {\it et al.}~\cite{Ikeno:2023wyh} believe the latest results for the $\Xi(1520)\,\overline{K}$~decay are consistent with a molecular state rather than a conventional orbital excitation and it is true that the most recent Belle signal appears to imply a larger branching fraction than the one predicted by the standard quark model of Ref.~\cite{Arifi:2022ntc}.
However, it should be noted that just as the analysis of the data depends critically on the lineshape of the $\Omega(2012)$, so does any theoretical calculation, and it is not clear if the calculations take into account the relatively large uncertainty on the $\Omega(2012)$ intrinsic width. 
We also note again that most theory approaches (quark models~\cite{Faustov:2015eba,Pervin:2007wa,Arifi:2022ntc}, Skyrme model~\cite{Oh:2007cr}, lattice studies~\cite{Edwards:2012fx,Engel:2013ig}) predict a $J^P=\frac{1}{2}^-$ partner at a lower mass, and this provides some guidance for experimental searches.

\subsection{Current and planned experiments on $\Xi^\ast$ and $\Omega^\ast$ spectroscopy}
\label{Subsubsection:PlannedExperiments}
The ongoing and planned near- to mid-term future experiments at Belle@KEKB, Jefferson Lab (GlueX, CLAS12), J-PARC, and PANDA@FAIR will be able (and continue) to provide a plethora of new information on multistrange hyperons. Although photoproduction is not ideal to search for excited states due to the lack of strangeness in the initial state, the accumulated data samples at JLab have already surpassed most of the older experiments that were performed until the 1990s in terms of statistics for the lower-lying $\Xi^\ast$~resonances. The upcoming experiments at JLab using the $K_L$~facility in conjunction with the GlueX exprimental setup and at J-PARC using the $K^-$~beam at the extended hadron facility will undoubtedly produce data samples of unprecedented quality. Some aspects of a spectroscopy program in $K_L\,p$~interactions using the beam of neutral Kaons at JLab are discussed in Ref.~\cite{Thiel:2020eqn}, for instance.

\subsubsection{Open questions in the spectroscopy of very strange hyperons}

\begin{figure}[t]
\caption{\label{Figure:RoperStates} (Colour online) Radial excitations (Roper-like states) for the octet members with $J^P = \frac{1}{2}^+$. Reprinted figure with permission from~\cite{Arifi:2022ntc}, Copyright (2022) by the American Physical Society.}
\vspace{2mm}
\begin{minipage}{1.0\textwidth}
\begin{center}
\begin{tabular}{c}
\includegraphics[width=0.8\textwidth]{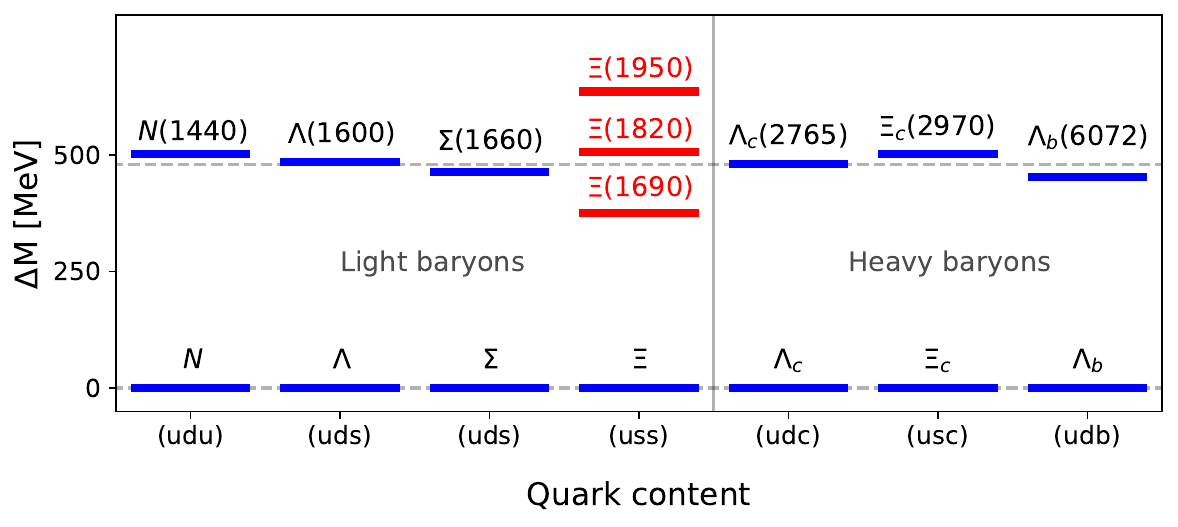}
\end{tabular}
\end{center}
\end{minipage}
\end{figure}

Experimentally, a simple phenomenological observation is that the masses of baryons increase when a light $u$ or $d$ quark is replaced with a strange quark. This effect can be nicely seen for the ground-state baryons in Tables~\ref{Table:BaryonOctets} and~\ref{Table:BaryonDecuplets}. Even for charmed and bottom baryons, the mass increases by about 120~MeV/$c^2$ with each additional strange quark, e.g. in the transitions
\begin{eqnarray}
    \Sigma_c^0(2520)\,3/2^+ \quad\to\quad \Xi_c^0(2645)\,3/2^+ \quad\to\quad \Omega_c^0(2770)\,3/2^+\nonumber\\[1ex]
    \Sigma_c^0(2455)\,1/2^+ \quad\to\quad \Xi_c^{\,\prime\, 0}\,1/2^+\quad\to\quad\Omega_c^0\,1/2^+\,.
\end{eqnarray}
The corresponding widths of these states is also observed to become smaller with an increasing number of $s$~quarks. At the same time, both $\Sigma$ and $\Xi$~baryons are octet and decuplet members, so the total number of states increases. However, $\Omega$~baryons are expected to have very narrow widths and the fewer states only appear in the decuplets. For this reason, $\Omega$~spectroscopy should provide a particularly clear picture of the systematics of the excitation spectrum.

The systematic search for excited~$\Xi$ and $\Omega$~resonances in various decay modes will be possible at the upcoming $K$-beam facilities. To understand the systematics of the baryon spectrum, the study of its full flavour structure is the next step in baryon spectroscopy. And very little is known about the strange partners of the nucleon and $\Delta$~states. For example, what is the flavour symmetry of the radial excitations, the Roper-like states, that have the same $J^P$~quantum numbers as their corresponding ground states? Figure~\ref{Figure:RoperStates} nicely illustrates the current situation for the baryon octet with $J^P = \frac{1}{2}^+$ (courtesy of the authors of Ref.~\cite{Arifi:2022ntc}). 
All members of the multiplet have a radial excitation energy of about 500~MeV. This also includes some of the heavy flavour states with the same quantum numbers that can be considered analogue states of the Roper resonance, i.e. the $\Lambda_c(2765)$ as well as the recently discovered resonances $\Xi_c(2970)$~\cite{Belle:2020tom} and $\Lambda_b(6072)$~\cite{LHCb:2020lzx}. Based on the pattern shown in Fig.~\ref{Figure:RoperStates}, we would expect the radial excitation of the $\Xi$~resonance around 1800~MeV/$c^2$, but no candidate is currently known. For this reason, finding the partner of the Roper resonance in the multi-strange sector and observing the similarities and differences will shed light on our understanding of the structure of excited baryon resonances.

In the search for excited states,
the investigation of the 1600--1700~MeV/$c^2$ mass region is of particular interest to address the $\Xi(1620)/\,\Xi(1690)$~situation and should have a priority in the upcoming experiments. Is the $\Xi(1620)$ just a single resonance? Its larger width suggests that the observed signal is a superposition of more than one resonance or that the state is exotic in nature. Where is the partner of the $\Omega(2012)^-$~resonance? If this state is the $J^P = \frac{3}{2}^-$~orbital excitation, the $J^P = \frac{1}{2}^-$~partner is expected at a lower mass.
Do we observe additional resonant structures based on meson-baryon hadronic degrees of freedom, similar to what has been proposed for other light-flavour baryons?

In addition to understanding the baryon mass spectrum, the unknown properties of established resonances and searching for hitherto unknown excited $\Xi$ and $\Omega$~states, studying these ground-state resonances in elastic scattering reactions from the nucleon is of fundamental importance. While nucleon-nucleon (NN) scattering is perhaps the most well studied of all nuclear reactions, less is known about the scattering of hyperons ($\Lambda$, $\Sigma$, $\Xi$, $\Omega$) from the proton. Such measurements will be interesting for astrophysical applications, for instance, where it is known that the presence of strange matter in the core of a neutron star can have a significant impact on its equation of state~\cite{Lonardoni:2014bwa}. Furthermore, low-energy $\Omega N$~scattering will be studied at J-PARC to search for possible $\Omega N$~bound states~\cite{J-PARC-P85}. In the following, we briefly highlight some more specific aspects of a spectroscopy program on multistrange hyperons. 

\subsubsection{Spin and parity}
The measurement of some spins and a parity measurement of the ground-state~$\Xi$ remain a strong possibility even in photoproduction. In the past, spin-parity quantum numbers of $\Xi$~resonances were studied by analysing the moments of their decay products, see for example Refs.~\cite{Biagi:1986vs,Teodoro:1978bu,Minnaert:1977td} for more details. However, this approach is limited to resonances above threshold with odd relative orbital angular momentum between the decay products~\cite{Nakayama:2012zp}.   
An alternative and model-independent determination of the parity of $\Xi$~hyperons was suggested in Ref.~\cite{Nakayama:2012zp} for the reactions $\overline{K}\,N\to K\,\Xi$ and $\gamma N\to K K \,\Xi$ based on reflection symmetry in the reaction plane. In $\overline{K}$\,-\,induced reactions, the transverse spin-transfer coefficient $K_{yy}$ directly determines the parity of the produced $\Xi$~hyperon as $\pi_\Xi = K_{yy}$, where $\pi_\Xi = \pm\,1$ is the parity. Experimentally in terms of the measured cross sections, the spin-transfer coefficient $K_{ii}$ is given by~\cite{Nakayama:2012zp}
\begin{eqnarray}
    K_{ii} = \frac{[d\sigma_i(++)+d\sigma_i(--)]-[d\sigma_i(+-)+d\sigma_i(-+)]}{[d\sigma_i(++)+d\sigma_i(--)]+[d\sigma_i(+-)+d\sigma_i(-+)]}\,,
\end{eqnarray}
where $d\sigma_i$ stands for the differential cross section with the polarisation of the target nucleon and of the produced~$\Xi$ along the $i$-direction. The first and second~$\pm$ argument of $d\sigma_i$ denotes the parallel~$(+)$ or anti-parallel~$(-)$ spin alignment along the $i$-direction of the target nucleon and the produced $\Xi$, respectively.
In photo-induced reactions, a similar approach is possible by measuring the transverse spin-transfer coefficient $K_{yy}$ using an unpolarised photon beam (double-polarisation observable) and the beam asymmetry $\Sigma$. The parity of the $\Xi$ is then given by 
\begin{eqnarray}
    \pi_\Xi = \frac{K_{yy}}{\Sigma}\,.
\end{eqnarray}
However, the measurement of spin observables in photoproduction, particularly of double-polarisation observables involving polarised targets, is statistically challenging owing to the much smaller production cross sections compared to hadronic reactions. The recipe outlined in Ref.~\cite{Nakayama:2012zp} can also be used for $\Omega$~production in reactions such as $\overline{K}\,N\to KK\,\Omega$ and $\gamma\,N\to KKK\,\Omega$, where $\pi_\Xi$ needs to be replaced with $-\pi_\Omega$ due to the presence of an additional Kaon. 

\subsubsection{Cross sections}
The measurement of cross sections and understanding the corresponding production mechanisms, even of the ground-state resonances, are interesting in their own right. How multistrange particles are produced in photo-induced reactions is only poorly understood. The production mechanism for $\Xi$~photoproduction in the reaction $\gamma p\to K^+ K^+\,\Xi^-$ was recently studied in Refs.~\cite{Nakayama:2006ty,Man:2011np} and studies for the reaction $\overline{K}\,N\to K\,\Xi$ have been discussed in Refs.~\cite{Sharov:2011xq,Shyam:2011ys}. The extraction of cross sections for the reaction $\gamma p\to K^+K^+K^0\,\Omega^-$ will be a first measurement. Current estimates for the photoproduction cross section come from vector-meson dominance and effective Lagrangian models. And overall, the estimates are consistent in that the cross section is small near threshold and then quickly rises to a few nanobarns. More information on these cross section estimates is available in Ref.~\cite{Afanasev:2012fh}. 

\subsubsection{Isospin splittings}
Measurements of the isospin-symmetry violating mass splittings $(M_{\Xi^{\ast -}} - M_{\Xi^{\ast 0}})$ in spatially excited $\Xi$~states are also very interesting and possible, for the first time in a spatially-excited hadron. Currently, mass splittings like $M_n - M_p$ or $M_{\Delta^0} - M_{\Delta^{++}}$ are only available for the octet and decuplet ground states, but these are hard to measure in excited $N,~\Delta$ and $\Sigma,~\Sigma^\ast$~states, which are broad. However, the lightest $\Xi$~baryons are expected to be narrower, and measuring the $M_{\Xi^-} - M_{\Xi^0}$~splitting of spatially-excited $\Xi$~states remains a strong possibility. Such measurements would allow an interesting probe of excited-hadron structure, and would provide important input for models which explain the isospin-symmetry violating mass splittings by the effects of the difference of the $u$- and $d$-quark masses and of the electromagnetic interactions between the quarks.

The ground-state octet $\Xi$~resonances are well established but it is interesting to note that just a few measurements of the $\Xi^0$~mass are listed in the RPP with only one measurement, reported by the NA48~Collaboration in 2000, based on more than 50~events~\cite{NA48:1999dxg}. Therefore, even the measurement of the $\Xi$~ground-state doublet mass splitting remains intriguing. The value of $(6.85\pm 0.7)$~MeV/$c^2$ given in the RPP (PDG~average)~\cite{ParticleDataGroup:2022pth} is higher than that of other baryon ground states. Recent lattice-QCD calculations give a value of $(5.68\pm 0.24)$~MeV/$c^2$~\cite{Duncan:1996xd}, whereas a higher value of about $6.10$~MeV/$c^2$~\cite{Delbourgo:1998qe} comes from a quark model calculation applying radiative corrections. A recent measurement of the $(\Xi^-,\Xi^0)$ mass splitting of $(5.4\pm 1.8)$~MeV/$c^2$ by the CLAS Collaboration has indicated a value that is lower than the PDG average but the uncertainty is large~\cite{Guo:2007dw}. The sole high-statistics measurement reported by the NA48~Collaboration needs to be urgently repeated in future experiments in order to access some of the fundamental parameters of QCD such as quark masses.            

\subsubsection{Hyperon transition form factors}
One of the fundamental goals of nuclear physics is to understand the structure and behavior of strongly interacting matter in terms of its basic constituents, quarks and gluons. Complementary to the nucleon spectrum, electromagnetic form factors are among the most basic quantities containing information about
the internal structure of the nucleon. These form factors describe the spatial distributions of electric charge and current inside the nucleon and thus are intimately related to its internal structure. For a very low momentum transfer between the scattered projectile and the target nucleon, $q^2$, the electric and magnetic form factors, $G_E$ and $G_M$, may be thought of as Fourier transforms of the charge and magnetization current densities inside the nucleon. The form factors for the proton can be accessed in the space-like region, $q^2 = \omega^2 - \vec{q}^{\,\,2} < 0$, where $\omega$ is the energy transfer and $\vec{q}$ is the three-momentum transfer, in elastic lepton-nucleon scattering and a huge experimental database has been accumulated over recent decades.

Unfortunately, hyperons are unstable and cannot be used as a target nucleon. For this reason, hyperons are very challenging to study in elastic scattering. Consequently, many low-energy quantities, such as magnetic and charge radii, are unknown for most ground-state hyperons. The corresponding time-like form factors for $q^2 > 0$ can be extracted in annihilation reactions such as $e^+ e^-\to Y_1 \,\overline{Y}_2$ for $q^2 > (m_{Y_1} + m_{Y_2})$ in a {\it hard} process for high-$q^2$ values. Complementary information for the low-$q^2$ region can be accessed in baryon Dalitz decays, $Y_1\to Y_2\,e^+ e^-$ for $q^2 < (m_{Y_1} - m_{Y_2})$. In such a decay, the form factors are either {\it direct} if $Y_1 = Y_2$, or if $Y_1 \neq Y_2$, they are {\it transition} form factors. The transition form factors are functions of the invariant mass of the dilepton, i.e. of the $e^+ e^-$ system. Resolving the shape of a form factor requires some range of invariant masses. For the Dalitz decay, $\Sigma^0 \to \Lambda\, e^+ e^-$, the upper limit of available invariant masses is just about $m_{\Sigma^0} - m_{\Lambda} \approx 77$~MeV/$c^2$ and thus, not very large as compared to typical hadronic scales. To extract even the electric or magnetic transition radius requires high experimental precision. Decuplet-octet transitions, e.g. from the decuplet~$\Sigma$ with $J^P = \frac{3}{2}^+$ to the $\Lambda$~hyperon, would be the next step. And also transitions involving doubly strange hyperons in decays such as $\Xi^0 \to \Lambda\,e^+ e^-$. The latter weak branching fraction is the only existing measurement in the strange sector and was reported by the NA48~Collaboration~\cite{NA48:2007smd}. Hyperons will be copiously produced at the upcoming $K_L$-, $K^-$\,-\,beam facilities and high-quality data on transition form factors will be available for the first time.

\subsection{Open questions in the spectroscopy of charmed and bottom baryons}
\label{Subsection:OpenQuestionsCharmedBottomBaryons}
Singly heavy baryons can be classified in terms of SU(4) flavour multiplets, but the symmetry is badly broken due to the significantly larger mass of the heavy charm or bottom quark. As discussed previously, this classification of states serves primarily for enumerating the possible states. In the singly heavy baryon sector, it is more useful to describe the three-quark system in terms of a quark-diquark picture. The spectrum can then be easily classified based on the flavour structure of the light diquark.

The multiplet structure of the diquark was already introduced in Section~\ref{Subsection:HeavyBaryonSpectroscopy}. The three-body system can be described in terms of the Jacobi coordinates $\vec{\rho}$ and $\vec{\lambda}$, where $l_\rho$ and $l_\lambda$ denote the orbital angular momentum between the two light quarks, and between the heavy quark and the light diquark, respectively. The total orbital angular momentum of the singly heavy baryon is then $L = l_\rho \otimes l_\lambda$. In the usual convention, $\vec{p}_\rho$ and $\vec{p}_\lambda$ are canonical conjugate variables of $\vec{\rho}$ and $\vec{\lambda}$, and the reduced masses of the $\rho$ and $\lambda$~oscillators are defined in terms of the kinetic energy of the three-particle system:
\begin{equation}
    T\,\propto\,\frac{\vec{p}^{\,\,\,2}_\rho}{2\mu_\rho}\,+\,\frac{\vec{p}^{\,\,\,2}_\lambda}{2\mu_\lambda}\,+\,\frac{\vec{P}^{\,\,2}}{2M}~=~\frac{\vec{p}^{\,\,\,2}_{\,1}}{2m_q}\,+\,\frac{\vec{p}^{\,\,\,2}_{\,2}}{2m_q}\,+\,\frac{\vec{p}^{\,\,\,2}_{\,3}}{2m_Q}~,
\end{equation}
where $M = 2m_q + m_Q$ and $\vec{P} = \vec{p}_q + \vec{p}_q + \vec{p}_Q$ are the total mass and momentum, respectively.
Similar to the discussion in Section~\ref{Subsection:LightBaryonSpectroscopy} for light-quark baryons, the reduced masses of the two oscillators are then given as:%\\[-1.8ex]
\begin{equation}
  \mu_\rho \,=\, m_q\quad{\rm and}\quad\mu_\lambda \,=\, \frac{3m_q\,m_Q}{2m_q + m_Q}~,
\end{equation}
where $q = s$ and $Q = u,d$ for the doubly strange $\Xi$~system, and $q = u,d$ and $Q = c,b$ for the singly heavy charmed or bottom baryons. 
The ratio of the harmonic oscillator frequencies is given by~\cite{Richard:2012xw}%\\[-1.6ex]
\begin{equation}
    \frac{\omega_\lambda}{\omega_\rho} \,=\, \sqrt{\frac{1}{3}\,(1+2m_q/m_Q})~\leq~1\,. 
\end{equation}
In the limit of $m_q\approx m_Q$, e.g. for $N^\ast$ and $\Delta^\ast$~resonances, the excitation energies in the $\rho$ and $\lambda$~oscillators are about the same, whereas the excitation energies in the $\lambda$~oscillator are reduced by a factor of $\sqrt{3}$ in the heavy-quark limit, $m_Q\to\infty$.

The light diquark obeys the Pauli Principle and the structure of this system is straightforward. The diquark has an anti-symmetric ${\bf{\bar{3}}}_C$ colour structure, and either a symmetric ${\bf{6}}_F$ or an anti-symmetric ${\bf{\bar{3}}}_F$ flavour structure, see Eq.~(\ref{Equation:diquarkStructure}). Moreover, the spin component of the diquark has either a symmetric $(s_{qq} = 1)$ or an anti-symmetric $(s_{qq} = 0)$ spin angular momentum structure. The orbital angular momentum $l_\rho$ of the diquark can then be combined with the spin angular momentum to yield either a scalar or an axial-vector diquark. These diquarks are often considered in the literature a {\it good} or a {\it bad} diquark~\cite{Gross:2022hyw,Chen:2016spr,Jaffe:2004ph}, respectively, since a scalar diquark has an attractive interaction making the system more tightly bound, whereas an axial-vector diquark has a repulsive interaction. Therefore, the $S$-, $P$-, and $D$-wave diquark  structure is:\\[-1.4ex]
\begin{eqnarray}
  l_\rho = 0~(S)=\left\{
    \begin{array}{@{} l c @{}}
      ~s_{qq} = 0~(A),~~{\bf{\bar{3}}}_F~(A) & j_{qq} = 0\,,\\[1ex]
      ~s_{qq} = 1~(S),~~{\bf{6}}_F~(S) & j_{qq} = 1\,,
    \end{array}\right.\\[1.5ex]
  l_\rho = 1~(A)=\left\{
    \begin{array}{@{} l c @{}}
      ~s_{qq} = 0~(A),~~{\bf{6}}_F~(S) & j_{qq} = 1\,,\\[1ex]
      ~s_{qq} = 1~(S),~~{\bf{\bar{3}}}_F~(A) & j_{qq} = 0/1/2\,,
    \end{array}\right.\\[1.5ex]
  l_\rho = 2~(S)=\left\{
    \begin{array}{@{} l c @{}}
      ~s_{qq} = 0~(A),~~{\bf{\bar{3}}}_F~(A) & j_{qq} = 2\,,\\[1ex]
      ~s_{qq} = 1~(S),~~{\bf{6}}_F~(S) & j_{qq} = 1/2/3\,,
    \end{array}\right.\\[-0.4ex] \nonumber
    %l_\rho = 0~(S) \bigl\left\{ ~s_{qq} = 0~(A),~~{\bf{\bar{3}}}_F~(A) & j_{qq} = 0\,,\\  
    %                             ~s_{qq} = 1~(S),~~{\bf{6}}_F~(S) & j_{qq} = 1\,, \\[1ex]
    %l_\rho = 1~(A) %~\begin{cases} ~s_{qq} = 0~(A),~~{\bf{6}}_F~(S) & j_{qq} = 1\,,\\  
                   %              ~s_{qq} = 1~(S),~~{\bf{\bar{3}}}_F~(A) & j_{qq} = 0/1/2\,, \end{cases}\\[1ex]
    %l_\rho = 2~(S) %~\begin{cases} ~s_{qq} = 0~(A),~~{\bf{\bar{3}}}_F~(A) & j_{qq} = 2\,,\\  
                   %              ~s_{qq} = 1~(S),~~{\bf{6}}_F~(S) & j_{qq} = 1/2/3\,, \end{cases}
\end{eqnarray}
where $j_{qq} = s_{qq} \otimes l_\rho$ denotes the total angular momentum of the diquark. Finally, the total angular momentum $J$ of the singly heavy baryon is then 
\begin{equation}
    J = s_Q \otimes (j_{qq} \otimes l_\lambda)\,.
\end{equation}

\begin{table}[t]
\begin{center}
\caption{\label{Table:CharmedBaryonClassification} Our $\lambda$ and $\rho$ classifications of $S$-wave (ground-state) charmed baryons and of the $P$- as well as $D$-wave excitations of singly charmed baryons, where $l_\rho,l_\lambda$ denote the orbital angular momentum of the two oscillators, $s_{qq}$ and $j_{qq}$ denote the spin and the total angular momentum of the diquark, respectively, and $L$ and $J^P$ are the total orbital angular momentum and the total spin of the baryon. All observed baryons~\cite{ParticleDataGroup:2022pth} are well established as defined by the PDG unless marked otherwise. The $J^P$~quantum numbers marked with $(?)^\dagger$ or \b{blue}-highlighted states are not listed in the RPP~\cite{ParticleDataGroup:2022pth}.}\footnotesize
\vspace{1mm}
\begin{tabular}{@{}c|ccccccc||c}
\br
 & $L$ & $l_\rho$ & $l_\lambda$ & $s_{qq}$ & $j_{qq}$ & $J^P$ & $(nL)$ & Observed Candidates\\
\br
 & 0 & 0 & 0 & 0 & 0 & $\frac{1}{2}^+$ & $(1S)$ & $\Lambda_c\,\frac{1}{2}^+$, $(\,\Xi_c^+,\Xi_c^0\,)\,\frac{1}{2}^+$\\[3ex]
 & 0 & 0 & 0 & 0 & 0 & $\frac{1}{2}^+$ & $(2S)$ & $\ast\,\Lambda_c(2765)\,\frac{1}{2}^+$\,(?)$^\dagger$, $\Xi_c(2970)\,\frac{1}{2}^+$\\[3ex]
 & 1 & 0 & 1 & 0 & 1 & $\frac{1}{2}^-$, $\frac{3}{2}^-$ & $(1P)$ & $\Lambda_c(2595)\,\frac{1}{2}^-$, $\Lambda_c(2625)\,\frac{3}{2}^-$\\[1ex]
 & & & & & & & & $\Xi_c(2790)\,\frac{1}{2}^-$, $\Xi_c(2815)\,\frac{3}{2}^-$\\[1ex]
 \rb{$\bf{\bar{3}}_F$} & 1 & 1 & 0 & 1 & 0 & $\frac{1}{2}^-$ & $(1P)$ &\\[1ex]
 $\Lambda_c^{(\ast)}$ & 1 & 1 & 0 & 1 & 1 & $\frac{1}{2}^-$, $\frac{3}{2}^-$ & $(1P)$ &\\[1ex]
 $\Xi_c^{(\ast)}$ & 1 & 1 & 0 & 1 & 2 & $\frac{3}{2}^-$, $\frac{5}{2}^-$ & $(1P)$ & \rbb{\hspace{0cm}$\biggr\}~~~\bigl(\,\Lambda_c(2940)\,\frac{3}{2}^-\,\bigr)^?$~~"possibly a $2P$ state"}\\[3ex]
 & 2 & 0 & 2 & 0 & 2 & $\frac{3}{2}^+$, $\frac{5}{2}^+$ & $(1D)$ & $\Lambda_c(2860)\,\frac{3}{2}^+$, $\Lambda_c(2880)\,\frac{5}{2}^+$\\[1ex]
 & & & & & & & & $\Xi_c(3055)\,\frac{3}{2}^+\,(?)^\dagger$, $\Xi_c(3080)\,\frac{5}{2}^+\,(?)^\dagger$\\[1ex]
 & 2 & 2 & 0 & 0 & 2 & $\frac{3}{2}^+$, $\frac{5}{2}^+$ & $(1D)$ & \\[0.5ex]
\br
 & 0 & 0 & 0 & 1 & 1 & $\frac{1}{2}^+$, $\frac{3}{2}^+$ & $(1S)$ & $\Sigma_c(2455)\,\frac{1}{2}^+$,~~$\Sigma_c(2520)\,\frac{3}{2}^+$\\[1ex]
 & & & & & & & & $(\,\Xi_c^{\,\prime\,+},\Xi_c^{\,\prime\,0}\,)\,\frac{1}{2}^+$,~~$\Xi_c(2645)\,\frac{3}{2}^+$\\[1ex]
 & & & & & & & & $\Omega_c^0\,\frac{1}{2}^+$, $\Omega_c(2770)^0\,\frac{3}{2}^+$\\[2ex]
 \rb{$\bf{6}_F$} & 1 & 0 & 1 & 1 & 0 & $\frac{1}{2}^-$ & $(1P)$ & \\[1ex]
 $\Sigma_c^{(\ast)}$ & 1 & 0 & 1 & 1 & 1 & $\frac{1}{2}^-$, $\frac{3}{2}^-$ & $(1P)$ & ~\multirow{3}{*}{
 $\left(
    \begin{array}{@{} c @{}}
      ~~\Omega_c(3000,~3050)^-,~~\Xi_c(\bl{2881},~\ast\ast~2923)^- ~~\\[1ex]
      ~~\Omega_c(3065,~3090)^-,~~\Xi_c(\bl{2938},~\bl{2965})^- ~~\\[1ex]
      \Omega_c(3120)^-
    \end{array}\right)^?$
 }\\[1ex]
 $\Xi_c^{\,\prime (\ast)}$ & 1 & 0 & 1 & 1 & 2 & $\frac{3}{2}^-$, $\frac{5}{2}^-$ & $(1P)$ & \\[1ex]
 $\Omega_c^{(\ast)}$ & 1 & 1 & 0 & 0 & 1 & $\frac{1}{2}^-$, $\frac{3}{2}^-$ & $(1P)$ & \\[2ex]
 & 2 & 0 & 2 & 1 & 1 & $\frac{1}{2}^+$, $\frac{3}{2}^+$ & $(1D)$ & \\[1ex]
 & 2 & 0 & 2 & 1 & 2 & $\frac{3}{2}^+$, $\frac{5}{2}^+$ & $(1D)$ & \\[1ex]
 & 2 & 0 & 2 & 1 & 3 & $\frac{5}{2}^+$, $\frac{7}{2}^+$ & $(1D)$ & \rb{~~~~States not assigned to any multiplet}\\[1ex]
 & 2 & 2 & 0 & 1 & 1 & $\frac{1}{2}^+$, $\frac{3}{2}^+$ & $(1D)$ & \rb{$\Sigma_c(2800)_{~l_\lambda = 1~{\rm "likely"}}$}\\[1ex]
 & 2 & 2 & 0 & 1 & 2 & $\frac{3}{2}^+$, $\frac{5}{2}^+$ & $(1D)$ & \rb{$\ast\,\Xi_c(3123)$}\\[1ex]
 & 2 & 2 & 0 & 1 & 3 & $\frac{5}{2}^+$, $\frac{7}{2}^+$ & $(1D)$ & \\[3ex]
 & 2 & 1 & 1 & & & & $(1D)$ & $l_\lambda \otimes l_\rho = 2$, $l_\lambda = l_\rho = 1$ omitted\\
\br
\end{tabular}
\end{center}
\end{table}

\begin{table}[t]
\begin{center}
\caption{\label{Table:BottomBaryonClassification} Our $\lambda$ and $\rho$ classifications of $S$-wave (ground-state) bottom baryons and of the $P$-, $D$-wave excitations of singly bottom baryons, where $l_\rho,l_\lambda$ denote the orbital angular momentum of the two oscillators, $s_{qq}$ and $j_{qq}$ denote the spin and the total angular momentum of the diquark, respectively, and $L$ and $J^P$ are the total orbital angular momentum and the total spin of the baryon. All observed states~\cite{ParticleDataGroup:2022pth} are well established as defined by the PDG; $J^P$'s marked with $(?)^\dagger$ are not listed in the RPP.}\footnotesize
\vspace{1mm}
\begin{tabular}{@{}c|ccccccc||c}
\br
 & $L$ & $l_\rho$ & $l_\lambda$ & $s_{qq}$ & $j_{qq}$ & $J^P$ & $(nL)$ & Observed Candidates\\
\br
 & 0 & 0 & 0 & 0 & 0 & $\frac{1}{2}^+$ & $(1S)$ & $\Lambda_b\,\frac{1}{2}^+$, $(\,\Xi_b^-,~\Xi_b^0\,)\,\frac{1}{2}^+$\\[3ex]
 & 0 & 0 & 0 & 0 & 0 & $\frac{1}{2}^+$ & $(2S)$ & $\Lambda_b(6070)\,\frac{1}{2}^+$\\[1ex]
 & & & & & & & & $\bigl(\,\Xi_b(6227)^-$,~~$\Xi_b(6227)^0\,\bigr)^?$\\[3ex]
 & 1 & 0 & 1 & 0 & 1 & $\frac{1}{2}^-$, $\frac{3}{2}^-$ & $(1P)$ & $\Lambda_b(5912)\,\frac{1}{2}^-$,~~$\Lambda_b(5920)\,\frac{3}{2}^-$\\[1ex]
 & & & & & & & & $\Xi_b(6100)^-\,\frac{3}{2}^-$\\[1ex]
 \rb{$\bf{\bar{3}}_F$} & 1 & 1 & 0 & 1 & 0 & $\frac{1}{2}^-$ & $(1P)$ & \\[1ex]
 $\Lambda_b^{(\ast)}$ & 1 & 1 & 0 & 1 & 1 & $\frac{1}{2}^-$, $\frac{3}{2}^-$ & $(1P)$ & \\[1ex]
 $\Xi_b^{(\ast)}$ & 1 & 1 & 0 & 1 & 2 & $\frac{3}{2}^-$, $\frac{5}{2}^-$ & $(1P)$ & \\[3ex]
 & 2 & 0 & 2 & 0 & 2 & $\frac{3}{2}^+$, $\frac{5}{2}^+$ & $(1D)$ & $\Lambda_b(6146)\,\frac{3}{2}^+$,~~$\Lambda_b(6152)\,\frac{5}{2}^+$\\[1ex]
 & & & & & & & & $\bigl(\,\Xi_b(6327)^0\,\frac{3}{2}^+\,(?)^\dagger$,~~$\Xi_b(6333)^0\,\frac{5}{2}^+\,(?)^\dagger\,\bigr)^?$\\[1ex]
 & 2 & 2 & 0 & 0 & 2 & $\frac{3}{2}^+$, $\frac{5}{2}^+$ & $(1D)$ & \\[0.5ex]
\br
 & 0 & 0 & 0 & 1 & 1 & $\frac{1}{2}^+$, $\frac{3}{2}^+$ & $(1S)$ & $\Sigma_b\,\frac{1}{2}^+$, $\Sigma_b^\ast\,\frac{3}{2}^+$\\[1ex]
 & & & & & & & & $\bigl(\,\Xi_b^{\,\prime}(5935)^-, \,?\,\bigr)\,\frac{1}{2}^+$,~$\bigl(\,\Xi_b(5945)^0,~\Xi_b(5955)^-\,\bigr)\,\frac{3}{2}^+$\\[1ex]
 & & & & & & & & $\Omega_b^-\,\frac{1}{2}^+$\\[2ex]
 \rb{$\bf{6}_F$} & 1 & 0 & 1 & 1 & 0 & $\frac{1}{2}^-$ & $(1P)$ & \\[1ex]
 $\Sigma_b^{(\ast)}$ & 1 & 0 & 1 & 1 & 1 & $\frac{1}{2}^-$, $\frac{3}{2}^-$ & $(1P)$ & \multirow{2}{*}{
  $\left(
     \begin{array}{@{} c @{}}
       ~~\Omega_b(6316)^-,~~\Omega_b(6330)^-~~\\[1ex]
       ~~\Omega_b(6340)^-,~~\Omega_b(6350)^-~~
    \end{array}\right)^?$
 }\\[1ex]
 $\Xi_b^{\,\prime (\ast)}$ & 1 & 0 & 1 & 1 & 2 & $\frac{3}{2}^-$, $\frac{5}{2}^-$ & $(1P)$ & \\[1ex]
 $\Omega_b^{(\ast)}$ & 1 & 1 & 0 & 0 & 1 & $\frac{1}{2}^-$, $\frac{3}{2}^-$ & $(1P)$ & \\[2ex]
 & 2 & 0 & 2 & 1 & 1 & $\frac{1}{2}^+$, $\frac{3}{2}^+$ & $(1D)$ & \\[1ex]
 & 2 & 0 & 2 & 1 & 2 & $\frac{3}{2}^+$, $\frac{5}{2}^+$ & $(1D)$ & \\[1ex]
 & 2 & 0 & 2 & 1 & 3 & $\frac{5}{2}^+$, $\frac{7}{2}^+$ & $(1D)$ & \rb{~~~~States not assigned to any multiplet}\\[1ex]
 & 2 & 2 & 0 & 1 & 1 & $\frac{1}{2}^+$, $\frac{3}{2}^+$ & $(1D)$ & \rb{$\Sigma_b(6097)^\pm$}\\[1ex]
 & 2 & 2 & 0 & 1 & 2 & $\frac{3}{2}^+$, $\frac{5}{2}^+$ & $(1D)$ & \\[1ex]
 & 2 & 2 & 0 & 1 & 3 & $\frac{5}{2}^+$, $\frac{7}{2}^+$ & $(1D)$ & \\[2ex]
 & 2 & 1 & 1 & & & & $(1D)$ & $l_\lambda \otimes l_\rho = 2$, $l_\lambda = l_\rho = 1$ omitted\\
\br
\end{tabular}
\end{center}
\end{table}

Tables~\ref{Table:CharmedBaryonClassification} and~\ref{Table:BottomBaryonClassification} show our attempt at the $\lambda$ and $\rho$ classifications of $S$-wave (ground-state) charmed baryons and of the $P$- as well as $D$-wave excitations of singly charmed and bottom baryons, respectively. All charmed and bottom baryons currently listed in the RPP~\cite{ParticleDataGroup:2022pth} are given as possible candidates. Blue-highlighted states have been observed but are not listed, yet. Moreover, we have replaced the RPP state $\ast\ast\,\Xi_c(2930)$ with $\Xi_c(2938)$ (see Section~\ref{Subsubsection:CharmedXiOmega} for more details). All the $J^P$~quantum numbers marked with $(?)^\dagger$ are conjectures and also not given in the RPP. For the sake of simplicity, we have omitted configurations with mixed excitations in $D$-wave baryons $(l_\lambda = l_\rho = 1)$ since they are less likely produced, see Section~9.2 in Ref.~\cite{Gross:2022hyw} for more details. For the ground states, we notice that there is only one flavour ${\bf{\bar{3}}}_F$ multiplet with $J^P = \frac{1}{2}^+$ due to $J = s_Q \otimes j_{qq}$ with $s_Q = \frac{1}{2},\, j_{qq} = 0$, whereas there are two ${\bf{6}}_F$~multiplets with $J^P = \frac{1}{2}^+,\,\frac{3}{2}^+$ due to $j_{qq} = 1$. With the exception of the $\Xi_b^{\,\prime \,0}\,\frac{1}{2}^+$ and $\Omega_b^\ast\,\frac{3}{2}^+$ resonances, all the $S$-wave charmed and bottom ground-state baryons have been found.

Any assignment of experimentally observed states to the multiplets of orbitally excited configurations is very suggestive and many states are still missing. In addition to their mass, the production and decay properties of these heavy baryons are important for understanding their nature and internal structure. The quartet of states, $\Lambda_c(2595)$, $\Lambda_c(2625)$, $\Xi_c(2790)$, and $\Xi_c(2815)$, provides excellent candidates for the $P$-wave charmed baryons ($l_\lambda = 1$) with $J^P = \frac{1}{2}^-$ and $J^P = \frac{3}{2}^-$. This assignment completes two ${\bf{\bar{3}}}_F$~flavour multiplets. Furthermore, the quartet of states, $\Lambda_c(2860)$, $\Lambda_c(2880)$, $\Xi_c(3055)$, and $\Xi_c(3080)$, presents good candidates for the $D$-wave charmed baryons ($l_\lambda = 2$) with $J^P = \frac{3}{2}^+$ and $J^P = \frac{5}{2}^+$, which again completes two ${\bf{\bar{3}}}_F$ flavour multiplets. However, some concerns with this assignment are discussed in the next section.
Unfortunately, all the quantum numbers of excited $\Sigma_c$ and $\Omega_c$~resonances remain experimentally unknown. In the singly bottom sector, the two pairs of $\Lambda_b$~states, $\Lambda_b(5912)$, $\Lambda_b(5920)$ and $\Lambda_b(6146)$, $\Lambda_b(6152)$, are good candidates for the $P$-wave ${\bf{\bar{3}}}_F$ flavour multiplets ($l_\lambda = 1$) with $J^P = \frac{1}{2}^-$ and $J^P = \frac{3}{2}^-$, and for the additional $D$-wave ${\bf{\bar{3}}}_F$ flavour multiplets ($l_\lambda = 2$) with $J^P = \frac{3}{2}^+$ and $J^P = \frac{5}{2}^+$. All other quantum numbers of excited bottom baryons are unknown for the most part.

The light-diquark correlations in singly heavy baryons and the role of the scalar ("good") diquark, as compared to the axial-vector ("bad") diquark, remain interesting topics that need to be addressed in future studies and experiments. Such correlations appear to be essential for the understanding of the singly heavy baryon spectrum.
Similar to the situation of the triply strange $\Omega$~baryon, it is slightly ironic that the $\Omega_c^\ast$~states turn out to be the best place to study the full family of states; ironic, because the $\Omega_c$~resonance was itself so difficult to find, and the production cross sections for excited $\Omega_c$~resonances are very small in $e^+e^-$~reactions. If they are produced however, they will be narrow and thus, easy to find because isospin violation limits the available decay chains. In contrast, excited $\Lambda_c$~resonances can decay via $\Sigma_c$ or $\Sigma_c^\ast$~states.

%\clearpage
\subsection{Concluding remarks}
Many other open questions need to be addressed in future (experimental) studies. For example, there is particular confusion in the mass range 2850--2950~MeV$/c^2$, where many 
$\Lambda_c^+$~states have been reported, some of which overlap in mass.
Assuming all the results for this mass region are basically correct, there is not a complete understanding of the nature of   
these states and there are many questions left unanswered:

\begin{itemize} 
\item 
The $\Lambda_c(2880)$ stands out well in both $pD^0$ and $\Lambda_c^+\pi^+\pi^-$ modes, 
with the latter resonating clearly through $\Sigma_c(2455)\pi$ but with little intensity
resonating through $\Sigma_c(2520)\pi$. There are consistent results for its mass and width. 
Its spin is measured, but the conventional assignment of a $J^P\ =\ \frac{5}{2}^+$ state with $l_\lambda = 2$ is problematic
because of the low level of the $\Sigma_c(2520)\pi$~decay mode.  
\item
The $\Lambda_c(2860)^+$ found by LHCb would conveniently fit into a 
heavy-quark, light-diquark picture as a $J^P=\frac{3}{2}^+$ state being $\approx 25$~MeV/$c^2$ below its spin-$\frac{5}{2}$ partner, the $\Lambda_c(2880)^+$.
However, this assumes the parity assignment of the latter is positive. We note that the LHCb analysis was made before the Belle report on the $\Lambda_c(2910)^+$ which could complicate the fits to the data. Although the $\Lambda_c(2860)^+$ has not been observed in, 
for instance, $e^+e^-$ continuum reactions, this particle 
could be ``hiding'' in the 
rapidly changing phase space close to the kinematic threshold. 
It has not been verified that it is a $\Lambda_c$ and not a $\Sigma_c$.

\item
The $\Lambda_c(2910)^+$ clearly needs confirmation. Whatever spin-parity is assigned to it, 
there needs to be an explanation why it is singled out in $B$ meson decays
but the other nearby states have not been observed there.

\item
The $\Lambda_c(2940)^+$ is particularly troubling. Clearly seen in several experiments, 
it decays to both $pD^0$ and $\Sigma_c(2455)\pi$ but apparently not to
$\Sigma_c(2520)\pi$, as there is no obvious peak in Fig.~\ref{Figure:Lambda2765-2880} which is taken from the Belle Conference report~\cite{Belle:2019bab}, 
though there are no hard numbers available. Its natural width would mean
that the state would extend above the threshold of decay into $pD^{*0}$, an experimentally difficult channel to investigate thoroughly 
(though BaBar has reported that they did not find the state
in their analysis~\cite{BaBar:2006itc}). 
It is possible that this nearby threshold explains why the observed mass spectrum of the $\Lambda_c(2940)$ 
does not always appear to be consistent with a Breit-Wigner 
function.
\end{itemize}

\begin{figure}[t]
\caption{\label{fig:cb}Comparison of the charmed baryons and bottom baryons spectra. Note that isospin splitting is ignored, and the base line is given as the weakly
decaying ground state for the $\Lambda$ and $\Xi$, and the spin-weighted average of the lowest two states for the $\Xi_c^{\prime}$ and $\Omega$. The position 
of the first excited state of the $\Omega_b$ (shown as a dashed line) is assumed.}
\vspace{2mm}
\centerline{\includegraphics[width=1.0\textwidth]{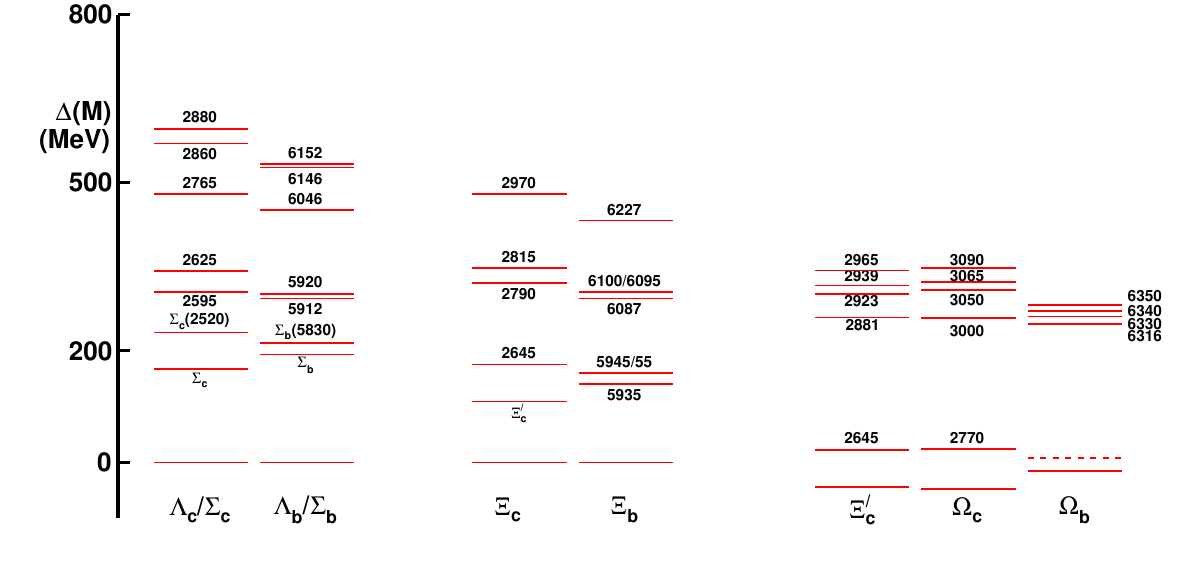}}
\end{figure}

\subsubsection{Summary of the charm-bottom comparison}
In Figure~\ref{fig:cb}, we demonstrate the symmetry found between the charm and bottom sectors. We note that the hyperfine splitting being inversely proportional to the heavy quark mass, is obeyed in a large number of different states. In addition, the expected rather small decrease in the excitation energy associated with a  $\lambda$~excitations is also apparent. These patterns appear to exist even for particles for which the precise spin-parity is not known - and thus, for any identification of a state, we must look at both spectra simultaneously. As shown in Fig.~\ref{fig:cb}, the masses of no fewer than 16 excited bottom baryons could have been well-predicted using these simple rules just by extrapolating from the charmed baryons. The bottom spectrum has understandably been slower to fill up than the charm one, and we will have to wait to see if there is a bottom analogue of, for instance, the $\Lambda_c(2940)$.